%% file: main.tex
\documentclass[10.5pt,twoside]{report}

\newcommand{\reporttitle}{DeepTrust: A Reliable Financial Knowledge Retrieval Framework For Explaining Extreme Pricing Anomalies}
\newcommand{\reportauthor}{CHAN Pok Wah}
\newcommand{\supervisor}{Dr. Ovidiu Șerban}
\newcommand{\secondsupervisor}{Prof. Francesca Toni}
\newcommand{\degreetype}{Computing~(Management~and~Finance)}

\input{includes}


\date{September 2021}

\begin{document}

\input{titlepage}

\pagenumbering{roman}
\clearpage{\pagestyle{empty}\cleardoublepage}
\setcounter{page}{1}
\pagestyle{fancy}

\begin{abstract}

\onehalfspacing 

Extreme pricing anomalies may occur unexpectedly without a trivial cause, and equity traders typically experience a meticulous process to source disparate information and analyze its reliability before integrating it into the trusted knowledge base. Traditionally, for financial data providers, their primary data sources are authoritative partnered news agencies with implicit trust, while there emerge needs to uncover insights from non-standard data sources like Twitter for its responsiveness and enrichment. We introduce DeepTrust, a reliable financial knowledge retrieval framework on Twitter to explain extreme price moves at speed, while ensuring data veracity using state-of-the-art NLP techniques. Our proposed framework consists of three modules, specialized for anomaly detection, information retrieval and reliability assessment. The workflow starts with identifying anomalous asset price changes using machine learning models trained with historical pricing data, and retrieve correlated unstructured data from Twitter using enhanced queries with dynamic search conditions. DeepTrust extrapolates information reliability from tweet features, traces of generative language model, argumentation structure, subjectivity and sentiment signals, and refine a concise collection of credible tweets for market insights. The framework is evaluated on two self-annotated financial anomalies, i.e., Twitter and Facebook stock price on 29 and 30 April 2021. The optimal setup outperforms the baseline classifier by $7.75\%$ and $15.77\%$ on F0.5-scores, and $10.55\%$ and $18.88\%$ on precision, respectively, proving its capability in screening unreliable information precisely. At the same time, information retrieval and reliability assessment modules are analyzed individually on their effectiveness and causes of limitations, with identified subjective and objective factors that influence the performance. As a collaborative project with Refinitiv, this framework paves a promising path towards building a scalable commercial solution that assists traders to reach investment decisions on pricing anomalies with authenticated knowledge from social media platforms in real-time. \smallbreak

\singlespacing 

\end{abstract}

\section*{Acknowledgments}

\onehalfspacing 

I would first like to express my sincere gratitude to my thesis supervisor Dr. Ovidiu Șerban of the Data Science Institute at Imperial College London. He is consistently being supportive of my project, and always welcomed me to raise questions whenever I was facing obstacles in project designs or challenges in implementations. He also generously offered me the flexibility in accomplishing this project in alignment with my research interest, and kindly steered me in the right direction throughout the project. Finally, I am grateful for all the invaluable insights and feedback he had provided for me to understand the field of natural language processing in greater depth, and it is my honour to be in his guidance to accomplish this exciting project. \bigbreak

I would also like to thank the experts from the industry partner Refinitiv, who were continuously supporting the project with curated historical news archives and other enriched unstructured data from industry-leading solutions Refinitiv Eikon and PermID, and offer regular feedback from end-users ensuring DeepTrust align with consensual objectives. The critical success factor behind project DeepTrust is the continuous collaboration between academia and corporation that facilitate best practice for building a practical customer-centric solution, which can truly enable institutional investors in discovering new insights from the market. \bigbreak

I would also like to acknowledge Prof. Francesca Toni of the Computing Department at Imperial College London as the second reader of this thesis, and I am gratefully benefited from her comments and suggestions on the project. Without her expertise and insightful input on the research area of argumentation, the proposed framework could not have been accomplished the milestone as of now. \bigbreak

Finally, I must gratefully acknowledge the continuous encouragement and unfailing support from my parents, my friends, and my partner Ziwi, through all the difficult times I have struggled during the project. The accomplishment I have made with DeepTrust would not have been possible without them. \bigbreak

\singlespacing 

\clearpage{\pagestyle{empty}\cleardoublepage}

\fancyhead[RE,LO]{\sffamily {Table of Contents}}
\tableofcontents

\chapter{Introduction}

\pagenumbering{arabic}
\setcounter{page}{1}
\fancyhead[LE,RO]{\slshape \rightmark}
\fancyhead[LO,RE]{\slshape \leftmark}

\section{Motivations}

\subsection*{Rationale of the DeepTrust initiative}
Refinitiv, as a global financial data provider, is a data-driven company that develops knowledge bases of financial information from trusted data streams. To maintain its commitment to provide authenticated market data to clients, it primarily gathers real-time data from reputable global news sources with a high degree of implicit trust, specialty franchises from professional financial institutions, and multimedia content from expert partners like Reuters, IFR, CNBC, etc~\cite{refinitiv-financial-news-coverage}. \smallbreak

Recently, non-standard data sources, such as Twitter, Facebook, and other social networking services have increasingly attracted more financial professionals and data providers to extract meaningful contents from the ocean of big data. These data sources constitute an additional input of business information, which can be used to evaluate stock price expectations, monitor market sentiment~\cite{cwynar2019social}. The speedy transfer of knowledge on these platforms can also be leveraged to explain anomalies in the pricing data, much faster than waiting for an analysis report from the traditional media. However, one intrinsic problem of non-standard data sources is data veracity, in which financial data providers like Refinitiv cannot directly include information extracted from these sources as a part of its trusted knowledge base. Besides, disinformation in the financial industry may cause severe economic consequences by using deception and frauds to mislead consumers in reaching financial decisions, thus requires prudent screening on information extracted from noisy social media data streams. \smallbreak

\subsection*{Challenges and Opportunities}

Regardless of the effort made by linguistics and professional natural language processing (NLP) experts in identifying fake news, the majority of detection mechanisms or frameworks such as SpotFake~\cite{singhal2019spotfake} and EANN~\cite{wang2018eann} aim at identifying a particular form of fake news by training a discriminator using multimodal features (e.g., textual contents and visuals), as information extracted from each modality can complement each other in evaluating information authenticity. However, when the model is learning visual representations, this knowledge is usually domain-specific, and limits the flexibility of applying to a different domain using these non-transferable feature representations. Besides, in the context of the financial industry, identifying unreliable financial information is more challenging because a post can be deceiving by embellishing an event by quoting out of context, and is also difficult to collect a sufficient amount of validated fake financial information from trusted data sources such as Securities and Exchange Commission (SEC). In addition, fake news is only a subset of unreliable information as will be discussed in Section~\ref{sec:overview}, whereas unconfirmed reported speech and subjective personal opinions should also be filtered out from the trusted knowledge base. Therefore, it is the responsibility of financial data providers to deploy additional validations, to ensure the extracted knowledge from Twitter is beyond \emph{"Not Fake"}, but is \emph{"Reliable"}. \smallbreak

Different from existing works, the present project proposes a reliable information extraction framework named DeepTrust. DeepTrust enables financial data providers to effectively source and analyze disparate financial information on Twitter that could explain a sudden price anomaly, and apply information processing and validation techniques to preserve only reliable knowledge that contains a high degree of trust. The prime novelty of DeepTrust is the integration of a series of state-of-the-art NLP techniques in retrieving information from a noisy Twitter data stream, and assessing information reliability from various aspects, including the argumentation structure, engagement metrics, neural generated text traces, and text subjectivity. The DeepTrust is comprised of three interconnected modules: \emph{(i)} Anomaly Detection module, \emph{(ii)} Information Retrieval module, \emph{(iii)} Reliability Assessment module. All modules function in sequential order within the DeepTrust framework, and jointly contribute to achieving an overall high level of precision in retrieving information from Twitter that constitutes a collection of trusted knowledge to explain financial anomalies. Solution effectiveness are evaluated both module-wise and framework-wise to empirically conclude the practicality of the DeepTrust framework in fulfilling its objective.  \smallbreak

\section{Objectives and Contributions} \label{sec:objectivies-and-contributions}

In this project, the author collaborated with Refinitiv in developing a financial information retrieval and assessment framework DeepTrust, to assist analysts and equity traders in understanding abnormal price movements in the securities market using information retrieved from social media at speed. The primary objective of DeepTrust is to establish a workflow that identifies unexpected price moves, explains the cause using unstructured data from non-standard data sources such as social media posts from Twitter, and ensures the basis of these information are reliable. In particular, DeepTrust has three essential purposes: \emph{(i)} discover anomalous price movements using anomaly detection algorithms and machine learning models trained with historical data \emph{(ii)} retrieve correlated information (e.g., industrial, company-specific, regional, etc.) from social media platform Twitter \emph{(iii)} analyze and refine an implicitly trusted collection of tweets that are reliable for explaining the financial anomaly. \smallbreak

The project has three main contributions. Firstly, the prime novelty of DeepTrust is to incorporate numerous state-of-the-art pre-trained language models of different specialties (e.g., few-shot learners, domain-specific language models, sequence labeling models), and extrapolate information reliability using textual and meta features, traces of language model footprints, argumentation structure and subjectivity. DeepTrust explores the topic of financial information credibility assessment on social media platforms, and goes beyond fighting against bot-generated spam to more complex forms of unreliable knowledge such as opinion pieces and disinformation. \smallbreak

In addition, though there exists a large number of datasets for downstream NLP tasks like synthetic text detection and argument mining, an annotated dataset on information reliability of tweets in the financial domain has never been published. Considering the peculiarities of Twitter-specific features and writing styles, the author annotated two collections of tweets from different anomalous price movement events based on their perceived credibility, and evaluate the proposed DeepTrust framework accordingly. Although these datasets only contain approximately $1,700$ tweets due to limited resources available, the established definition of financial information reliability and annotation guideline paves the way for future researchers in designing their own enhanced reliable knowledge screening system, or benchmarking their framework on the task of tweet reliability analysis. \smallbreak

Lastly, as a collaborative project with Refinitiv, the proposed DeepTrust framework architecture and experimental results contribute valuable input to Project Mosaic\footnote{Project Mosaic from Refinitiv Labs: \url{https://www.refinitiv.com/en/labs/projects/mosaic}.}, which is an ongoing prototype developed by Refinitiv Labs that uncover insights into extreme asset price movements. Unlike DeepTrust, project Mosaic focuses on generating actionable knowledge on anomalous financial events instantly using data from multiple data sources, while DeepTrust goes beyond information retrieval and applies reliability assessment techniques using state-of-the-art language models, to guarantee only credible information is presented to users. DeepTrust is implemented in adherence to standard software engineering principles, including the adoption of modular programming and microservices architecture for code maintainability and testability. In addition, experimental results in Chapter~\ref{cha:evaluation-and-discussion} are insightful for Refinitiv Labs to further improve the existing framework following analytical results and proposed suggestions, and continue training classifiers and language models for better performance, with the large amount of proprietary financial datasets (e.g., machine readable news) at their disposal. \smallbreak

\chapter{Background}

The present study forms a pragmatic approach in assessing financial information reliability on Twitter. Hence, in the following, an overview of terminologies, concepts, and forms of financial data unreliability on Twitter is provided in Section \ref{sec:overview}. A review of common anomaly detection and information retrieval techniques is discussed in Section~\ref{sec:adfm} and~\ref{sec:ced}. Section~\ref{sec:ra} reviews a series of reliability assessment techniques, particularly focuses on Twitter-specific benchmarks that are composed of short text with informal languages. \smallbreak

\section{Overview of Financial Information Reliability} \label{sec:overview}

Non-standard data sources, with their representatives such as Twitter, Facebook, and other online social networking services, have provided a wider channel for non-traditional journalists to share information and engage with a mass audience~\cite{wall2015citizen}. One prominent example is Twitter, which claimed its users sent more than $500$ million tweets daily as of December 2014, the latest release of its official usage statistics~\cite{twitter}. Regardless of the benefit brought by Twitter such as speedy information exchange and free information dissemination, this channel of information blurs the conceptualization of an authenticated, verified information source, as well as facilitates the spread of unverified or even falsified information~\cite{shu2017fake}. The following sections will firstly define unreliable and fake in the finance industry, along with some common forms of disinformation on Twitter. \smallbreak

\subsection{Definition and Categories of Unreliable Financial Information}

For traditional news from an agency in authority, a trusted piece of information is commonly defined as an accurate representation of a real event~\cite{kershner2012elements}. Whereas reliable information on social media, such as a character-limited post or multimedia content, should also adhere to this standard. For financial data providers, the knowledge in association with a financial event should be authenticated, verified, and objectively described, and it is their responsibility to guarantee the data reliability for their clients. \smallbreak

Unreliable information, on the other hand, represents an instance of misinformation that can be either unverified or contradicted with the reality~\cite{wardle_2018}, while the term "unreliable" is frequently misused interchangeably with "fake" \footnote{Unreliable usually means misleading or not credible, aiming at a broader scope, while fake can be perceived as a subset of unreliable that refers to the state of fraudulent or not genuine. For DeepTrust, the objective is to identify fake information, while also extending the framework to cover certain forms of unreliable information.}. Tandoc et al. further subdivide the term "fake" into six "operationalizations" based on its facticity and intention, in which categories \emph{Fabrication} and \emph{Advertising} are closely associated with the finance industry~\cite{tandoc2018defining}. The fabrication refers to information with intention of misleading or deceiving, contains no factual basis but written in a style similar to authoritative news articles to establish legitimacy. Fabricated information is usually accompanied by a bot-based network that automatically spread the identical or a variant of these fake news using bogus accounts to reinforce the illusion that it is authenticated~\cite{albright_2017}. For instance, fabricated financial statements with erroneous content can be used to manipulate stock prices by encouraging retail investors to sell promising equities based on misleading information. On the other hand, advertising means promotional materials whose true intent is under the guise of genuine news reports. For instance, undisclosed sponsored articles are commonly used by malicious actors to execute the pump-and-dump schemes, causing a huge deficit for investors who trusted this misinformation. Overall, both operationalizations, as discussed in the typology of "fake"~\cite{tandoc2018defining}, originated from the author's immediate intention to deceive audiences, but differs from the level of facticity in its content. \smallbreak

\subsection{Forms of Unreliable Financial Information}\label{sec:foufi}

\subsubsection*{Neural Disinformation - Fabrication}

Neural disinformation refers to fraudulent content created by a neural language model, which aims at generating realistic-looking content using large-scale generative models such as GPT-2, GROVER and BERT. The objective is to mimics the style of an authenticated information from an authoritative party, whereas the content itself is either viral or deceiving with a malicious intention. \smallbreak

For instance, in the Suez Canal traffic incident caused by cargo ship blockage starting from 21 March 2021, the crude oil price has surged $4.54$\% possibly due to fears of crude oil supply disruptions~\cite{al_jazeera_2021}. This financial anomaly was discussed by Bloomberg Economics on 23 March 2021, stating that \emph{"A realignment of global crude flows has seen westbound shipments from the Gulf tumble, making the Suez Canal far less important"} \footnote{Twitter post URL: \url{https://twitter.com/economics/status/1375177958197956613}.}. Neural fake news can be simultaneously generated using solely this information, and result in another compelling story regarding this anomalous event, such as \emph{"Suez Canal also opening up the possibility that new oil-by-rail routes to Canada or the US ,"}\footnote{Generated using GPT-2 XLarge model with 1.5B parameters, with 40-words diversity and 0.8 temperature. Selected content is the most compelling piece out of 40 generated samples.}. Without decent domain knowledge and rigorous fact-checking, it can be challenging even for a well-educated person to distinguish which one is the authenticated piece of information. \smallbreak

Intuitively speaking, neural fake news generally works better without length constraint, such as a news article or an interview script, because more fabricated contextual information can be included in the passage to build up a realistic-looking story. However, for Twitter, neural disinformation can be identified easily due to the character-limit restriction, as most often neural fake tweets are not compelling enough to deceive a human. \smallbreak

\subsubsection*{Human-written Disinformation - Fabrication}

Human-written disinformation refers to untruthful information written by human beings with malicious intentions, and is commonly used in the financial market to manipulate stock prices via misleading investors with falsified company-specific information. This type of disinformation is by far the primary source of fake news existing on social media platforms~\cite{vargo2018agenda}. \smallbreak

\begin{figure}[!htbp]
\centering
\includegraphics[width = 0.6\hsize]{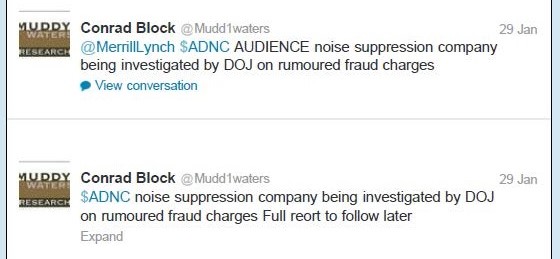}
\caption{Craig's fabricated financial information on Twitter. He maliciously used the logo and style of the securities research firm Muddy Waters, and falsely accused Audience Inc. on 29 January 2013 (Source:~\cite{sec_emblem_2015})}
\label{fig:fake_tweets}
\end{figure}

For instance, SEC filed securities fraud charges against James Alan Craig in 2014 about publishing a series of tweets that falsely accusing Audience Inc. was under investigation, as shown in Figure \ref{fig:fake_tweets}. The direct consequence is the Audience Inc.'s stock price crashed and triggered a trading halt shortly after these tweets. This type of disinformation is hard to identify because it is nearly impossible to distinguish this with an authenticated posts from the real Muddy Waters unless examining the account profile. In fact, the SEC also acknowledged that it is challenging to identify this perpetrator~\cite{sec_emblem_2015}. \smallbreak

Therefore, for the consideration of feasibility, DeepTrust aims to primarily detect human-written disinformation that is \emph{(i)} from a personal account without disguise and \emph{(ii)} written in his/her own linguistic style. \smallbreak

\subsubsection*{Opinion Piece - Subjective Statement} 

An opinion piece is a subjective expression that represents an individual's perspectives and judgements~\cite{xuan2012linguistic}. Regardless it is riddled with logical fallacies or polarized subjective statements, an opinion piece alone does not render itself fake information~\cite{Ireland2018}. It is merely unreliable information that lacks an adequate chain of evidence to reinforce its validity. In other words, if the opinion piece has clearly identified the author and its responsibility to be held accountable for its statement, it should only be treated as knowledge that needs more scrutiny before being deemed reliable. \smallbreak

One example of misusing opinion pieces for malicious intention is the article \emph{"Partisan, Moro pushes Brazil’s democracy in the abyss"}~\cite{weisbrot_2018}. This article was published by Mark Weisbrot in New York Times (NYT) in January 2018, which is an opinion piece that does not reflect the newspaper's political stance. However, left-wing media quickly leveraged this opinion piece by taking it out of context and used it as a representation of NYT's editorial line, portraying a negative image of Judge Sérgio Moro~\cite{valente2018}. \smallbreak

Besides, it is worth noticing that subjectivity is not equivalent to opinionated, which refers to sentences explicitly or implicitly express a sentiment, such as \emph{"The earphone malfunctioned after two days"} is an objective statement but expressing negative sentiment~\cite{liu2012sentiment}. Therefore, in this project, an opinion piece is treated as unreliable only if \emph{(i)} the claim directly contradicts evidence from trusted sources (e.g., news agencies) or \emph{(ii)} primarily composed of subjective statements regardless of the sentiment. \smallbreak

\section{Anomaly Detection in Financial Market}\label{sec:adfm}

In an ideal world, the market should constantly operate under the efficient market hypothesis, implying that all securities (e.g., stocks, options, commodities) are priced perfectly based on their inherent investment properties, which are knowledge available to all participants. In other words, securities should always be traded at their fair value regardless of market timing or stock selection, and investors should never be able to create an arbitrage situation by reselling undervalued securities for higher prices without incurring additional risks. \smallbreak

However, in reality, the market is rarely operating at its perfect efficiency, and the presence of anomalies can lead to distortions and inefficiency in the market. In~\cite{yavrumyan2015efficient}, Yavrumyan defined three categories of financial anomalies in the securities market: \emph{(i)} technical anomalies - anomalies that relate to abnormal trading behaviors, that enable investors to leverage technical analysis of historical pricing data to predict the pattern of stock price changes \emph{(ii)} fundamental anomalies - anomalies in association with fundamental information of the listed company that allows investors to forecast the stock price using fundamental analysis \emph{(iii)} calendar anomalies - systematic anomalies that caused by the seasonality of stock price movements that allow investors to identify the price changing pattern. In addition, in the domain of data science, the anomaly can be classified into three categories: \emph{(i)} global anomaly - a single data point that significantly deviates from the rest of data points. \emph{(ii)} collective anomaly - a subset of data points that exhibit peculiar behavior comparing to the whole dataset. \emph{(iii)} contextual anomaly - individual or group of data points that deviates within a specific context (e.g., a time period)~\cite{han2011data}. In DeepTrust, the fundamental anomaly is the focus because the objective is to identify abnormal pricing changes that can be explained by information retrieved from non-standard data sources like Twitter. Besides, our model emphasizes identifying contextual anomaly, because generally, breaking news would lead to a sudden price change within a time period, and this anomaly is difficult to detect when searching for global or collective anomalies.  \smallbreak 

Anomaly detection refers to algorithms that identify abnormal data points from a given dataset. In the context of this project, the anomaly can either be detected by financial experts or by anomaly detection algorithms. Anomaly detection algorithms can be classification-based, clustering-based, nearest neighbor-based, statistical-based and others~\cite{golmohammadi2016time}. In an overview of present anomaly detection algorithms in the stock market, the supervised algorithms (e.g., Naive Bayes, K-Nearest Neighbour KNN, Support Vector Machine SVM and Class Outlier Factor COF) achieved a significantly better result comparing to unsupervised algorithms such as Local Outlier Factor (LOF) and other density-based or distance-based methods. Rajeswari et al. further evaluated these algorithms against different types of anomalies, and concluded that \emph{(i)} all supervised algorithms can detect 80\% anomalies, and KNN achieved the best F-measure comparing to others \emph{(ii)} density-based and LOF algorithms performed identically well in the unsupervised category, but is only able to detect 58\% anomalies correctly~\cite{rajeswari2018comparative}. \smallbreak

In most recent literature, unsupervised anomaly detection algorithm becomes the mainstream due to its capability in handling previously unseen anomalies, and does not need an annotated data set for training. These algorithms are effective in identifying global and collective anomalies. In~\cite{ahmed2014outlier}, a detailed evaluation of unsupervised anomaly detection algorithms is discussed, ranging from clustering-based nearest neighbor approaches like k-Means and Local Density Cluster-based Outlier Factor (LDCOF)~\cite{amer2012nearest}, to SVM-based approaches such as C-SVM and one-class SVM~\cite{lin2008anomaly}. Among all these algorithms, Ahmed and Uddin evaluated their performance on financial anomalies detection~\cite{ahmed2017anomaly}. Algorithms were evaluated using precision, recall, F-measure and false-positive rates (FPR), then revealed that LOF and Clustering-based Multivariate Gaussian Outlier Score (CMGOS) achieved the highest F-measure comparing to all competitors. However, none of these unsupervised anomaly detection approaches outperform a random classifier, indicating the trade-off between robustness and accuracy. \smallbreak

Another variant of the anomaly detection algorithm is prediction-based Contextual Anomaly Detection (CAD) proposed by Golmohammadi and Koosha~\cite{golmohammadi2015time}. The system firstly extracts a subset of time series and calculates the centroid (i.e., average stock value over this time period), as the expected value of this subset of data. This centroid is then used along with the correlation score between each time window to predict the expected values of the next time window. Comparing with supervised clustering algorithms, CAD improved the recall by 26\% while maintaining the same level of precision. \smallbreak

In addition, the Early Aberration Reporting System (EARS) is a widely adopted enhanced surveillance system in the infectious disease domain that can analyze syndrome monitoring data in public health to identify bioterrorism in its early phase~\cite{hutwagner2003bioterrorism}. EARS aims to detect aberration, which in the context of public health, refers to the change in the frequency distribution when comparing to historical data (i.e., ranging from 9-days to 3-years). EARS consists of algorithms targeting different scenarios, including long-term (i.e., an extended baseline for \numrange{3}{5} years, a limited baseline for 7 days to 5 years) and short-term(i.e., recent data for less than 30 days and short-term for \numrange{1}{6} days). The system also allows user-specified sensitivity and specificity estimates to configure the system that is customized for each scenario. EARS is available in forms of its variants C1, C2, C3 and NB, which based on cumulative sum (CUSUM) algorithms that identify an anomaly if the observed count exceeds the baseline count mean by a multiple of standard deviations. \smallbreak

\section{Correlated Events Detection}\label{sec:ced}

Starting from a financial anomaly in the pricing data, the objective is to identify correlated posts on Twitter that can serve as supporting evidence of this anomalous event, or commonly known as specified event detection. Posts may be deemed correlated based on \emph{(i)} textual-content analysis to identify relevant keywords \emph{(ii)} social network analysis on the network topology with follower-followee patterns \emph{(iii)} behavior analysis on individuals and the metadata of his/her posts. The present project focuses on the analysis of textual content and behavior data to infer information correlation with the interested anomalous event. \smallbreak

Two main categories of event detection are namely: \emph{(i)} Retrospective Event Detection (RED) by clustering accumulated set of documents into events, commonly known as offline event detection \emph{(ii)} First Story Detection (FSD) by allocating documents from a live stream of data into the closest event, known as online event detection. In the context of the DeepTrust, both RED and FSD are applicable because the system can either start from known knowledge from a trusted knowledge base or starts from filtering a live data stream so only tweets that are related to the event are recorded. The following sections review topic-detection and tracing mechanisms and event-oriented text retrieval techniques for gathering correlated knowledge of a given financial anomaly from the Twitter data stream. \smallbreak

\subsection{Topic-Detection and Tracking (TDT)} \label{sec:tdt}

Topic-Detection and Tracking (TDT) is the widely adopted framework to establish a timeline of textual contents that are centered around a single event, and to associate events with textual content by analyzing the temporal pattern of keywords. It involves three stages \emph{(i)} segmentation by splitting texts into chunks \emph{(ii)} detection by identifying the first occurrence of a novel event \emph{(iii)} tracking by constructing an evolution path of the detected event over the temporal dimension \cite{allan1998topic}. TDT algorithms can be classified broadly into document-pivot and feature-pivot paradigms. The document-pivot approach refers to document-level clustering based on the pairwise semantic distance among documents~\cite{bennett2002exploiting}, and the feature-pivot approach detects events using word and topic distributions per document, results in a cluster of features~\cite{yang1998study}. 

\subsubsection*{Document-Pivot Paradigm (DPP)}

DPP aims to cluster documents into separate events based on pairwise document similarity scores. For DeepTrust, DPP can rapidly cluster documents into different events, and then locate the correlated cluster of events by calculating the similarity score between the given financial anomaly and clusters. In early research, a single-pass incremental algorithm with a document frequency threshold model is used to cluster broadcast news stories into different events~\cite{allan1998line}. Brants et al. further improved the model by using incremental Term Frequency-Inverse Document Frequency (TF-IDF) and normalized similarity scores among documents, which results in greater recall and precision~\cite{brants2003system}. Zhang and Li applied cosine distance on Vector Space Model (VSM) topics representation model, but improvement (i.e., $38.378$\% increase in topic detection rate) is only significant on large-scale corpus~\cite{zhang2011topic}. All these approaches have achieved an overall good performance on traditional media. \smallbreak

Besides, extensive research efforts were made on improving DPP performance on social media such as Twitter with multimodal analysis. Becker et al. proposed clustering documents using multiple types of unique Twitter features such as time, location and other metadata, then assign specific similarity metrics (e.g., cosine similarity for content and Haversine distance for metadata) on each feature~\cite{becker2009event}. In their later research, the performance was improved by leveraging similarity metrics across multimodal documents as learnable parameters, and use logistic regression to classify if a pair of documents belong to the same event or not~\cite{becker2010learning}. In other words, the single-pass incremental algorithm is used in conjunction with the multimodal similarity metrics. Reuter and Cimiano simplified the TDT algorithm by reducing the number of comparisons needed when evaluating pair-wise document similarity~\cite{reuter2012event}. Originally, the number of comparisons grows exponentially with more modalities used as a feature. Instead of computing a combined pair-wise similarity score, \cite{reuter2012event} proposed to retrieve a set of closest events with respect to each feature and compare them separately. A Support Vector Machine (SVM) classifier was also used to pre-classify incoming documents into \emph{(i)} belongs to top bursty event \emph{(ii)} belongs to another novel event. The scalability of DPP is further enhanced in Hossny et al., which amplified keyword signals by mapping semantically correlated words with similar temporal patterns into a single word via Singular Value Decomposition with K-means (K-SVD), with an average $0.38$ increase on the document-level correlation scores~\cite{hossny2018enhancing}.  \smallbreak

\begin{figure}[!htbp]
\centering
\includegraphics[width = 1\hsize]{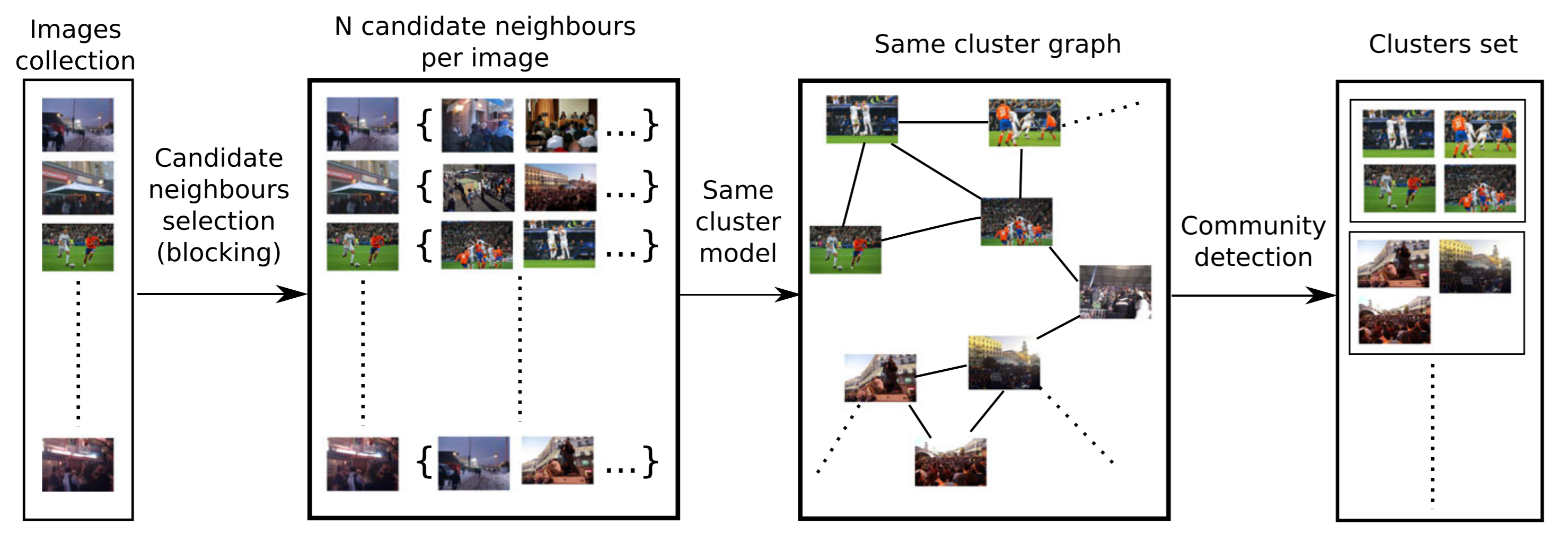}
\caption{Schematic diagram of the graph-based multimodal clustering technique using a Same-Cluster model (Source:~\cite{petkos2017graph}).}
\label{fig:graph-based-clustering}
\end{figure}

Other than incremental clustering, recent researches also utilized the graph-based multimodal clustering technique, by representing documents as vertices and pairwise similarity score as weighted edges. Clustering is achieved by identifying an optimal partition scheme based on an objective function, which determines edges to be removed. Additional modalities such as visual contents, uploader's identity are included to aid the analysis of similarity among documents. For instance, in~\cite{petkos2017graph}, as shown in Figure~\ref{fig:graph-based-clustering}, candidates are firstly retrieved from the collection of documents represented as a multimodal item. The example clustering technique was used to train an SVM classifier, known as the Same-Cluster (SC) model, in determining if sets of multimodal documents belong to the same cluster, based on per modality distance. With the generated SC graph, two popular graph partitioning algorithms, batch-based algorithm Structural Clustering Algorithm for Networks (SCAN) and incremental-based algorithm Quick Community Adaption (QCA) were used to extract finalized clusters. \smallbreak

The latest DPP-based event detection framework was proposed by Hasan et al. as TwitterNews+, an online event detection system consists of a Search component and EventCluster component~\cite{hasan2019real}. The Search module combines Random Indexing based term vector model~\cite{petrovic2010streaming} and Locality Sensitive Hashing (LSH) scheme~\cite{sahlgren2005introduction}, to quickly hash similar tweets into the same buckets in a reduced dimension, then determine if an incoming tweet relates to a known topic or not. The EventCluster module applies the classic TDT technique (i.e., single-pass incremental clustering) to group tweets into events. A filter based on the user-specified topic is then applied to retrieve relevant tweets. The TwitterNews+ application has achieved a significant improvement in terms of precision and recall after an exhaustive hyper-parameter tuning. \smallbreak

\subsubsection*{Feature-Pivot Paradigm (FPP)}

FPP is another TDT technique based on the analysis of feature patterns, in which the feature can either be content features or contextual features. For Twitter, the content features refer to words, sentiments and linguistic styles in the 280 characters-long unstructured textual data. The contextual feature refers to the metadata of the tweet, including the uploader's identity and network information, geographical and temporal information, etc. The conceptualization of FPP is that events are composed of prominent text features that exhibit abnormalities in temporal patterns, and the occurrence of features would increase when correlated events had happened. For example, the term "interest rate" shall occur frequently when discussing the drastic tech stocks correction caused by a bond yield spike in March 2021.  \smallbreak

Before extracting features, Tembhurnikar and Patil discussed the importance of appropriate text preprocessing in achieving a high-quality topic detection~\cite{tembhurnikar2015topic}. A combination of tokenization, slang word translation, normalization and temporal aggregation through merging correlated tweets were used, primarily focused on improving feature extraction performance on information short text on Twitter. These techniques were proved effective by alleviating ambiguity caused by the informal nature of the language used in tweets. \smallbreak

\begin{figure}[!htbp]
\centering
\includegraphics[width = 1\hsize]{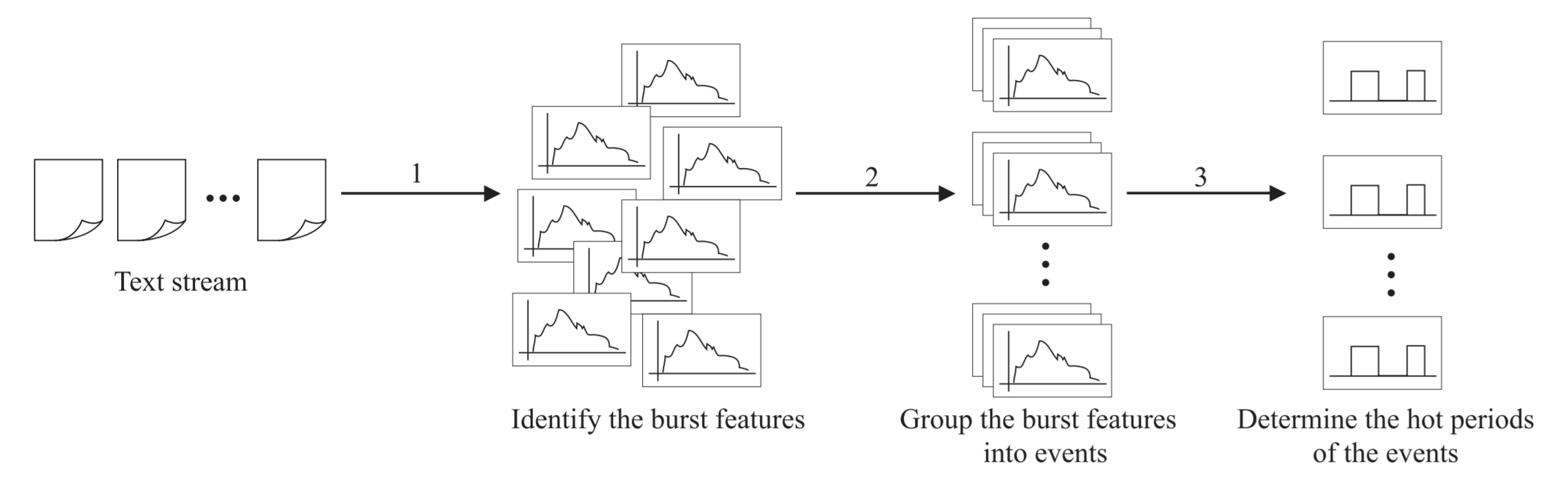}
\caption{Feature-Pivot clustering process of an enduring event based on bursty feature identification (Source:~\cite{fung2005parameter}).}
\label{fig:feature-pivot-clustering}
\end{figure}

In~\cite{fung2005parameter}, Fung et al. proposed a feature-pivot clustering algorithm to detect bursty features using time-series feature distributions from a sequence of chronologically ordered text documents. The overview of the event detection pipeline is described in Figure~\ref{fig:feature-pivot-clustering}. Bursty features are identified using a generative probabilistic model on a binomial distribution. In short, a feature is concluded bursty if the number of tweets containing it is greater than the expected value of the modeled distribution. Although this research focused on traditional media TDT, it served as the foundation of bursty feature identification in later research. \smallbreak

Sakaki et al. was recognized as the pioneering research of FSD-based FPP on Twitter, by using statistical and contextual features (e.g., length of the tweet, word choices, and neighboring words) to train an SVM classifier to determine if an incoming tweet belongs to natural disaster events or not~\cite{sakaki2010earthquake}. However, this approach is limited in detecting solely earthquake-related tweets, heavily relies on common lexical features correlate to natural disasters. Whereas, for financial anomalies, the scope of potential word choices is much larger, thus making this approach ineffective. \smallbreak

\begin{algorithm}[!htbp]
  \caption{Topic Clustering and Tweet Retrieval Algorithm. Source: \cite{benny2015keyword}.}\label{alg:TCTR}
  \begin{algorithmic}[1]
      \State Sort all topics $t_{i} \in TOPICS$ in descending order on keywords per topic.
      \State Calculate pairwise common words $com = no.common words(t_i, t_j)$ for $i\neq j$.
      \For{\texttt{$t_{i} \in TOPICS$}}
        \For{\texttt{$t_{j} \in TOPICS$} where $j>i$}
        \State $w_j = $ keywords in $t_j$.
         \If {$com > (w_j + 1) / 2$}
            \State Cluster $t_i$ and $t_j$.
         \EndIf
        \EndFor
      \EndFor
  \end{algorithmic}
\end{algorithm}
\smallbreak

The framework proposed by Chen et al. is an insightful technique for DeepTrust. Once users specified keywords that might be related to an anomaly (e.g., oil price drop, interest rate, bond yield, etc.), these keywords are expanded by fusing importance, penalty factors and relevance~\cite{chen2013user}. Incoming tweets are filtered using a logistic regression classifier, and bursty words in each event are determined using the Affinity Propagation (AP) algorithm. A variant of the user-specified topic detection algorithm is proposed by Benny and Philip through evaluating the Associate Gravity Force (AGF) between each frequent pattern originated from a user-defined keyword~\cite{benny2015keyword}. Relevant tweets are retrieved using Topic Clustering and Tweet Retrieval (TCTR) algorithm as described in Algorithm~\ref{alg:TCTR}, by evaluating the number of pairwise common words among event topics. \smallbreak

Instead of detecting topics on a per-tweet basis, recent researches focused on high-granularity TDT algorithms. In~\cite{cataldi2010emerging,cataldi2014personalized}, Calteldi et al. proposed a five-stage keyword life cycle using term aging theory, to determine if a keyword is emerging or not based on its occurrence frequency in the different time window. The model also formulates a directed uploader-based graph based on the social relationships of the uploader, then evaluates his/her authority using the PageRank algorithm. For each term at each time interval, the \emph{term energy} is calculated using a nutrition formula on its frequency signal. A threshold-based model is used to filter less energetic terms. For the remaining \emph{emerging terms}, they are used to construct a navigable topic graph based on the correlation vectors calculated based on a set of weighted factors. Lastly, topics are retrieved using a graph-based clustering algorithm. For DeepTrust, these topics can then be filtered, so that only those containing keywords mentioned in the given financial anomaly are preserved for later reliability assessment. \smallbreak

Another relevant research was proposed by Guille and Farve, a novel Mention-Anomaly-Based Event Detection (MABED) approach that detects both anomalies and their magnitude of impact based on the uploader's authority~\cite{guille2014mention}. For each event, a main bursty term is identified as the event theme, while other terms with similar temporal dynamics are used for describing the event in detail. The key innovation of MABED is its ability to dynamically model the bursting time of an anomaly, instead of using a predefined fixed duration. Specifically, the burst interval is determined by formulating a Maximum Contiguous Subsequence Sum \emph{(MCSS)} problem that aims at maximizing the impact of each word. For each word $t \in V$ within $i^{th} \in [1: n]$ time window:

\begin{equation} 
Anomaly(t, i) = N_{@t}^{i} - E[t|i] \label{eq:1}
\end{equation}
\begin{equation} 
Magnitude(t, I) = max\{\sum_{i=a}^{b} anomaly(t, i)|1\leq a\leq b \leq n\} \label{eq:2}
\end{equation}

where $N_{@t}^{i}$ is the number of times word $t$ is mentioned  and $E[t|i]$ is the expected number of tweets containing word $t$. By solving this problem in linear-time, each detected event can then be described in three dimensions \emph{(i)} event theme word \emph{(ii)} bursting duration $I \in [a: b]$ \emph{(iii)} event impact magnitude $Magnitude(t, I)$, thus offers equity traders with additional insights on the anomaly and its severity based on bursting magnitude.  \smallbreak

\subsubsection*{Topic-Detection and Tracking Limitations}

\begin{figure}[!htbp]
\centering
\subfloat[Tweet from Financial Times]{{\includegraphics[width=6cm]{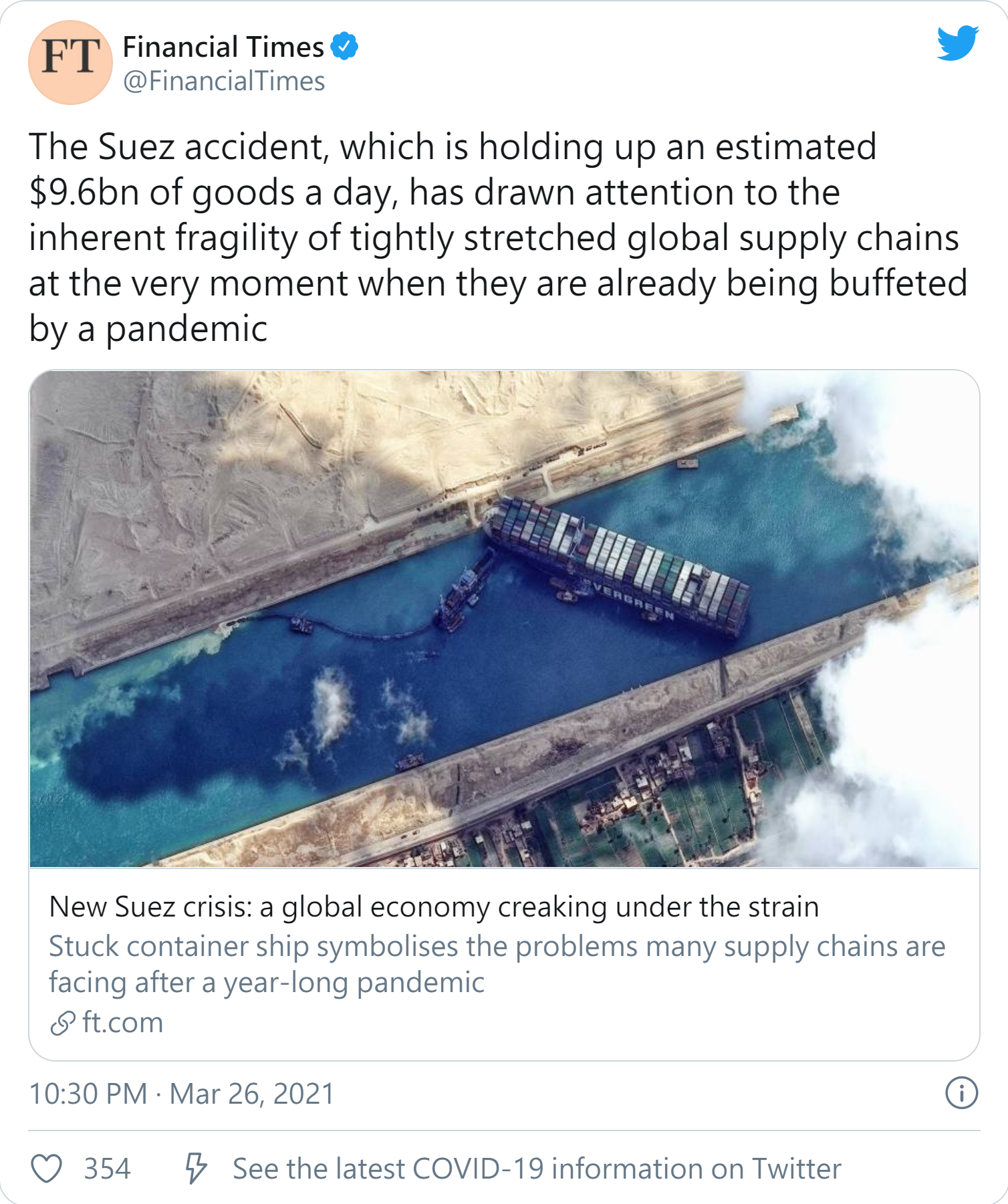}}}
\qquad
\subfloat[Tweet from Bloomberg]{{\includegraphics[width=6cm]{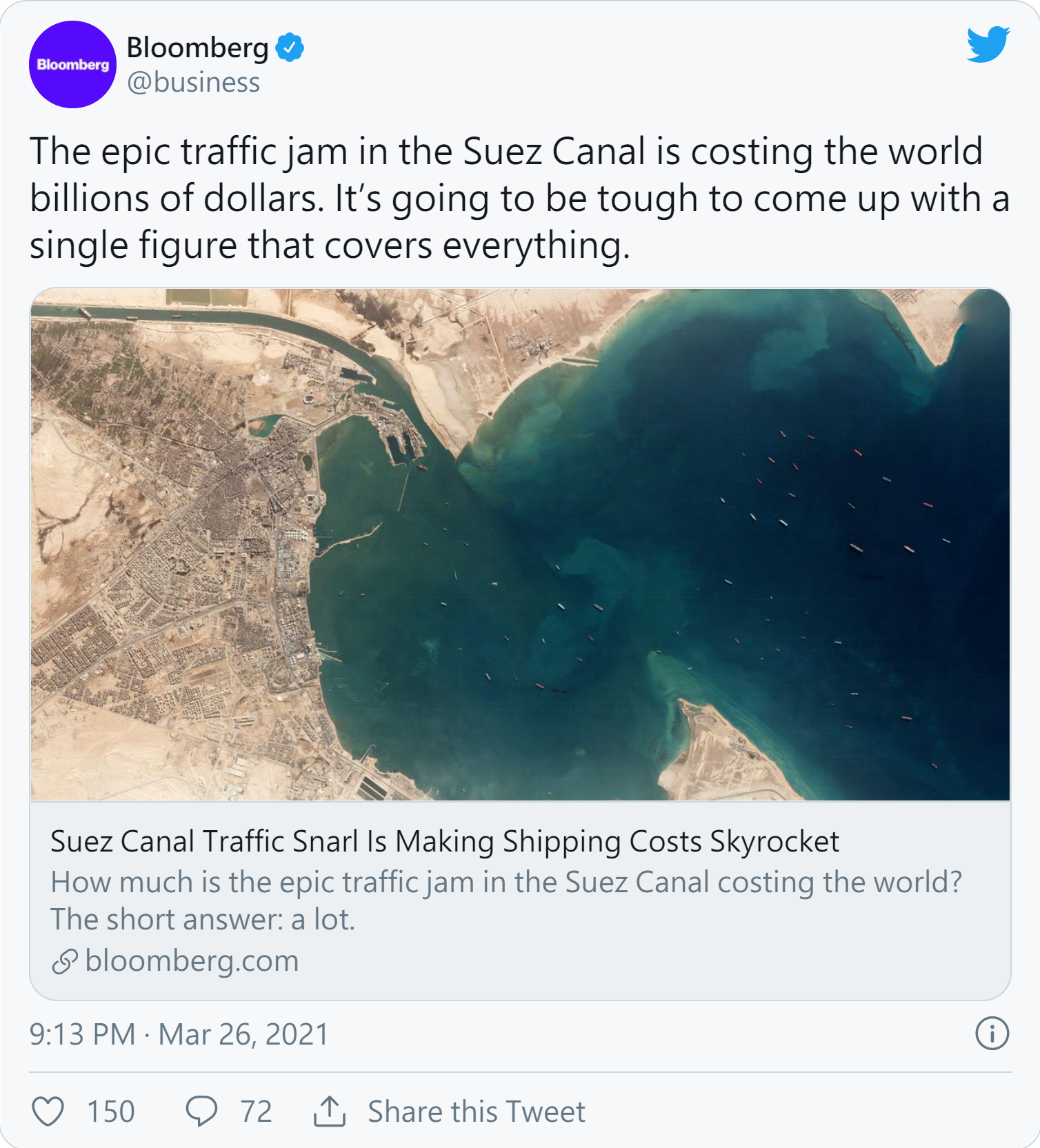}}}
\caption{Strongly-related tweets about Suez Canal traffic jam incident posted by Financial Times and Bloomberg, with low similarity score using document-pivot clustering approach}
\label{fig:DPP-tweets}
\end{figure}

Regardless of the popularity of TDT (i.e., DPP and FPP) approaches in detecting events, several limitations of DPP are identified in prior research. In~\cite{fung2005text}, experiments revealed that similarity scores are sensitive to noises, thus documents with large similarity scores are not necessarily describing the same event. Besides, as observed in~\cite{yang1998study}, the DPP technique is prone to misclassify tweets if the time span of the anomalous event is lengthy. Lastly, it is difficult to justify an appropriate threshold to decide which event is correlated to incoming data points. For instance in~\cite{becker2009event}, a threshold $\mu$ is defined to cluster documents across modalities. The document is assigned to the closest matching cluster if the similarity score is greater than $\mu$, or created a new cluster instead. However, the defined threshold may not be robust enough for handling diverse linguistic styles of individuals when describing the same subject. One example is Figure~\ref{fig:DPP-tweets}, both tweets are discussing how the Suez Canal has impacted the world financially, thus can be reasonably concluded as correlated by a human evaluator. Whereas with DPP, the n-gram similarity distance might be smaller than the defined threshold, thus be clustered into different events under threshold-based models. \smallbreak

In addition, one critical issue with FPP is its capability in detecting less-trendy events. The FPP relies on bursty features with abnormalities when a trending event is happening, whereas for less attentive events, FPP is likely to perform poorly in identifying correlated documents. Besides, the fundamental prerequisite of FPP is to obtain knowledge of the frequency pattern of features over time, and for initial documents, there is no prior knowledge available. Therefore, for FPP approaches to support FSD-based TDT tasks, an incremental-based model is necessary. \smallbreak

\subsection{Event-oriented Text Retrieval (ETR)} \label{sec:eor}

Event-oriented Text Retrieval (ETR) is a supervised learning task, using a keyword-based model to locate relevant messages based on the user-supplied structured query, represented as a set of keywords. Formally speaking, given a list of $n$ event-specific queries $Q = \{q_1, q_2, ..., q_n\}$ and $m$ tweets $T = \{t_1, t_2, ..., t_m\}$, the objective is to identify a list of $i$ correlated tweets $D = \{d_1, d_2, ..., d_i\} \in T$ that are related to $q_j \in Q$. Comparing to TDT, this approach can directly locate relevant tweets from a noisy data stream, but is also challenging to sanitize user-supplied input and match them with event representations. For the textual content in each tweet, the usage of abbreviations, typos, cyberspeak and nonstandard expressions create considerable obstacles to match the query with the tweet~\cite{hossny2018event}. On the other hand, user-supplied keywords can also be either too general (i.e., ambiguous) or too specific as a query string, making the event information retrieval process difficult~\cite{becker2011automatic}. \smallbreak

In general, the ETR process contains three main components: \emph{(i)} query modeling - using query expansion techniques to enhance user-specified query \emph{(ii)} event representation - formalizing and representing tweets by machine-interpretable knowledge \emph{(iii)} event retrieval - retrieving and ranking relevant documents~\cite{zhao2020framework}. Unlike the TDT approach that clusters tweets into groups, the ETR process directly identifies correlated tweets of a given anomalous event. \smallbreak

\subsubsection*{Query Modeling} \label{sec:eorqm}

The simplest query modeling technique is by aggregating user-specified keywords and contextual information (e.g., geographical information, time period, event description), then retrieving relevant tweets using high-precision strategies~\cite{becker2011automatic}. A more dynamic solution is Pseudo Relevance Feedback (PRF) mechanism, which is an iterative query expansion technique that assumes the top $\gamma$ ranked documents in the initial retrieval results are correlated and can be used to expand the original query~\cite{efthimiadis1996query}. The process typically starts with searching documents using the user-specified query, and the system uses the retrieved documents to expand the query set in order to produce a new query, so to achieve a greater recall. In each iteration, retrieved documents are ranked based on the textual score $\phi k$, calculated as the cosine distance between the TF-IDF feature vector of the keyword and each document. The entire PRF process requires no user interactions, commonly denoted as the blind relevance feedback mechanism. \smallbreak

Metzler et al. further incorporate the temporal dimension to the event retrieval process, based on the proposed temporal query expansion technique~\cite{metzler2012structured}. The system firstly filters time intervals when the user-specified keywords are intensely discussed. For each filtered time interval, the system calculates the bursty score for all terms other than the keyword, then identifies similarly trendy terms as correlated keywords, similar to the bursty feature identification in FPP as proposed by~\cite{fung2005parameter}. An aggregated bursty score is also calculated by summarizing the per-term bursty score over all time intervals. Lastly, the PRF mechanism is applied by expanding the user-specified keywords with the additional trendy terms identified, and reiterate the entire process. A timeline of correlated tweets in response to the user-specified query is returned as the final output. \smallbreak

Tekeuchi et al. proposed a spatio-temporal query expansion technique STT-PRF, by including spatial, temporal and textual information in each iteration of the PRF process~\cite{takeuchi2017spatio}. In addition to calculating the traditional text-based similarity score using cosine distance, a normal distribution is firstly used to approximate the spatial and temporal information of a data point. Bhattacharyya Distance, commonly used to measure the distance between two probability distributions, is then used to estimate the distance between two data points considering all three dimensions. \smallbreak

\subsubsection*{Event Representation} \label{sec:event-representation}

Event, defined as the objective representation of reality, usually consists of complex features that need structuralization and transformation into machine-readable knowledge for downstream applications such as ETR. The goal of event representation is to derive distributed representations of events using word embedding~\cite{ding2019event}, then compare with the user-supplied query for correlation detection. \smallbreak

\begin{figure}[!htbp]
\centering
\includegraphics[width = 1\hsize]{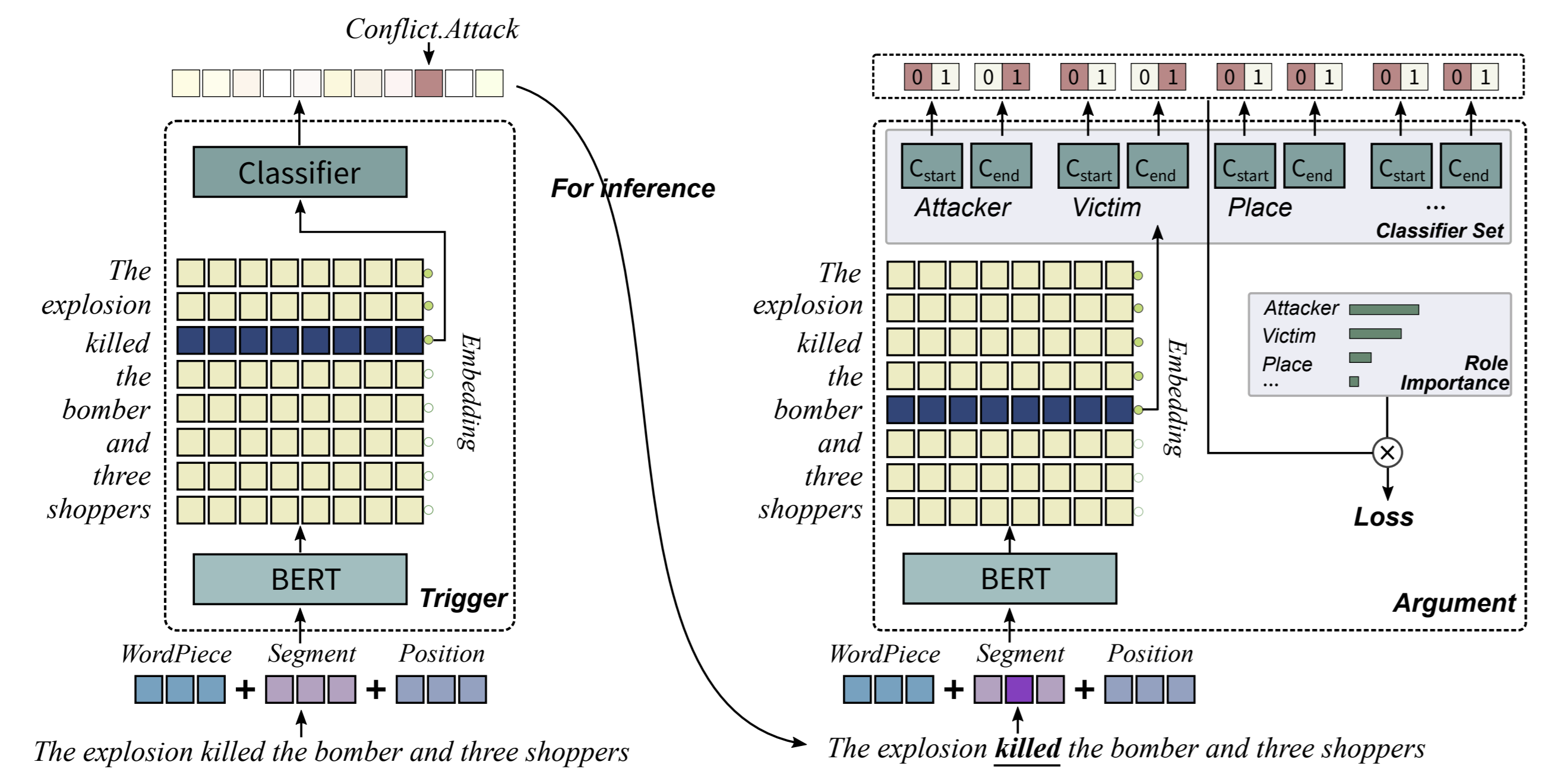}
\caption{Pre-trained Language Model based Event
Extractor (PLMEE) Architecture. This graph demonstrates triggering word identification (i.e., \emph{"killed"} and event category classification (Source:~\cite{yang2019exploring}). }
\label{fig:plmee}
\end{figure}

The neural network is commonly used to generate robust word embeddings as event representation, and skip-gram with negative sampling (SGNS) word2vec model is a widely adopted unsupervised learning technique for representing each word in the event as a vector in an N-dimension space~\cite{mikolov2013distributed}. In recent researches, more complex neural networks are used to learn contextualized word embeddings from sentences, such as BERT. BERT leverages transformer, a self-attention-based mechanism along with Masked Language Modeling (MLM) and Next-Sentence Prediction (NSP) techniques to assist learning contextual relationships among words in an unsupervised manner~\cite{vaswani2017attention}. For BERT-base model, it uses 12 hidden layers of transformer encoders. Word embedding per token can be extracted from any one of the 12 layers, or use a sum of the last four layers depending on the task. These learned word embeddings can then used to generate event representations in different forms. For instance in Figure~\ref{fig:plmee}, Yang et al. proposed a framework named PLMEE for representing sentences using extracted important named-entity from the plain text~\cite{yang2019exploring}. It leverages the contextualized word embeddings learned by BERT to perform a "trigger word identification", which is a token-level classification task that predicts if a token triggers an event. The event representation in~\cite{yang2019exploring} is the summation of three embeddings of these trigger words, which are WordPiece\footnote{WordPiece is a sub-word segmentation algorithm that breaks rare words into characters and preserves frequent words in its original form.}, segment and position embeddings. This representation can be further used for other downstream NLP tasks such as sentiment analysis or classification. \smallbreak

However, the word embedding based event representations fail to capture semantic or syntactic similarity among events. In other words, the lexical similarity is the primary factor that determines the similarity between two event representations. For example, "I love cat" and "kitten is my favorite animal" are referring to the same event, whereas event embeddings cannot capture this relationship correctly. To tackle this issue, knowledge graph embedding (KGE) was proposed as a low-dimensional representation of an event using directed heterogeneous multigraph. KGE encodes knowledge and semantics in a graph in which entities (e.g., objects, persons, etc.) are connected via different relations. Based on KGE, Ding et al. proposed a joint embedding model which, in addition to the traditional entity-centric event embeddings, simultaneously encode event participants' intention and emotions into event vectors based on "why" and "how" does the actor perform the event~\cite{ding2019event}. This technique was evaluated using the stock market prediction task against baseline methods. The experiment results indicate that by adding emotion into the embedding vector, the performance was better than all competitive methods, including the previously best-performing knowledge-graph-based event representation approach~\cite{ding2016knowledge} by 2.4\% in accuracy. \smallbreak

\begin{figure}[!htbp]
\centering
\includegraphics[width = 0.3\hsize]{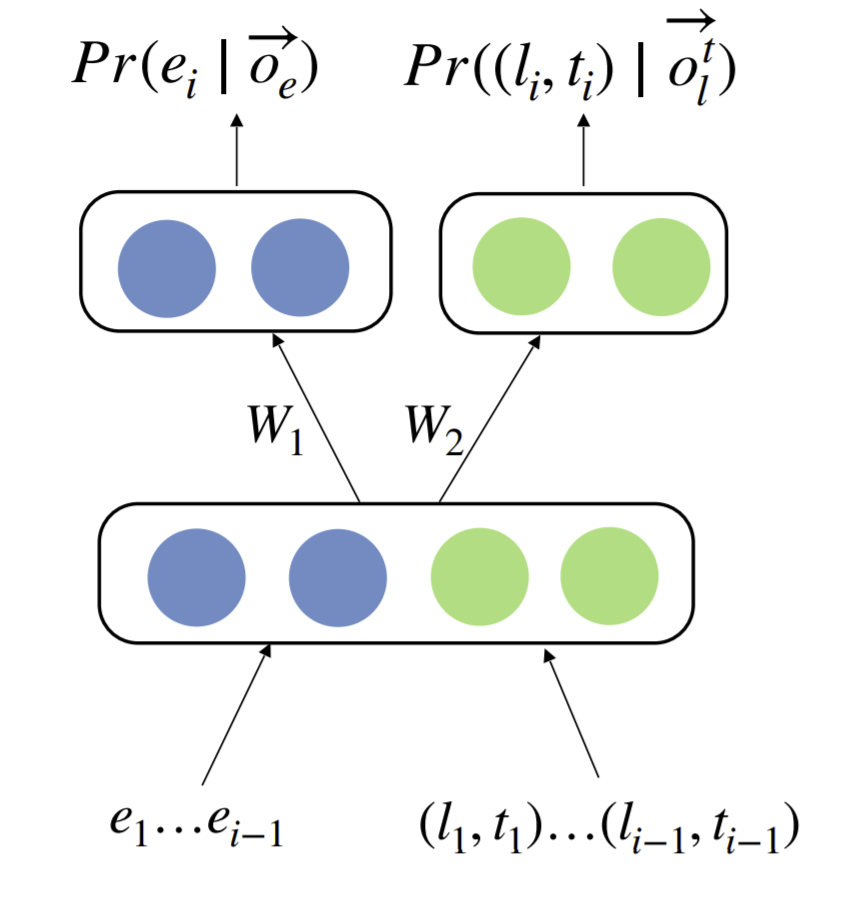}
\caption{Event2Vec-2 Architecture that derives latent representation using multiple modalities from spatial and temporal dimensions (Source:~\cite{wang2019event2vec}).}
\label{fig:event2vec}
\end{figure}

In recent literature, models also incorporate multiple modalities for the rich semantic information in the spatial and temporal dimensions. For instance, Want and Tang proposed the Event2Vec model, which encodes events into a lower dimension for learning the latent representations of event, time and locations simultaneously~\cite{wang2019event2vec}. The best-performing model is Event2Vec-2 as presented in Figure~\ref{fig:event2vec}, which learns the embedding for both event and spatial-temporal pair while considering the interaction between spatial and temporal elements. The multimask self-supervised learning is used along with a feed-forward neural network (FFNN) to learn the embedding by naturally leveraging all available information in each document. \smallbreak

\subsubsection*{Event Retrieval and Ranking}

The event retrieval component usually ranks retrieved documents based on their quality or other user-specified metrics. For an ordered document collection, only the top $\gamma$ documents are selected as the output. Traditionally, the ranking technique uses solely the relevancy of documents in determining the document quality. Most recent approaches leverage additional dimensions, such as recency, temporal profile and authority to assist in evaluating document relevancy from diverse aspects.\smallbreak

To rank documents based on relevancy, Shokouhi proposed the central-rank based collection selection (CRCS) algorithm to rank and determine the top $\gamma$ documents as the collection $D$~\cite{shokouhi2007central}. The relevancy of $i^{th}$ ranked document $d_i$ is $R(d_i) = \alpha exp(-\beta \times i)$, in which $\alpha$ and $\beta$ are user-specified constants and $d_i \in D$. The objective is to maximize the collective relevancy $R(D) = \sum_{d_i\in D} R(d_i)$ of the top $\gamma$ retrieved documents' collection $D$, by assigning the optimal rank to each document. CRCS is widely used in later research as a foundation to rank documents by relevancy. \smallbreak

Recent ranking models incorporate the temporal profile as a factor, which tackles the issue of retrieving irrelevant documents from the past. For instance, for users interested in knowing why the crude oil prices rose significantly during March 2021, only information in association with the Suez Canel incident should be retrieved instead of other possible causes that happened years ago. Campos et al. proposed a time re-ranking algorithm that adds temporal similarity scores to measure both the topical and temporal relevancy of a document~\cite{campos2016gte}. The concept behind this algorithm is trivial, which is the greater the similarity between the document and keywords/dates in the query, the more correlation there is. The relevant dates of a query are identified using a generic temporal similarity measure named GTE, and in the context of Twitter, the ease to access the temporal metadata of each tweet makes calculating GTE even more convenient. \smallbreak

\begin{figure}[!htbp]
\centering
\includegraphics[width = 1\hsize]{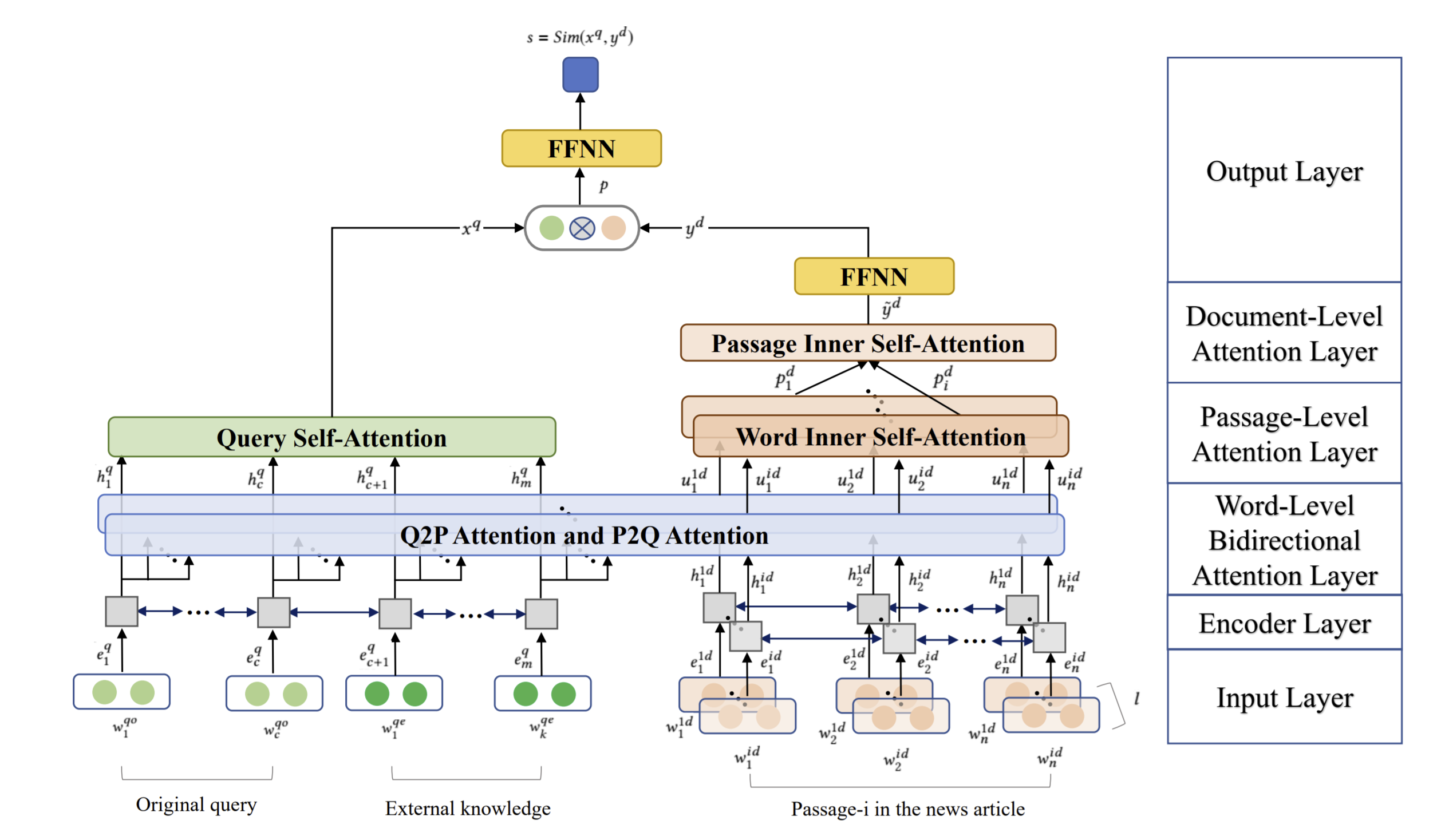}
\caption{Event-oriented  Neural  Ranking  Model  for  News  Retrieval (ENRMNR) Architecture with a  hierarchical attention mechanism to evaluate the similarity between the query and event (Source:~\cite{zhao2020event}).}
\label{fig:enrmnr}
\end{figure}

Besides a ranking model, the neural network is an alternative to directly assign weights to retrieved documents and return the top $\gamma$ correlated documents. For instance in Figure~\ref{fig:enrmnr}, Zhao et al. proposed an event-oriented neural ranking model for news retrieval (ENRMNR) that uses a hierarchical attention mechanism to capture the dynamics and coupling nature of events~\cite{zhao2020event}. It uses three attention layers with different granularity: \emph{(i)} word-level bidirectional attention layer to compute the similarity between each query keywords and words in the document (i.e., query-to-passage) \emph{(ii)} passage-level attention layer using query self-attention and passage inner-words self-attention techniques to grant more weights to important words in queries and documents \emph{(iii)} document-level attention layer to assign greater weights to relevant documents as a whole. The combined weights are then used to determine rankings of documents in the retrieval process. Lastly, documents are retrieved using a 3-layers FFNN as defined in the Output Layer, and the similarity score $s$ between the query representation $x^q \in Q$ and the document representation $y^d \in D$ is calculated as follows:

\begin{equation}
s = w_{p}^T p + b_p \text{where } p = x^q \odot y^d
\end{equation}

\subsubsection{Event-oriented Text Retrieval Performance Limitations}

Regardless of the effectiveness of ETR in identifying correlated tweets comparing to clustering-based approaches, ETR has several limitations as well. Firstly, the retrieval performance is strongly associated with the popularity of the event category. Since ETR is a keyword-based model, and keywords are identified by the number of mentions in the time span, event popularity can have a direct impact on ETR performance. As proved in~\cite{metzler2012structured}, the correlation coefficient between event popularity and retrieval precision on Twitter corpus is 0.63 ($p$ $<$ 0.01) using sole keyword-based approach, and 0.61 ($p$ $<$ 0.01) even with the PRF mechanism deployed. Therefore, for less trendy events, ETR may not be able to retrieve events with high precision. \smallbreak

Secondly, modern variants of PRF mechanism such as STT-PRF demonstrate a significant improvement in terms of recall, whereas the precision also drops considerably compared to the traditional PRF mechanism~\cite{takeuchi2017spatio}. In other words, more related tweets are retrieved by incorporating more modalities, at the expense of a reduced quality on the document collection. \smallbreak

In addition, incorporating intention and semantic vectors into event representation can improve the performance, whereas a large-scale annotated dataset that covers the financial domain is required~\cite{ding2019event}. Most commonsense knowledge dataset such as ATOMIC\footnote{ATOMIC covers 877k textual descriptions of everyday commonsense knowledge, aiming to generate an inferential knowledge repository based on if-then knowledge.}~\cite{sap2019atomic} only covers daily-life scenarios like "drinking coffee" or "break a window", but lacks coverage on events related to the financial domain. Therefore, if DeepTrust uses multi-modal event representations, additional efforts are needed to annotate a special dataset consists of the intention and semantic elements of financial-related tweets. \smallbreak

\section{Reliability Assessment} \label{sec:ra}

The goal of reliability assessment is to filter out fabricated or unvalidated information before integrating it into the trusted knowledge base. Since DeepTrust focuses on evaluating reliability using only textual and behavior features, other language-independent detection techniques such as diffusion pattern, user-based features will not be included in the scope. The below sections summarize relevant literature that assesses information reliability using various state-of-the-art techniques, with the aim to exclude known fake tweets, neural fake tweets, non-argumentative tweets, and subjective tweets. \smallbreak

\subsection{Feature-based Credibility Assessment} \label{sec:fbca}

Tweet features constitute a content representation of itself, which potentially is a useful indicator of credibility. These features can be used as the input of a neural network for classification, as a supervised learning task. For example, O'Donovan et al. identified textual features (e.g., URLs), behavior features (e.g., @mentions, retweets) and statistical features (e.g., length of Tweet, number of question marks and emojis) are key identifiers that distinguish credible and non-credible posts~\cite{odonovan2012credibility}. Krishnan further refined the feature set into two categories, namely user features and content features~\cite{krishnan2018identifying}. For content features, it includes URLs, image, sentiment, length, word count, the usage of exclamation and question marks, and the number of upper case letters and tags. User features are account-based metadata such as number of posts, retweets, likes, etc. \smallbreak

More recent approaches leverage the automatic feature extraction capability of neural network models. For instance, Ajao et al. used a hybrid Convolutional Neural Network (CNN) and Long-Short Term Memory (LSTM) system to learn the dependencies among words in fake posts without explicitly stating a list of pre-defined rules~\cite{ajao2018fake}. Singhal et al. used the BERT-base model to represent each token in the sequence of text as a contextual embedding (i.e., a feature vector of length 32), and this vector has already captured the underlying contextual and semantic association amongst tokens~\cite{singhal2019spotfake}. This embedding, along with visual feature embedding collected from VGG-16, is concatenated as the input of another fully connected neural network for classification. \smallbreak

One issue with feature-based reliability assessment is that it requires an expert-annotated dataset for supervised learning. For instance, Twitter Media Eval\footnote{Source: \url{https://github.com/MKLab-ITI/image-verification-corpus/tree/master/mediaeval2016}.} and FakeNewsNet\footnote{Source: \url{https://github.com/KaiDMML/FakeNewsNet}.} are two widely adopted datasets used by previous researches. However, most available datasets are related to politics or natural disaster, and there are limited resources to create one associated to the financial domain from scratch. Therefore, existing feature-based assessment techniques can only be used as a complementary reliability assessment tool (e.g., cross-validating the credibility of URL against known fake URLs), instead of a conclusive factor of information reliability. \smallbreak

\subsection{Neural Fake News Verifier}

With the success of Pre-trained Language Models (PLMs) in delivering promising results for downstream NLP tasks, these deep neural networks may potentially be leveraged to generate fake information that is indistinguishable from an authentic text. To a retail investor that does not possess in-depth financial knowledge, financial disinformation fabricated by the neural network can be abused for malicious intentions. The following sections discuss forensic techniques for identifying machine-generated text that is fabricated by neural networks without a factual basis.  \smallbreak

\subsubsection*{Language Model (LM)}\label{sec:lm}

Language modeling is the fundamental principle of natural language generation, and it is a statistical language model that represents a sequence of text by the conditional probability of each word given the context (i.e., all previous seen words)~\cite{bengio2003neural}. For a document $D$ that begins with \emph{\textless start\textgreater} and ends with \emph{\textless end\textgreater} special tokens, the probability of generating such document is the product of generating each word $x_i$ given the full context:
\begin{equation}
    p(D) = \prod_{i=1}^{N}p(x_i | x_{1:i-1})
\end{equation}
The document is terminated whenever the \emph{\textless end\textgreater} is predicted as the $N^{th}$ word, and the natural language generation process is completed. The objective function of the LM is to maximize the log-likelihood of the conditional probabilities over the sequence of words:
\begin{equation}
    \Theta^* = \argmax_{\Theta} \sum_{i=1}^N logP(x_i | x_{1:i-1};\Theta)
\end{equation}
where the parameter set $\Theta$ can be learned using different architectures such as LSTM or other RNN-based nerual networks. \smallbreak

\subsubsection*{Giant Language model Test Room (GLTR)} \label{sec:gltr}

GLTR is a statistical forensic inspection tool that identifies machine-generated fake text by matching the reviewed text and the text generated by a language model following a similar sampling assumption~\cite{gehrmann2019gltr}. In other words, GLTR tries to use the identical model (e.g., BERT, GPT-2) that is used to generate fake content as the detector. For GLTR, three main factors are evaluated when concluding the authenticity of the text: For $i^{th}$ word $X_i$ in the input text and $Y_i$ in the GLTR-generated neural text \emph{(i)} the probability of these two words are identical given the full context $p(X_i = Y_i | X_{1:i-1})$ \emph{(ii)} the ranked order of $X_i$ in the predicted word distribution $p(X_i | X_{1:i-1})$ \emph{(iii)} the entropy of $Y_i$ distribution. For instance, for an input text that each word matches with the most confident predicted word from using a GPT-2 model, this input text is very likely generated by another GPT-2 or a model with similar architecture. GLTR is commonly used as a visualization aid for expert evaluators, and was proved to assist in improving 18\% in neural fake text detection rate~\cite{gehrmann2019gltr}. \smallbreak

However, GLTR is limited by its core assumption that the adversarial model tends to sample words from the top $\gamma$ predicted words in each iteration. Therefore, GLTR can be easily deceived by adjusting the sampling parameters with a degree of randomness (e.g., sampling from top 20-50 instead of from top 10 predicted words). Besides, GLTR was proved to be effective in identifying fake text generated by BERT and GPT-2, but the performance worsens when detecting against other language models. The reason is trivial as GLTR relies on known language models to generate a similar fake text for detection. Therefore, in the context of this project, GLTR can be adopted as supplementary assistance in neural fake text detection, or as a reference to aid other models in judging the reliability. \smallbreak

\subsubsection*{Generating aRticles by Only Viewing mEtadata Records (GROVER)} \label{sec:grover}

GROVER model is an effective system for detecting neural fake news, which converts fake news detection into an adversarial game between two players, adversary and verifier~\cite{zellers2020defending}. The adversarial model strives to create convincing disinformation based on controllable parameters, whereas the verifier model aims to correctly distinguish generated fake paragraphs from the adversarial model using unlimited authenticated information from trusted sources. This creates an escalating competition between two participants, which eventually yield a pair of strong disinformation generator and verifier. \smallbreak

One novelty of GROVER is its capability of controlled text generation. Most transformer-based language models such as GPT-2 and BERT are proved to generate realistic-looking fabricated content that can even deceive human experts~\cite{Radford2019LanguageMA}, but the generated text cannot be restricted into a specified domain by using controllable parameters~\cite{hu2017toward}. Instead, GROVER models the neural fake news as a joint distribution of \emph{p(domain, date, authors, headline, body)}, and the adversarial player can specify either all five parameters or partially to guide the neural text generator conditionally producing fabricated content. Besides, the adversarial can easily adopt enhancing techniques such as NeuralLogic Decoding~\cite{lu2020neurologic} to generate fluent realistic-looking text while following complex lexical rules defined by linguistic experts. These techniques can expose the verifier model to more deceiving neural disinformation, thus improve its performance throughout iterations. \smallbreak

\begin{table}[!htbp]
\begin{adjustbox}{center}
\begin{tabular}{@{}c|ccc|@{}}
                                       & \multicolumn{3}{c|}{\textbf{Generator Size}}                          \\
\textbf{Detector}                  & \textbf{1.5B}                      & \textbf{335M}                      & \textbf{124M} \\ \midrule
\multicolumn{1}{|l|}{GROVER-Base (124M)} & \multicolumn{1}{l|}{\textbf{71.3}} & \multicolumn{1}{l|}{\textbf{79.4}} & \textbf{90.0} \\
\multicolumn{1}{|l|}{BERT-Base (110M)} & \multicolumn{1}{l|}{67.2} & \multicolumn{1}{l|}{75.0} & 82.0 \\
\multicolumn{1}{|l|}{GPT-2 (124M)}     & \multicolumn{1}{l|}{67.7} & \multicolumn{1}{l|}{73.2} & 81.8
\end{tabular}
\end{adjustbox}

\caption{Comparison between detectors (BERT-base, GPT-2, and GROVER-Base) versus GROVER generators with unpaired settings across different architecture sizes. GROVER has achieved the best 90\% accuracy in identifying neural fake news from its paired adversarial model (Source:~\cite{zellers2020defending}) }
\label{tab:GROVER-performance}
\end{table}

The performance of GROVER detector is evaluated against two competitors with similar architecture size, BERT-base, and GPT-2 124M. The experiment is conducted under an unpaired setting, in which the verifier model needs to distinguish human-written or machine-generated text independently without additional referencing materials. The experiment results are presented in Table~\ref{tab:GROVER-performance}, and is clear that GROVER-base outperforms other models significantly across all GROVER's generations. \smallbreak

GROVER is not only effective in identifying fake text generated by its paired adversarial model, but also achieved an excellent performance against other models. This is because of the limitation of the predicted word sampling mechanism, which is also a common weakness for all natural language generation systems. Human speech is a combination of both anticipated and unexpected sequences of words, which introduces certain randomness that cannot be modeled by a probability distribution. Whereas for language models, each word is predicted based on the contextual information and probability distribution, which is a finite set of candidate words. Therefore, GROVER essentially leverages this weakness of text generation systems, and identity the unique pattern of neural-generated text that is unlikely to be seen in a human-written text. \smallbreak

\subsubsection*{Generative Pre-trained Transformer (GPT)} \label{sec:gpt}

Generative Pre-trained Transformer (GPT) is an autoregressive language model trained by OpenAI, and GPT-2 is the second version in GPT generations that is pre-trained on a massive corpus of web-based English resources using self-supervised learning~\cite{Radford2019LanguageMA}. The training process uses the MLM technique to predict masked words, similar to the mechanism discussed in Section~\ref{sec:lm}. The WebText\footnote{WebText is an internal OpenAI corpus based on web-based scarped content with over 8 million high-quality documents and their textual contents.} dataset is used to train the GPT-2, and the model can learn the linguistic patterns of the English language represented by low-dimension vectors. GPT-2 is specialized in generating compelling content from a given prompt, which usually is the topic sentence or the headline that the user wants the model to emphasize. \smallbreak

\begin{table}[!htbp]
\begin{adjustbox}{center}
\begin{tabular}{@{}llll@{}}
\toprule
\textbf{Detector}  & \textbf{Amazon} & \textbf{Yelp} & \textbf{Overall} \\ \midrule
GLTR               & 40.9\%          & 35.9\%        & 38.5\%           \\
GPT-2D             & 20.9\%          & 25.8\%        & \textbf{23.5\%}  \\
GROVER             & 43.6\%          & 36.9\%        & 40.7\%           \\ \midrule
GLTR+GROVER        & 35.3\%          & 34.6\%        & 34.9\%           \\
GPT-2D+GROVER      & 24.9\%          & 22.2\%        & 23.4\%           \\
GLTR+GPT-2D        & 25.0\%          & 19.6\%        & 22.5\%           \\
GROVER+GLTR+GPT-2D & 25.0\%          & 19.6\%        & \textbf{22.5\%}  \\ \bottomrule
\end{tabular}
\end{adjustbox}
\caption{Equal Error Rate in distinguishing neural fake product reviews on Amazon and Yelp consists of 80 real reviews and 160 fake reviews (Source:~\cite{adelani2020generating})}
\label{tab:ensemble-performance}
\end{table}

GPT-2 Detector (GPT-2D), on the other hand, uses a combination of RoBERTa-BASE and RoBERTa-LARGE models with 125M and 355M parameters size to classify if a given text is generated by a neural network. The performance of GPT-2D was evaluated against GROVER and GLTR on assessing the credibility of product reviews on Amazon and Yelp in~\cite{adelani2020generating}, and the detection results are mentioned in Table~\ref{tab:ensemble-performance}. Equal Error Rate (EER), which is a performance metric commonly used for measuring the overall accuracy of biometrics authentication reliability~\cite{teh2016survey}, is used for evaluating the model performance. Although performance-wise, the GPT-2D outperforms GLTR and GROVER considerably, the testing dataset only includes 240 samples, which is not sufficient to reach a conclusion that GPT-2D is the superior neural fake text detection model. \smallbreak

\subsubsection*{Ensemble Learning}

Ensemble Learning works by running a base algorithm or a set of algorithms multiple times, then reaches a consensus on a finalized decision via voting mechanisms~\cite{dietterich2002ensemble}. As shown in Table~\ref{tab:ensemble-performance}, a combination of GROVER, GLTR, and GPT-2D greatly improved the overall EER comparing with individual detectors by 1.0-18.2\%, justifies its benefit in improving the classification performance by \emph{(i)} reducing variance in predictions from member models and \emph{(ii)} reducing the dispersion while generalizing learned knowledge. The predictions from these three models were fused together using a logistic regression model, and were trained on 4M+ reviews collected from Amazon and Yelp review databases~\cite{adelani2020generating}. Other simpler fusing techniques such as plurality voting can also be used to reach the consensus among member models. \smallbreak

However, while the ensemble technique introduces improvements to the overall system performance, the additional complexity also hinders the explainability and interpretability. This is because the relation between features and the label is harder to identify after applying the fusing algorithm. Besides, the overall performance improvement using ensemble learning is not guaranteed, and such improvement is only significant when models are independent of one another (i.e., preferably with high variance in predictions). Lastly, whether adopting ensemble learning in this project or not should not solely depending on the improvement on evaluation metrics, but should also ensure the model is not overfitting to the training data. \smallbreak

\subsection{Argument Mining (AM)} \label{sec:argument-mining}

Argument mining (AM) is commonly defined as the automated detection and extraction of the argumentation structure (i.e., inference and reasoning) and named-entity recognition on argumentation components~\cite{moens2018argumentation, lawrence2020argument}. Traditionally, an argumentation structure consists of six elements as proposed by Toulmin~\cite{toulmin2003uses}, namely claim, qualifier, grounds, rebuttal, warrant, and backing\footnote{\emph{Claim} is the key argument of a topic, or the assertion made by the author. \emph{Qualifier} adds exceptions which the claim may not hold. \emph{Grounds} is the factual basis of the claim, if available. \emph{Rebuttal} refers to alternative views on the claim, which usually are disagreements. \emph{Warrant} and \emph{Backing} are additional hypotheses associated with the \emph{Grounds} that might hinder the reliability of the original \emph{claim}.}. The following sections provide an overview of AM in the context of Twitter, and discuss some insights in incorporating AM to assess information reliability. \smallbreak

\subsubsection*{Overview of Twitter-specific Argumentation Structure}\label{sec:twitter-specific-as}

\begin{table}[!htbp]
\begin{adjustbox}{center}
\begin{tabular}{@{}ll@{}}
\toprule
\textbf{Twitter ID}          & \textbf{Tweet Content}                              \\ \midrule
Bloomberg@business &
  \multicolumn{1}{p{10cm}}{\#Bitcoin declined for the sixth time in seven days, extending losses after President Biden was said to propose almost doubling the capital-gains tax for the wealthy} \\ \midrule
The Wolf Of All Streets@scottmelker &
  \multicolumn{1}{p{10cm}}{\#Bitcoin daily RSI has not hit oversold at 30 since March of 2020 when the market crashed. It is currently about 32, the lowest it’s been since that day. Would love to hit oversold again and be done with it.} \\ \midrule
Altcoin Daily@AltcoinDailyio & \multicolumn{1}{p{10cm}}{\#Bitcoin is doing great. 100k is inevitable. HODL!} \\ \midrule
Mira Christanto@asiahodl &
  \multicolumn{1}{p{10cm}}{\#Bitcoin is down -23\% from the \$65,000 peak. Let's put that into context where the average dip was -35\% in 2017 This is all part of the plan \url{https://bitcoininsider.org/article/101910/...}} \\ \bottomrule
\end{tabular}
\end{adjustbox}
\caption{Various forms of Twitter-specific argumentation in relation to the trending topic \#Bitcoin on 23 April, 2021, right after a drastic Bitcoin price correction over the weekend.}
\label{tab:tweet-am-examples}
\end{table}

Twitter as a microblogging service provider, impose strict character-limit restriction on the tweet, which causes complex argumentation structure cannot be formulated appropriately in the majority of tweets. Therefore, recent researches with a focus on simplifying AM formulation into two components: \emph{claim} and \emph{evidence}, where the evidence can be either supporting or contradicting~\cite{habernal2017argumentation}. Besides, the form of Twitter-specific argumentation can varies drastically, as demonstrated in Table~\ref{tab:tweet-am-examples}\footnote{Tweet from Bloomberg@business: \url{https://twitter.com/business/status/1385375837055246336}.}\footnote{Tweet from The Wolf Of All Streets@scottmelker: \url{https://twitter.com/scottmelker/status/1385416836683116544}.}\footnote{Tweet from Altcoin Daily@AltcoinDailyio: \url{https://twitter.com/AltcoinDailyio/status/1385411871692365831}.}\footnote{Tweet from Mira Christanto@asiahodl: \url{https://twitter.com/asiahodl/status/1385425967913472005}.}. The first tweet from Bloomberg consists of a clear claim-evidence structure, with a verifiable factual basis \emph{"doubling the capital-gains tax for the wealthy"} to support its assertion. A similar strategy is adopted by the second tweet, which is also an author in authority (i.e., with a large audience base and a certain degree of credibility). However, the vast majority of tweets published by civilians are similar to the latter two examples. For instance, the third tweet is sarcastic, and yet still follows a claim-evidence structure\footnote{The claim is \emph{"100k is inevitable"} and its supporting evidence is \emph{"Bitcoin is doing great"}, though the evidence is unverified. }, makes AM even more challenging. The last tweet is a typical argumentation structure in which the evidence is included as external links. The validity of these URLs is not trivial, and examining them will significantly increase the complexity of AM. Therefore, it is essential to acknowledge the peculiarities existing in Twitter-specific argumentation structure, and adjust the AM technique to handle information retrieved from user-generated data. \smallbreak

\subsubsection*{Argument Detection} \label{sec:argument-detection-literature}

The first step of argument detection is to define the source of arguments. In prior research, many different definitions arose without a finalized standard. For instance, Bosc et al. stated that argumentation, or a claim towards a certain topic, should primarily present in the format of independent tweets~\cite{bosc2016dart}. Other researchers, such as Schaefer and Stede, argued that replying tweets should also be evaluated alongside the main tweet for argument and relation detection~\cite{schaefer2020annotation}. However, using both independent and replying tweets in discourse as the source of argumentation introduces significant complexity to the AM system, as the argument potentially consists of rebuttal and elements other than a simplified claim-evidence structure. Therefore, as a compromise to the complexity, the present project only focuses on using independent tweets as the sole source of argumentation. \smallbreak

With a defined source of information, the argument detection aims to classify tweets into argumentative or non-argumentative based on a list of criteria such as claim-evidence structure discussed in Section~\ref{sec:twitter-specific-as}. Specifically, the most commonly adopted standard is defined in DART\footnote{DART refers to Dataset of Arguments and their Relations on Twitter annotated by Bosc et al.~\cite{bosc2016dart}.}. A tweet is argumentative if \emph{(i)} contains explicitly expressed opinionated statements \emph{(ii)} contains rhetorical questions or claims in the form of questions \emph{(iii)} contains a claim-evidence pair regardless of the validity of the evidence. This standard was further clarified in~\cite{dusmanu2017argument}, by dividing argumentative tweets into two categories, namely \emph{factual} and \emph{opinion}. In the context of DeepTrust, only factual tweets are considered as reliable financial knowledge that can be incorporated into the Refinitiv knowledge base, while opinionated tweets are further evaluated using subjectivity analysis techniques mentioned in Section~\ref{sec:se}. \smallbreak

Argument detection is usually defined as a binary classification task, to identify the presence of argumentation structure. For instance, Addawood and Masooda trained an SVM model to classify tweets based on lexical (i.e., n-grams), statistical (e.g., part-of-speech POS percentages) and Twitter-specific features (e.g., URLs, hashtags, emojis)~\cite{addawood2016your}. Later, Dasmanu et al. further incorporated syntactic and semantic features to train a logistic regression model, which yielded the best score on the DART dataset with 0.89 F1 score~\cite{dusmanu2017argument}. An alternative approach was used in~\cite{schaefer2020annotation}, which leveraged XGBoost\footnote{eXtreme Gradient Boosting (XGBoost) is an ensemble learning algorithm based on a gradient boosting framework~\cite{chen2016xgboost}.}
model to classify tweets using the pre-trained BERT embeddings. \smallbreak

Although mainstream AM models use supervised learning to achieve a better performance, it requires an expert annotated dataset with identified argument structure and relation. In the absence of an annotated dataset for financial-related tweets, the unsupervised AM technique proposed by Persing and Ng~\cite{persing2020unsupervised} might be inspirational. The core idea is to automatically identify and weakly label argument components using both linguistic features and heuristics, and use self-training to improve the performance with unlabelled data. However, this method relies on a well-defined document structure to heuristically identify major claims and argumentative components, such as the premise of \emph{"The major claim is assumed to appear only in the first and last paragraph, as these are typical places for a thesis statement"} for an essay. These assumptions do not apply to tweets, thus the effectiveness of this approach is questionable when applying to DeepTrust. \smallbreak

\subsubsection*{Evidence Detection and Validation}

Evidence detection aims to identify associated evidence of a claim, while the evidence can either be validated against a factual basis or invalidated. The source of evidence also varies, for instance, \emph{News}, \emph{Blogs}, \emph{Newspaper}, \emph{Magazines} as used in~\cite{dusmanu2017argument, addawood2016your}. To detect evidence, the simplest approach is to train a multi-class classifier such as SVM using an annotated dataset on evidence types. However, a rule-based evidence detection system yields similar performance without needing an annotated dataset. For example, a string matching system that detects if known trusted Twitter users or news agencies (e.g., Bloomberg@business, CNN@cnn) exist in a tweet can be used to infer tweet reliability, in which such tweet is likely to be a reported speech from authorized parties. Besides, named-entity recognition (NER), which identifies and classifies mentions of entities into predefined categories (e.g., news agencies, persons), can be used along with knowledge bases like DBpedia\footnote{DBpedia is data catalog of extracted structured information from Wikipedia \url{https://www.dbpedia.org/}.} and Refinitiv Eikon to validate the evidence. Models such as Bidirectional LSTM (Bi-LSTM) were proved effective in Twitter NER tasks by training on both word-based embeddings and orthographic features\footnote{Orthographic features refer to norms of linguistic styles, such as \emph{"1st Game has END!!"} can be represented as \emph{"ncc~Cccc~ccc~CCCpp"}.}~\cite{limsopatham2016bidirectional}, and DBpedia contains adequate structured information that covers the financial domain for evidence validation. \smallbreak

Nevertheless, the effectiveness of evidence validation relies on certain assumptions. Firstly, both reported speech from news agencies in authority and argumentative tweets with trusted named entities are assumed to be reliable, whereas this chain of trust is easily exploitable. For instance, a malicious player can add \emph{"CNN stated:"} in each fake tweet so to bypass the evidence validation system. To tackle this issue, one solution is to apply the ETR technique on the mentioned entity, such as CNN's recent posts, then evaluate if the reported speech is authenticated or not. However, drastic increase in time complexity is the trade-off for this enhanced validation scheme, and for DeepTrust, to hypothetically assume most Twitter users are decent by nature is a compromise made to the efficiency. \smallbreak

\subsection{Subjectivity Analysis (SA)}\label{sec:se}

Subjectivity analysis (SA) quantifies the degree of subjectivity, expressed in personal opinion or claims without factual evidence. SA can be a binary classification task by distinguishing subjective and objective tweets, or can also be a regression task, which assigns a per-tweet subjectivity score and uses a threshold-based model to evaluate the intensity of subjectivity and decide if the tweet is subjective. The following sections review various SA techniques experimented in literature, and evaluates their suitability for Twitter. \smallbreak

\subsubsection*{Syntax-based Subjectivity Analysis}

Syntax-based patterns can be an effective indicator of subjectivity, as subjective text often contains identifiable linguistic features expressed in the way of using different combinations of phrases. For instance, Xuan et al. investigated and extracted 22 syntax-based patterns including adjectives, adverbs, verbs, and nouns. These patterns were used as the input feature for classifying subjective movie reviews, achieved a 92.1\% accuracy along with a Maximum Entropy Model~\cite{xuan2012linguistic}. For each tweet, the Stanford Parser\footnote{Stanford Parser is probabilistic-based NLP tool to generate parse trees: \url{http://nlp.stanford.edu:8080/parser/}.} is used to extract linguistic features from the raw textual content, such as \emph{(ROOT (S (NP (PRP)) (VP (VBP) (SBAR (S (NP (NNP)) (VP (VBZ) (NP (DT) (JJS))))))))} for the phrase \emph{"I believe Bitcoin is the best"}. \smallbreak

In the context of Twitter, the main challenge is the colloquial languages and Twitter-specific elements~\cite{limsopatham2016bidirectional}, making syntactic-based SA less effective comparing to lexicon-based and semantic-based approaches. For instance, users often make grammatical mistakes or include misspell words in the tweet, which can confuse the syntax-based classifier in reaching an accurate decision on the text subjectivity. Besides, Twitter users often not following proper grammatical rules due to the character-limit constraint, thus subjective syntax patterns learned from other mediums such as newspapers or blogs cannot be directly reused for SA on Twitter. \smallbreak 

\subsubsection*{Lexicon-based Subjectivity Analysis} \label{sec:textblob}

Lexicon-based SA is an effective approach for evaluating subjectivity in tweets, as it not relying on a structured syntactic pattern. Instead, it leverages word-based or n-grams subjectivity, along with contextual information to estimate a score for each token. The sentence-level subjectivity score can then be calculated either using summation or averaging techniques. One application is TextBlob\footnote{TextBlob is a Python library that provides NLP tools to process textual data: \url{https://textblob.readthedocs.io/en/dev/}.}, which provides a subjectivity lexicon for English adjectives. Each adjective may contains multiple entries for each unique \emph{"Sense" (Context)} and \emph{"POS tags"}. For example, the word \emph{"entire"} can either means \emph{"constituting the full quantity or extent"} with $1.0$ subjectivity score or \emph{"(of leaves or petals) having a smooth edge"} with a score of $0.0$. Besides, the POS tag can be used to infer the subjectivity tagging. One possible rule proposed by Riloff et al. is that objective sentences usually consist of interjections or adjectives, whereas verbs and pronouns are more common in subjective sentences~\cite{riloff2003learning}. Another subjectivity lexicon is MPQA\footnote{Multi-Perspective Question Answering (MPQA) is a part of OpinionFinder that defines lexicon-based subjectivity clues: \url{https://mpqa.cs.pitt.edu/lexicons/subj\_lexicon/}.}, which also consists of a database of pre-trained subjectivity score for tokens. When evaluating tweets, each document is assigned with a cumulative sum of subjectivity scores of tokens it comprises, and each token has a weighted MPQA score adjusted for its polarity and intensity, similar to the mechanism in TextBlob. \smallbreak

Besides pre-trained subjectivity score, statistical information on tokens can indicate the degree of subjectivity. For instance, Hatzivassiloglou and Wiebe identified that the presence of dynamic adjectives, semantically-oriented adjectives, and gradable adjectives are statistically correlated with subjectivity~\cite{hatzivassiloglou2000effects}, while Dey et al. improved such subjective indicators by incorporating the knowledge in Wordnet and SemEval 2016 stance detection dataset. The concept behind this improvement is that instead of focusing on the POS tag of tokens in the document, all associated POS tags of a given token defined in Wordnet are examined to check if it has an entry as an adjective. This essentially transforms a pure lexicon-based SA to a combination of lexicon-based and syntactic-based SA technique, and reduces the inconsistencies and lack of accuracy of lexicon-based SA as proved in~\cite{dey2017twitter}. \smallbreak

However, lexicon-based SA suffers from various limitations. Firstly, the threshold on cumulative subjectivity score should provide flexibility for different scenarios, especially for special subjective contents such as irony, which generates additional complexity to SA as evident in~\cite{bosco2013developing}. For example, a subjective sentence can be an objective statement with a rhetorical question at the end, like \emph{"The CNN stated that the Suez Canal incident is over...is it?"}, in which a lexicon-based SA model cannot capture the subjective nature of this statement if using a simple threshold-based rule. Besides, traditional lexicon-based SA models are limited by the scope of their knowledge source, as they only learn information via text representation models such as Bag-of-Words (BOW) or Word2Vec. Whereas Twitter metadata (e.g., Hashtags, Likes, Replies, Mentions), commonly denoted as structured information, contains a considerable amount of additional resources that can be used as reinforcement to the SA model. For instance, Barbosa and Junlan used structured information and statistical information to assess subjectivity~\cite{barbosa2010robust}. Five prominent features were exploited for subjectivity classification, namely the number of verbs, URLs and upper-case, positive polarity, and prior subjectivity scores\footnote{Prior Subjectivity means the subjectivity score of the word preceding the target word, and a comprehensive collection of subjectivity lexicon can be found in~\cite{wiebe2005creating}.}. For instance, the sentence \emph{"@Tesla's \$1.5bn Bitcoin holding could influence other corporations seeking to diversify their investments. Could this be what makes enterprise \#DLT a reality? https://buff.ly/..."} is a typical objective tweet that contains the aforementioned features. The most recent research is from Sixto et al., which combined both lexical features and structured information to train a classifier for SA~\cite{sixto2018analysis}. A stacked generalization ensemble model with five different classifiers yielded the best result on the TASS Spanish SA dataset with 89.80\% accuracy, and each classifier is trained using both meta-information and Bag-of-Words (BOW) representations. These experiments concluded that combine multiple sources of knowledge, such as contextual information (e.g., tweet metadata) and statistical information (e.g., number of dynamic adjectives) can benefit lexicon-based SA models in classifying subjectivity in greater accuracy. \smallbreak

\subsubsection*{Pre-trained LM Subjectivity Analysis} \label{sec:lm-sa}

Inspired by the capability of pre-trained models such as BERT and GPT-2, SA as a downstream NLP task can also be solved using these models with some fine-tuning. For instance, Huo and Iwaihara experimented with pre-trained BERT in SA tasks, with a customized LSTM classifier, layer-wise discriminative learning rate, and multi-task learning, which has proved its effectiveness in SA with 95.23\% accuracy on SUBJ\footnote{SUBJ contains 5,000 subjective and objective sentences extracted from IMDB movie reviews annotated by~\cite{pang2004sentimental}.} dataset~\cite{huoutilizing}. \smallbreak

However, as BERT was trained on general corpora like Wikipedia and BookCorpus, the performance is poor when re-applying the BERT-base model to a specific domain. Hence, BERT-based domain-specific models such as BioBERT and SciBERT were proposed to counter this limitation. Domain-specific models can either be continually pre-trained by initialized with BERT-base weights, or pre-trained from scratch using a specialized dataset. For Twitter, Nguyen et al. proposed BERTweet, a large-scale pre-trained language model trained using 850M English Tweets~\cite{nguyen2020bertweet}. BERTweets was originally designed to perform three downstream NLP tasks, namely POS tagging, NER, and sentiment classification, while its learned embedding can be leveraged for other tasks such as subjectivity analysis. Another enhanced model TweetBERTv2 was proposed by Qudar and Mago, which was continually trained using 10.1B Twitter word tokens in addition to the general corpora~\cite{qudar2020tweetbert}. Similar to BERTweets, this model was trained for various text analytics tasks, including sentiment analysis, question answering, NLP inferences, and paraphrase. However, to leverage BERTweets and TweetBERTv2 for SA workloads, an annotated SA dataset on tweets is required to fine-tune the model for this additional task. Although available datasets such as SUBJ are great sources for training, these datasets contain general topics extracted from information platforms other than Twitter, thus might be ineffective when used for SA tasks on financial tweets from Twitter. \smallbreak 

\chapter{DeepTrust Framework}

\section{Overview of DeepTrust Framework}

\begin{figure}[!htbp]
\centering
\includegraphics[width = 1\hsize]{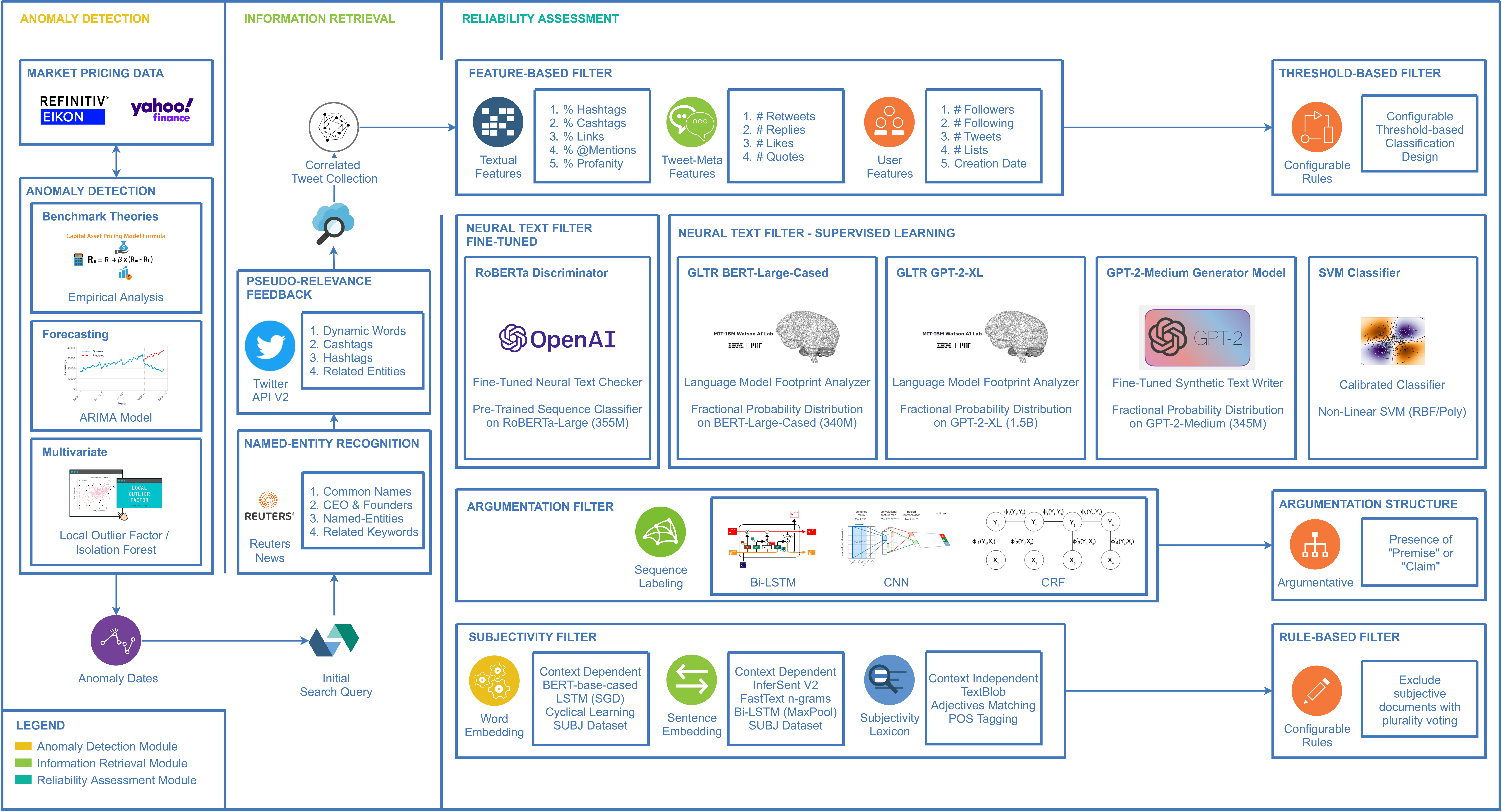}
\caption{The schematic diagram of the reliable financial knowledge retrieval framework DeepTrust.}
\label{fig:deeptrust-system-architecture}
\end{figure}

DeepTrust is a framework for retrieving reliable financial information from the Twitter data stream, which aims at discovering explanatory knowledge regarding an anomalous financial event faster than traditional news agencies. The overview of DeepTrust and its components is shown in Figure~\ref{fig:deeptrust-system-architecture}. It consists of three modules that are executed in a sequential order to ensure information reliability. \smallbreak

The anomaly detection module defines the origin of an unusual event, typically represented by an anomalous pricing movement in the market. The cause of such a financial anomaly should not be trivial. For instance, a small fluctuation in price (e.g., $\pm5\%$) after releasing a quarterly financial statement is considered trivial, whereas a large-scale stock price correction that appeared without basis is the type of event DeepTrust aims to explain using non-standard data sources. The financial anomaly can either be a spike (i.e., a sudden rise in price followed by another large crash) or a change point (i.e., a persistent pricing movement trend). To identify a pricing anomaly, either a financial expert with solid domain knowledge or a time-series anomaly detection algorithm can fulfill the requirement. For DeepTrust, various AD algorithms are applied and compared for their effectiveness in recognizing contextual anomalies, and more details are discussed in Section~\ref{sec:anomaly-detection}. \smallbreak

The information retrieval module defines the source of knowledge, which can be extremely noisy compared to traditional implicitly trusted data sources like newspapers. The financial anomaly identified in the previous module is typically described in plain text, which mentions the targeted entity \emph{"Twitter Inc. TWTR"} and date \emph{"30 April 2021"}. To construct the initial search query, various query modeling techniques are employed to initialize a list of keywords and search parameters for information retrieval. Besides, information from relevant news articles or other documents from the Refinitiv knowledge base is used to enhance the query. In particular, these trusted knowledge are analyzed using Refinitiv Intelligent Tagging API to identify relevant keywords that exhibit a similar temporal pattern as the original keywords, and then integrated into the query for information retrieval. The pseudo-relevance feedback mechanism is used to iteratively improve the retrieval performance by including dynamic keywords into the search query, and more details are discussed in Section~\ref{sec:irm}. \smallbreak

The reliability assessment is the core module of DeepTrust, to screen unreliabile information from the Twitter data stream and preserve only trusted knowledge related to the financial anomaly. Retrieved documents from the information retrieval module are analyzed and screened using various NLP techniques, including \emph{(i)} feature-based filter that leverages textual, tweet-meta and author features to infer reliability \emph{(ii)} neural fake text detector that identifies if a given text is fabricated content generated by a neural network \emph{(iii)} argumentation filter that removes non-argumentative tweets that lack a valid argumentation structure \emph{(iv)} subjectivity filter that analyzes opinionated tweets on its reliability to be considered as trusted information. All these modules work collaboratively to exclude unreliabile information from entering the trusted knowledge base. More details are discussed in Section~\ref{sec:ram}. \smallbreak

\section{Anomaly Detection module} \label{sec:anomaly-detection}

The AD module in DeepTrust focuses on detecting equity pricing anomalies, particularly on famous or trending stocks that their management teams are proactively communicating with the public on social networking services such as Elon Musk from Tesla Inc. and Jack Dorsey from Twitter Inc. Section~\ref{sec:adeabt} outlines mainstream benchmark theories for financial analysts to reference when detecting equity pricing anomalies using empirical analysis, while Section~\ref{sec:uadarima} and Section~\ref{sec:multi-variate} presents three AD algorithms implemented in DeepTrust to identify anomalies using univariate or multivariate analysis. \smallbreak

\subsection{Anomaly Detection using Empirical Analysis and Benchmark Theory}\label{sec:adeabt}

For empirical analysis, detecting equity pricing anomalies is equivalent to test the null hypothesis, which is the efficient market hypothesis discussed in Section~\ref{sec:adfm}. In other words, the equity market should \emph{(i)} consistently operate as an informationally efficient market and \emph{(ii)} the fair value of equities is expressed by an equilibrium model based on economic theories. If the null hypothesis is rejected, the cause can be either an abnormal price movement that is not predicted by cross-sectional analysis or time-series analysis using benchmark theories, or an incompatible equilibrium model is used to derive the fair value. Since 1972, Blume and Husic have started to establish a standard benchmark theory on modeling stock prices using previous rates of returns~\cite{blume1973price}, while in later researches, hundreds of different return predictors are proposed using different return generating functions such as payout yield or tail risk beta. Nevertheless, there is no standardized formulation on the return generating function, hence the equity pricing AD is inherently subjective by its choice of underlying benchmark theory. For DeepTrust, one of the sources of curated anomaly list is domain experts. These specialists possess solid understandings of both the company itself and different pricing models, which are key factors of AD using empirical analysis. As introductory information, some most commonly acknowledged benchmark theories are described in the following paragraphs as examples, and can be adopted by financial experts to empirically detect equity pricing anomalies. \smallbreak

The cross-sectional analysis is a comparison-based method to match a target company with its competitors or peers with similar positioning. The representative benchmark theory in this category is Cross-Sectional Return Patterns (CSRP), which is based on the well-known Capital Asset Pricing Model (CAPM). In general, CSRP assumes the security return should conform to a linear relation with its non-diversifiable risk (i.e., beta $\beta$) and other security-specific characteristics. Formally, the expected return $R_i$ of security $i$ in the market portfolio that consists of all marketable securities is
\begin{equation}
R_i = a_0 + a_1 \beta_i + \sum a_j c_{ij} + e_i
\end{equation}
where $a_0$ and $a_1$ are constant coefficients, $\beta_i$ is systematic risks that measure covariance with the market, $c_{ij}$ are all security-specific attributes such as accounting-based measures, and $e_i$ is the risk-free return. However, security-specific attributes cannot holistically represent a company, thus ineffective in cross-sectional analysis. For instance, market sentiment and intangible fundamental factors like the quality of management teams are examples of intangible factors. These potential equity price driven forces are not captured by any attribute that can be quantified in the CAPM model. Besides, cross-sectional analysis is effective in identifying point anomalies (e.g., an unusual operational cost on balance sheet comparing to the financial report of peers) instead of contextual anomalies that DeepTrust is focused on. \smallbreak

Meanwhile, time-series analysis is another solution of equity price forecasting and AD by extracting meaningful information from the time-series data. Typically, the expected return is governed by time-varying factors that are influenced by previous returns or seasonal milestones. For instance, the momentum of prior semiannual price movements can have explanatory power in the subsequent quarterly pricing trajectory, and evidence shown that trading strategies that leverage the momentum power can yield greater profits than the baseline trading strategy~\cite{jegadeesh1993returns}. However, though stocks may be seasonal with cyclical demands (e.g., sunscreen companies may face drastically different demands in different seasons), the majority of companies are time-independent, especially in the long run. Therefore, the equity price should still fairly represents the overall productivity of its company regardless of market timing, and time-series analysis that captures the momentum or trend of price movements may work effectively in the short-term, but less reliable for long-term equity pricing AD. \smallbreak

Overall, throughout observations, equity price movement is constantly changing in an unpredictable manner, and defying economic interpretations and benchmark theories. Hence, there is no consensus on a standardized paradigm to precisely model the expected returns in the future. Nevertheless, well-established pricing paradigms can still help predicting the expected return with greater probability than tossing a coin, allow detecting anomalous stock prices when it deviates from its normal expectation. The aforementioned models or more complex benchmark theories can, therefore, be adopted by financial data providers as referencing tools of AD, and generate a list of equity pricing anomalies accordingly. \smallbreak

\subsection{Univariate Anomaly Detection using ARIMA model}\label{sec:uadarima}

For unsupervised AD on equity pricing data, one novel approach adopted in DeepTrust is via time-series forecasting. Anomalies are detected by measuring the difference between the actual price and the predicted price. Detailed methodology and parameter tuning strategies are described in following paragraphs.  \smallbreak

For time series forecasting in econometrics, the first step is to model the equity pricing data into a univariate time series.  Formally speaking, a univariate time series is a collection of observations recorded sequentially with a constant interval (e.g., stock prices with intervals of ticks, minutes, or longer periods), and analysis on univariate time series involves only a single variable. For the equity pricing data, the sole independent variable involved is the timestamp, and the assumption made is that the equity price can be formulated as a function of time. In DeepTrust, the AutoRegressive Integrated Moving Average (ARIMA) model is used to perform time-series forecasting on equity prices, and detect price anomalies by measuring the degree of deviation of the actual price from the expected price. Mathematically, the ARIMA$(p, d, q)$ models the expected value $Y_t$ at time $t$ as
\begin{equation}
Y_t = \mu + \alpha_1 Y_{t-1} + \alpha_2 Y_{t-2} + ... + \alpha_p Y_{t-p} + \epsilon_t + \beta_1 \epsilon_{t-1} + \beta_2 \epsilon_{t-2} + ... + \beta_q \epsilon_{t-q}
\end{equation}
where $\mu$ is the average value, $\alpha$ and $\beta$ are correlation coefficients and $\epsilon$ is an independent and identically distributed random variable from $\mathcal{N}(\mu,\,\sigma^{2})$ functioned as an error term of the forecast. The ARIMA$(p, d, q)$ is governed by three parameters, namely $p$ and $q$ for the order of autoregressive and moving average process, respectively, and $d$ for the degree of differentiation. Specifically, $p$ and $q$ are equivalent to sliding windows that define the number of past values that are influential to the current value, whereas parameter $d$ is only needed for non-stationary data with a varying mean, variance, and auto-correlation. To determine the stationarity of time series data, the Augmented Dickey-Fuller (ADF) is used to test the null hypothesis of non-stationarity. Mathematically, a stationary time-series should have $\gamma \neq 0$ in 
\begin{equation}
\Delta Y_t = Y_t - Y_{t-1} = \alpha + \beta t + \gamma Y_{t-1} + \delta_1 \Delta Y_{t-1} + \delta_2 \Delta Y_{t-2} + ... \delta_p \Delta Y_{t-p}
\end{equation}
so to reject the null hypothesis, and the result should be statistically significant with $p < 0.05$. The benefit of stationary time series data is that its easier to model as its statistical characteristics are time-independent, and seasonality and trend do not affect the value at different observation times. In DeepTrust, each equity is evaluated individually using the ADF test for its stationarity, then decide if differencing (i.e., the parameter $d \neq 0$) is required. \smallbreak

In order to obtain the best parameter set of ARIMA$(p, d, q)$ for each equity, an efficient stepwise grid search strategy proposed in~\cite{hyndman2008automatic} is applied. The objective is to minimize Akaike Information Criterion (AIC), which equals to
\begin{equation}
AIC = -2 log(L) + 2(p + q + k + 1)
\end{equation}
where $L$ is the log-likelihood estimate and $k$ is a binary variable correlated to model coefficients. In DeepTrust, the fine-tuned ARIMA model with minimum AIC of each equity is serialized into a pickle object and saved as a file, in which the model is trained using all historical pricing data extracted from publically available stock data APIs like Yahoo Finance and Alpha Vantage. \smallbreak 

To detect equity price anomalies, the find-tune ARIMA model is utilized to iteratively predict data within a user-specified time period using prior observations. When generating rolling statistics with a window size of $n$, the initial $n-1$ observations do not have any rolling statistics, as the first $n$ observations are all used to calculate the initial rolling statistics starts from the $n^{th}$ observation. Therefore, the user-specified time is automatically adjusted to include the previous $n-1$ stock prices for rolling statistics. In addition, as generating a rolling mean and standard deviation over the entire time period since Initial Public Offering is time-consuming, it is suggested that the user should specify a smaller time interval using domain knowledge. When comparing the actual value against the predicted value, observations that defy the expected price movements concluded by the ARIMA model are recorded and quantified by its degree of deviation. At each time $t$, the predicted price $Y^{'}_t$ and actual price $Y_t$ are used to compute the error term, denoted as \emph{anomalous severity}. A rolling standard deviation with a user-specific time window is used to quantify the degree of deviation in terms of $\sigma^2_t$. Each error term is tagged with a severity category in $6$ levels, by the degree of $1.5$, $1.75$ and $2$ standard deviations of errors in positive or negative scale. For DeepTrust, a confidence interval of $\pm 2\sigma^2$ with $95\%$ certainty is used to establish a boundary between normal and abnormal prices, while anomalies and \emph{anomalous severity} are assigned accordingly based on its distance to the confidence interval. \smallbreak

\subsection{Multivariate Anomaly Detection using Unsupervised Learning Algorithms} \label{sec:multi-variate}

Since univariate AD algorithms do not capture interactions between different security-specific attributes such as equity price and trade volume per timestamp, while the inter-relationship among different variables might brings additional insights in understanding the nature of price movements, DeepTrust incorporates multivariate AD algorithms as an alternative solution. Based on evaluations in Section~\ref{sec:adfm}, LOF and Isolation Forest are selected for their effectiveness in identifying contextual anomalies. \smallbreak

Firstly, LOF is a density-based unsupervised AD algorithm that can be applied on time series data, and it detects anomalies by comparing Local Reachability Density (LRD) between the point and its $k$ neighbors, as specified by users. Mathematically, For each point $A$ that has $K$ neighbors $X_j$ in collection $K(A)$, the LOF is calculated as a ratio of the average LRD of k-neighbors to the LRD value of point $A$:

\begin{equation}
LRD_k(A) = \frac{1}{\sum_{X_j \in K(A)} \frac{max(K - d(A, X_j), d(A, X_j))}{\norm*{K(A)}}}
\end{equation}
\begin{equation}
LOF_k(A) = \frac{\sum_{X_j \in K(A)} LRD_k(X_j)}{\norm*{K(A)}} \times \frac{1}{LRD_k(A)}
\end{equation}

where the metric of distance function $d(A, X_j)$ is selected based on empirical researches, such as the traditional Euclidean distance or recently proposed Maximum shifting correlation distance (MSCD) in~\cite{jiang2019novel}. For each point, three correlated variables are used to define the coordinate, namely timestamp, equity price and trade volume. $K$ nearest neighbors are automatically selected using either the Ball Tree,  KD Tree, or brute force algorithms by partitioning points into different spatial clusters in the feature space. \smallbreak

To identify anomalies using LOF, expected data points are firstly defined using average LRD value, and anomalies are identified as points that defy the expectation. The intuition in density-based AD algorithms is that the average density (i.e., LRD value for LOF algorithm) of the cluster consists of $K$ neighbors should be similar to the density of a given point. In other words, a normal equity price should have a LOF value closer to $1$, vice versa. The LOF value can also rank the severity of anomalies, similar to the one in ARIMA approach as mentioned in Section~\ref{sec:uadarima}. In addition, as LOF detects anomalies by evaluating points locally against its neighbors instead of on a global scale, this approach is effective in identifying contextual anomalies by definition. \smallbreak

Unlike LOF that profiles normal behaviors, Isolation Forest identifies anomalies explicitly using a spatial partition scheme. The intuition behind Isolation Forest is that anomalies tend to scatter sparsely across the feature space, whereas normal points are closer to each other and located within a certain region. When splitting the feature map into regions, it is expected that less number of splits (i.e., shorter observation path) is needed for anomalies, whereas to isolate a normal point from its neighbors in a dense region, more splits are required. As the term "Forest" suggests, Isolation Forest is an ensemble learning method that consists of hundreds of different decision trees. Within each tree, a split means branching a parent node into child nodes, while the split threshold and the number of features taken into account are chosen randomly within a user-specified range. For each data point $A$, an anomaly score is assigned to indicate its severity, similar to the \emph{anomalous severity} value mentioned in Section~\ref{sec:uadarima}. The anomaly score of $i^{th}$ decision tree is calculated as

\begin{equation}
    IF_i(A, n) = 2^{-\frac{E(h(A)}{c(n)}}
\end{equation}

where $E(h(A))$ is the average number of splits to isolate $A$, $n$ is the number of external nodes, and $c(n)$ is the averaged path length that failed to search in the binary decision tree. To interpret the $IF$ score, a value closer to $1$ indicates the given point is anomalous, whereas for values smaller than $0.5$, the point is likely to be following its normal behavior. \smallbreak

Theoretically, the multivariate AD algorithms are more robust than the univariate approach, and has great potential to be extended into other more complex markets such as futures and options market. Additional equity-specific metrics such as delta, expiry date, the strike price with enriched information can also be taken into account when detecting anomalies. Nevertheless, careful feature engineering should be conducted by experts with domain knowledge so to ensure irrelevant variables are properly excluded.  \smallbreak

\section{Information Retrieval Module} \label{sec:irm}

\begin{figure}[!htbp]
\centering
\includegraphics[width = 0.8\hsize]{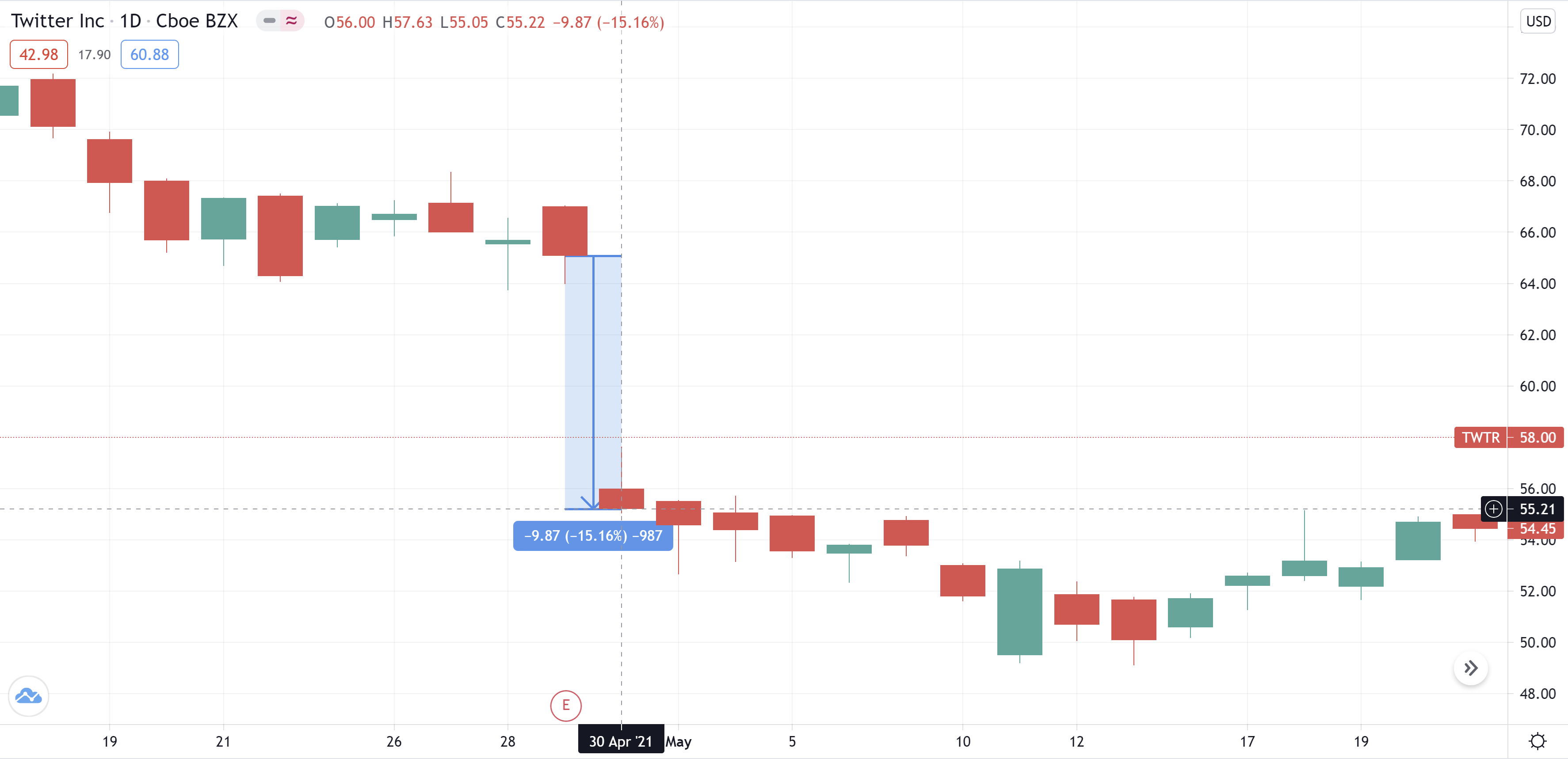}
\caption{Twitter Inc. stock price between 29 April and 3 May, 2021. The stock price dropped 12.85\% right after the market starts on 30 April, 2021, and closed with 15.16\% decrease before after-hours market.}
\label{fig:twitter-stock}
\end{figure}

The information retrieval (IR) module in DeepTrust aims to analyze, process and retrieve a collection of documents that matches a set of requirements that define a financial anomaly. In the below sections, the drastic tanking of Twitter Inc. stock on 30 April 2021 is used for demonstration purposes of how the query expansion works in a real-life anomalous price movement. As indicated in Figure~\ref{fig:twitter-stock}, the Twitter stock price sank over $12\%$ right after the Friday market starts and closed with a $15\%$ price drop due to its recently released financial reports with revealed first-quarter revenue and monthly active users (MAU) figures. Specifically, the performance of Twitter in 2021 is merely aligned with the estimate of financial analysts, and the company also announced weak future guidance, indicating its expected worse performance in the following quarters. The earliest analyst report published by Reuters\footnote{Reuters News: https://www.reuters.com/technology/twitter-may-struggle-replicate-bumper-2020-growth-analysts-2021-04-30/.} was at 18:50 Eastern Time, whereas thousands of users or retail investors were already widely discussing this anomaly on Twitter. The goal of the IR module is to recognize these anomaly-related tweets via learning unique features that distinguish irrelevant tweets from targeted ones. For DeepTrust, the IR module follows the ETR design mentioned in Section~\ref{sec:eor}, and the below sections outline the design and implementation details using this anomaly as an example. \smallbreak

\subsection{Twitter Search Query Modeling}

As mentioned in Section~\ref{sec:eorqm}, the objective of query modeling is to formulate an event-specific query $Q$ with $n$ lexical features $\{w_1, w_2, ... w_n\}$ for retrieving correlated tweets from the Twitter data stream. Query modeling begins from constructing the initial query using the output of the Anomaly Detection module, then applying various query enhancing techniques such as NER and PRF to improve the query quality by enriching $Q$ with strongly correlated textual or behavior information. The following sections outline steps of query modeling. \smallbreak

\subsubsection*{Query Initialization with Lexical and Temporal Features} \label{sec:iqfre}

Ideally, the query for the first retrieval should contain a \emph{(i)} temporal condition such as a date range to exclude outdated events \emph{(ii)} enriched lexical features that can precisely trace to the interested entity \emph{(iii)} dynamic query condition that is appropriately relaxed, so to avoid mistakenly exclude correlated tweets from the retrieval scope. Since the source of financial anomalies in this project has been restricted equity pricing data, it is reasonable to expect that the target equity is associated with a stock symbol or other standardized identifiers that are uniquely identifiable in major knowledge bases. This stock symbol, along with its variants such as company names or stock symbols in different formats, are essential elements that define the correlation of a tweet with the interested equity. In DeepTrust, the initial query is derived from the detected anomalies output by the Anomaly Detection module by aggregating key pieces of information, namely the \emph{(i)} anomaly date - the date when anomalous pricing movement occurred \emph{(ii)} entity - the stock symbol (e.g., ticker or RIC code\footnote{Ticker is an abbreviation to uniquely identify traded equity of a company in the exchange market, whereas RIC refers to Reuters Instrument Code, a ticker-like index for Refinitiv knowledge base. For example, Twitter Inc. has ticker \emph{TWTR} and RIC \emph{TWTR.K}.}) expressed as a unique sequence of characters. These two inputs collectively define the initial query with both lexical and temporal conditions, and query enhancement techniques are employed to make every endeavour including all potentially correlated tweets (i.e., improve recall of IR). \smallbreak

Firstly, since financial data providers often use different formats of stock symbols to represent the same equity (e.g., exchange-level ticker in Yahoo Finance, RIC in Reuters, regional-level Committee on Uniform Securities Identification Procedures (CUSIP) number, and global-level International Securities Identification Number (ISIN) in Bloomberg), it is necessary to detect and convert all symbols into the standardized RIC format for IR in the Refinitiv knowledge base. A rule-based detection algorithm is first applied to identify the format of stock symbol, and public financial knowledge bases such as Yahoo Finance and Eikon Data API are used for symbol conversion (i.e., symbology function in Eikon Data API and web scrapper in Yahoo Finance). Besides, equity fundamentals are retrieved from Eikon Data Item Browser, such as \code{TR.CompanyName}, \code{TR.CommonName} and \code{TR.AssetName}. These names are either official or frequently used terms on social media by the public to refer to the entity, such as \emph{"Twitter"}, \emph{"Twitter Inc."}, \emph{"TWTR"} for the company Twitter Inc. \smallbreak

For anomaly date, the range should be properly extended to cover the delay of information propagation on social media platforms. For instance, the Bitcoin (BTC) price plunged $30\%$ to $30,000$ USD/BTC on 19 May 2021, and this anomalous price movement is a collective consequence of various negative announcements released earlier over the past two weeks. Firstly, Elon Musk, the Key Opinion Leader and the CEO of Tesla Inc., has suspended BTC payment channels on Tesla cars due to environmental concerns of the crypto mining process on 12 May 2021. In addition, multiple nations, including the United States and China, have tightened regulations over BTC transactions to fight against potential financial crimes, and three state-backed Chinese financial organizations have announced official warnings regarding the risk of BTC on 18 May 2021. In order to capture all these factors causing the sharp drop in BTC price, news from the past $7$ days should all be included, whereas the number of days should be determined prudently to filter out outdated news from the retrieved collection of documents. In DeepTrust, the anomaly date is expanded to a range of $T-D$ days where the parameter $D$ can be either user-specified or derived from information diffusion models. To determine $D$ mathematically, a Twitter network can firstly be represented as a directed graph $G = (V, E)$ with $V$ represents users and $E$ represents follower-followee relationships. Upon receiving an external influence (e.g., negative news on BTC), the external influence metric $\theta$ can be estimated using multi-source Breadth-first search (BFS) algorithms as proposed by Sadikov and Martinez~\cite{sadikov2009information}. Formally, for each step $i$ of the influence from an external source $x$, the external influence metric is
\begin{equation}
    \theta_{ix} = \sum_{t=2}^{i} \min_{1 \leq k \leq t} \{d(v_t, v_k), d_{max} + \epsilon \}
\end{equation}
where $d(v_t, v_k)$ is the shortest path from user $t$ to user $k$ and $d_{max}$ is the longest relationship between two users in the network. On an aggregated level, the $\theta$ can be represented as a function of \% of infected nodes, and normalized to an appropriate scale to derive the value of $D$ for compensating the delay of information propagation. However, this approach requires extensive efforts in both data collection and analysis, and it requires language-independent features such as social network structure, which is out of the scope of DeepTrust. Therefore, it is recommended for users to manually specify the parameter $D$ to \numrange{3}{14} days considering the popularity of the company and empirical knowledge. \smallbreak

\subsubsection*{Query Enhancement using Named-Entity Recognition and Pseudo-Relevance Feedback} \label{sec:query-enhancement}

\begin{table}[!htbp]
\begin{adjustbox}{center}
\begin{tabular}{@{}ll@{}}
\toprule
\textbf{Steps}          & \textbf{Query}                                          \\ \midrule
Initial Search Query    & \{'ticker': 'TWTR', 'date': '2020-04-30' (YYYY-MM-DD)\} \\ \midrule
NER on Reuters News     & \{'ric': 'TWTR.K', 'dates': ['2020-04-30', '2020-04-29', ...]\}  \\ \midrule
PRF on Twitter &
  \begin{tabular}[c]{@{}l@{}}(stock OR price) (TWTR OR Twitter OR Twitter Inc. ...) \\ (\$TWTR OR \#TWTR OR \#TWTR.K...)\\ (Facebook OR India OR Jack Dorsey...)\\ lang:en -is:retweet is:verified -is:nullcast ...\end{tabular} \\ \midrule
Final Search Query &
  \begin{tabular}[c]{@{}l@{}}(stock OR price) (TWTR OR Twitter OR Twitter Inc. ...) \\ (\$TWTR OR \#TWTR OR \#business...)\\ (Facebook OR Amazon OR Apple...)\\ ('tank' OR 'consensus' OR 'P/E'...)\\ lang:en -is:retweet -is:nullcast ...\end{tabular} \\ \bottomrule
\end{tabular}
\end{adjustbox}
\caption{An example of search query evolution during each stage of the query enhancement process.}
\label{tab:ner-prf}
\end{table}

For query enhancement, two implicitly trusted knowledge sources are used, namely Refinitiv Eikon and Twitter Verified knowledge bases. In particular, a one-pass NER technique is applied on Refinitiv Eikon and an iterative PRF mechanism is used on the Twitter data stream. The objective is to identify correlated entities or keywords that are frequently mentioned collectively with the target equity, while curated information such as news articles, media posts from trusted businesses are sources of expansion. Ideally, the enhanced query with additional keywords can improve the recall, while the precision of IR is governed by the degree of granularity of lexical and temporal conditions. An overview of the formulated query for each step and its corresponding output is presented in Table~\ref{tab:ner-prf}. \smallbreak

In Refinitiv Eikon, a wide range of trusted financial news is collected from partnered news agencies and financial analysis corporations, including company and economic fundamental data, breaking news and analytical reports. A Question Answering (QA) system is used by Refinitiv to expeditiously locate and return relevant information upon receiving user queries. Although the actual Eikon QA system implementation details remain proprietary and confidential, it should fall into the category of Open-book QA, in which the answer to a question is explicitly stated somewhere in the database (e.g., the answer to query \emph{"Nvidia ARM"} should be a deal tearsheet of Nvidia Corp's acquisition on ARM Ltd.). A transformer-based Reader Model (i.e., reading comprehension) or an Information Retrieval model (i.e., locating related context in the repository of knowledge) are common architectures used to implement high-performance QA systems under the open-book condition. For DeepTrust, the QA task is even easier, in which a keyword-based matching system is sufficient to locate relevant financial news on a given entity. For instance, for an analytical report on the Acquisition proposal from Nvidia to chip designer ARM, it should be returned as a valid answer whenever correlated tags such as \emph{"Nvidia"}, \emph{"ARM"}, \emph{"Morgan Stanley\footnote{Morgan Stanley is the chief financial advisor of Nvidia Corp in the deal of acquiring ARM Ltd. for expanding its market in the electronics market.}"} are searched. Since the ultimate objective of DeepTrust is to relevant information that can explain a financial anomaly in real-time, only historical news published before the anomaly date are referenced for query enhancement. \smallbreak

Once all relevant Reuters news is retrieved from Refinitiv Eikon, an enhanced NER algorithm named Intelligent Tagging from OpenPermID API is applied to extract expansion words for the search query. In short, it analyzes unstructured data and extracts named entities and key phrases such as company names, geographical locations and persons from it. A Permanent Identifier (PermID) is assigned to each entity for relationship and relevance analysis. These PermIDs are further mapped with Refinitiv proprietary PermID taxonomy database maintained by experts in the financial domain, and generate entity resolution and relevance scores based on documents with similar context. Henceforth, for each identified entity, it can either be \emph{(i)} included as a part of the Entity Markup that is semantically identical to a known entity \emph{(ii)} resolved into a standardized form that can unambiguously be indexed in the knowledge base, and DeepTrust uses the resolved entity for query expansion whenever feasible. In addition, user-specified thresholds (e.g., minimum relevance threshold $>0.5$ and minimum importance threshold $>0.8$) are used to exclude irrelevant entities from contaminating the query, along with other common filtering rules (e.g., non-URL and non-numeric entities). Lastly, only the top $N$ frequent entities are selected, whereas the parameter $N$ is either specified by users or fine-tuned using an evaluation dataset. An example of the top $5$ entities returned for query expansion in the Twitter case aforementioned is:

\begin{quote}
\emph{Reuters NER Keywords}: ['Twitter', 'India', 'S\&P 500 Index - CBOE', 'Facebook', 'United States']
\end{quote}

These keywords from diverse domains are included in the final search query as optional matching keywords, for the purpose of improving recall at the expense of a lower precision. Nevertheless, an optimal value of parameter $N$ should balance these two performance metrics. For DeepTrust, the $N$ is tentatively set to $10$ based on empirical analysis, and the value is manually adjusted for different financial anomalies during evaluation. \smallbreak

\begin{algorithm}[!htbp]
  \caption{Improved Twitter Pseudo-Relevance Feedback Mechanism with Tags and Entities.}\label{alg:tprf}
  \begin{algorithmic}[1]
  \State Initial query $Q$ = $q_1, q_2, \ldots, q_k$ enters the PRF process.
  \While{True}
      \State Retrieve top $\gamma$ documents $d_1, d_2, \dots, d_j$ from Twitter using query $Q$ and store as collection $D$.
      \State Initialize a collection of tags and entities $H$ for query expansion.
      \For{$d_i \in D$}
        \State Extracts tags and entities in the tweet content as collection $H_i$.
        \State For $h_j \in H_i$ increment the counter in $H$ by 1.
      \EndFor
      \State Sorts the collection $H$ and appends $Q$ using the top $\gamma$ tags and entities.
      \If{all top $\gamma$ tags and entities already in $Q$}
          \State Query Expansion is completed, and $Q$ has stabilized. 
      \EndIf
  \EndWhile
  \end{algorithmic}
\end{algorithm}
\smallbreak

For Twitter, the unsupervised PRF method is used to expand the query iteratively until the set of keywords is stablized. The pseudocode of PRF is presented in Algorithm~\ref{alg:tprf}. Generally, in each iteration, topic-related words in the top $\gamma$ retrieved tweets are assumed correlated with the entity, and are selected for relevance feedback in the next iteration, while these words can be in the form of lexical features (e.g., individual words, phrases) or Twitter-specific features (e.g., hashtags, emojis). In DeepTrust, an enhanced PRF method with entity and context recognition and Twitter verification is utilized for query enhancement. Firstly, different from the NER algorithm used in Refinitiv Eikon, the entity and context recognition algorithm is employed to extract both lexical features and its underlying contextual meanings (e.g., domains such as \emph{trading}, \emph{finance} or \emph{money}) that can be used to directly exclude tweets from an irrelevant domain. For instance, for a tweet that discusses the stock performance of Twitter Inc., its contextual domain should not be primarily composed of words from the Sports domain, but should contain terminologies or common layman terms from the financial domain like \emph{"price"} and \emph{"money"}. In addition, the search query is also refined with Twitter-specific filters (e.g., verified flags, hashtag flags, media duration, etc.) to extract information with high granularity. In particular, verified Twitter accounts refer to individuals or organizations that are recognized as \emph{authentic}, \emph{notable} and \emph{active} by Twitter, and are validated using a government-issued identification document or a company-registered domain, which screens the majority of bot and spam accounts that may corrupt the query with falsified content. The majority of those verified accounts are news agencies with an established reputation, in which their published tweets can be implicitly trusted. \smallbreak

As an example, the below list consists of keywords and tags extracted from the top $10$ tweets of the Twitter financial anomaly in its first retrieval: 

\begin{quote}

\emph{Cashtags}: [\$TWTR, \$FB, \$twtr] \smallbreak
\emph{Annotations}: [Microsoft, U.S., Google, Amazon, \$TWTR, Refinitiv] \smallbreak
\emph{Hashtags}: [\#risks, \#priceaction, \#makingmoney, \#stockmarket, \#marketalert] \smallbreak

\end{quote}

Based on the experimental results discussed in~\cite{bektemirov2015tweetement}, the PRF method using top $2$ frequently mentioned hashtags as feedback yielded the greatest level of originality and retrieval efficiency comparing to non-expanded query or keyword-based expansion. To take advantage of both Twitter verification and Twitter-specific features, the top $2$ hashtags, cashtags, and annotations mentioned in tweets from Twitter verified accounts are selected and used for query expansion iteratively, and the entire query enhancement process terminates when no new hashtags appear in the retrieval results. \smallbreak

\subsubsection*{Query Finalization and Composition Analysis} \label{sec:query-finalization}

The finalized Twitter search query (e.g., the Final Search Query in Table~\ref{tab:ner-prf}) consists of multiple components that collectively strive to cover as many relevant tweets as possible, and can be visualized as a combination of $6$ components:

\begin{quote}
    \emph{Query} = (\textcolor{MidnightBlue!60}{domain}, \textcolor{BurntOrange!60}{names} or \textcolor{red!60}{tags} or \textcolor{Plum!60}{related entities} or \textcolor{Sepia!60}{descriptive keywords}, \textcolor{ForestGreen!60}{parameters})
\end{quote}

The \emph{domain} refers to the category of the interested entity in the exchange market, typically as an investible and tradable asset. Common categories include stock, bond and commodity markets, and each asset within these domains has a market price, as the function of demand and supply dynamics. \smallbreak

The \emph{names} are frequently mentioned titles of the interested entity or correlated key decision makers that are often implicitly linked with the entity, which are the most popular approaches for users referring to an entity. For instance, a user may use a ticker with hashtag or cashtag (i.e., \$TSLA or \#TSLA.O), or the company name Tesla to refer to Tesla Inc. The collection of correlated names of the interested entity are retrieved during both the PRF process in Twitter and NER process in Reuters News. In addition, key decision makers that are affiliated with the interested equity are also included, as users may address a financial anomaly using solely the name of the founder or CEO. For instance, Bill and Melinda Gates announced their divorce in May 2021, and thousands of tweets were discussing this news and its possible impact on the market. As a potential influential factor of MSFT stock price, conditionally restrict the search criteria to the entity name may mistakenly filter out relevant tweets that are discussing the divorce but not addressing the company by its name. Therefore, DeepTrust leverages both names of the company and key decision makers as matching criteria.  \smallbreak

The \emph{tags} are Twitter-specific features named Hashtag and Cashtag. The hashtag (i.e., words start with \# symbol) is used to index topics on Twitter, and cashtag (i.e., ticker starts with \$ symbol) is used to index ticker symbols. Both tags are frequently used by users on Twitter to promote their posts to a wider population who follow these tags, and these features reduce the complexity of locating or filtering information from a specific theme. \smallbreak

The \emph{related entities} are named entities that are deemed correlated as appearing in relevant financial articles in Reuters News. These entities are usually mentioned alongside the interested entity, and are potentially relevant to the cause of the financial anomaly. For instance, the company Facebook, as a competitor in the social networking service market, is constantly mentioned in financial news of Twitter Inc. during \numrange{27}{30} April 2021, in which the expected performance of Facebook is used by analysts to approximate the expectation of Twitter in later quarters, and partially cause the price crash after the weak future guidance. \smallbreak

The \emph{descriptive keywords} describe the anomalous pricing movement and its cause using financial terminologies. For instance, to describe a decrease in stock price, professional traders usually use \emph{"price tanking"}, \emph{"crash"} or \emph{"decline"}. Similarly, fundamental indicators like \emph{"P/E ratio"} and \emph{"cash flow"} may frequently appear when discussing company performance. These keywords may help DeepTrust to capture related entities that are exhibiting similar pricing movements, which may constitute explanatory evidence of the financial anomaly. \smallbreak

The \emph{parameters} in the search query governs the search scope in Twitter using content-dependent flags and language filters. For instance, using \emph{"\-is:nullcast"} can remove advertising tweets from the search result, and \emph{"\-is:retweet"} to exclude retweets and only retrieve independent tweets from its author. Other parameters such as language filters and data range filters are also used to restrict the search results within a narrowed range. \smallbreak

\subsection{Tweets Retrieval and Ranking} \label{sec:tweets-retrieval-and-ranking}

Apart from the Query Modeling component, the remaining two ETR components, Event Representation and Event Ranking and Retrieval are already implemented by Twitter on the server-side. Although Twitter has never officially released any information on how its social media algorithm (i.e., \emph{"In case you missed it (ICYMI)"} feature in Twitter) works on responding to user queries, the underlying principle is similar to the traditional ranking and retrieval model. Based on empirical evidence, the ranking of tweets when responding to a non-registered user depends only on its content, including recency, relevancy, confidence, multi-media and other confidential factors. For instance, tweets with implicit trust (i.e., Twitter verified account) or with excellent public metrics (e.g., number of likes, retweets, etc.) are anticipated to be returned as top results, whereas tweets with fewer interactions are often neglected from the first page. This ranking mechanism is particularly useful during the PRF process, because the objective of PRF is to retrieve correlated entities only in the top $\gamma$ documents, and the Twitter ranking algorithm is able to rank tweets using its proprietary algorithm with high quality. Nevertheless, the ranking model from Twitter is only useful for queries with no date range is specified, whereas for the final search with a specified date range, the ranking algorithm is implicitly disabled as all tweets that satisfy the search query condition are returned iteratively using search pagination. \smallbreak

In addition, it is noticed that the dynamics of the Twitter ranking algorithm lead to inconsistent results of identical queries. In particular, if two identical queries are submitted simultaneously using two clients, the retrieval results of the top 500 tweets might be different in terms of order and content. Although this inconsistency is not frequently observed throughout the project, it may affect the PRF process as expansion entities and tags are only extracted from the top-ranked tweets. To mitigate this issue, DeepTrust implemented a buffering mechanism that allows a stabilization period of $5$ iterations once the expansion keywords list has converged to a fixed set. \smallbreak

Lastly, regardless of the benefit brought by using the Twitter verification system to improve the implicit trust of the expansion keywords list, the number of available tweets is significantly reduced depending on the popularity of the entity. This is primarily due to the privacy concern, as the Twitter verification system requires users to submit official government-issued documents for identity validation, and most users are unwilling to upload their IDs to Twitter for the verified status. For less popular equities, the number of verified Twitter accounts that are discussing the anomaly is drastically low, approximately only \numrange{5}{10} verified users with less than $50$ tweets during a period of $7$ days. Therefore, DeepTrust provides a parameter to specify if the verified flag is enabled or not depending on the judgment call of financial analysts on entity popularity, and allocating more responsibilities to the reliability assessment module to filter out irrelevant information. \smallbreak

\section{Reliability Assessment Module} \label{sec:ram}

\subsection{General Tweet Pre-Processing for Reliability Assessment} \label{sec:general-preprocess}

While analyzing tweets collected by the Information Retrieval module, it is observed that each tweet contains several noisy features that may obstruct the reliability assessment components in reaching a justified result. For instance, the following is a typical example of tweets retrieved by the IR module:

\begin{quote}
    \emph{\$TWTR Is Twitter a Buy or Sell After Its Post-Earnings Plunge? 
        \#technicalanalysis \#investment :-) https://t.co/...}
\end{quote}

The first feature to be removed is the URL link (\emph{"https://t.co/...}) appended by Twitter for identifying the dependencies associated with the tweet. The link is typically used by Twitter to track its associations, either linked to its quoted content or the tweet itself if it is an original post. In practice, this link may confuse synthetic text verifier and argumentative detector as including the link may mislead the model to believe the text is trustworthy, whereas the link itself contains no useful information for any reliability assessment components. For instance, for a tweet containing only hashtags and emojis, the RoBERTa neural fake text detector predicts its real probability only as $58\%$ whereas with Twitter embedded link included, its estimated real probability immediately increases to $86\%$ while no substantial information is indeed added. Therefore, a regular expression is used to remove these links from tweets before supplying them to any reliability assessment components. \smallbreak

In addition, emojis or emoticons are converted to interpretative words for better semantic interpretation and context understanding. For instance, an emoticon \emph{":-)"} that transforms to \emph{":smiley-faces"} can provide useful insights into the emotional context of the tweet, and a similar argument can be applied to emojis which instead of converting to standard Unicode tokens, enriched information stored in interpretive words like \emph{"blinking-stars"} can be beneficial for contextual-related tasks. However, this pre-processing technique should only be applied for filters other than synthetic text filter, as these additional words may mislead the neural detector into falsely believing the tweet is written coherently and was not generated by a language model. Therefore, in DeepTrust, emoji-word conversion is by default disabled, and is only applied when manually specified by users. \smallbreak

\subsection{Feature-based Screening using Heuristics and Threshold-based Classifiers} \label{sec:feature-filter}

The first phase of reliability assessment is conducting a preliminary feature-based microblog credibility assessment on the collection of retrieved tweets. The study from Krishnan and Chen is referenced for selecting prominent features that are correlated to tweet reliability. In DeepTrust, features are separated into three groups, namely \emph{(i)} textual features - statistical information on lexicon features \emph{(ii)} tweet-meta features - tweet metadata that captures interactions within the social network such as \emph{"Likes"}, \emph{"Retweets"} and \emph{"Replies"} \emph{(iii)} user features - attributes that describes the social networking profile of a Twitter user. Among all these types of features, the survey results from Morris et al. are used to rank feature importance by its impact on perceived credibility~\cite{morris2012tweeting}. Specifically, the $3$ most prominent features that affect the perceived reliability of a tweet are \emph{Twitter verified account / validated expertise in the topic}, \emph{contained URLs from a trusted domain}, \emph{published by an author with an active footprint in Twitter}\footnote{DeepTrust only examines features that are user-account independent. Features that involve social network analysis are excluded in DeepTrust, such as authors that users are following.}, and other metrics such as the number of followers or the impression of author bio have also received considerable attention from users when evaluating tweet credibility. However, due to the absence of an annotated tweet dataset from the financial domain, the feature-based screening system in DeepTrust is not trained using supervised learning with classifiers such as SVM to determine reliability. Instead, it constitutes a complementary filter in reliability assessment as discussed in~\ref{sec:fbca}, with relaxed conditions that solely exclude trivially unreliable tweets from the collection. Therefore, instead of applying sophisticated feature extraction or defining complex rules on feature-based reliability assessment, simple heuristics are used to perform a quick round of preliminary filtering, and this component is only used for screening the simplest forms of unreliable information. \smallbreak

\begin{quote}
    \emph{Textual Features}: [\% Hashtags, \% Cashtags, \% Links, \% @Mentions, Profanities] \smallbreak
    \emph{Tweet-Meta Features}: [\# Retweets, \# Replies, \# Likes, \# Quotes] \smallbreak
    \emph{User Features}: [\# Followers, \# Following, \# Tweets, \# Lists, Creation Date] \smallbreak
\end{quote}

For textual features, DeepTrust focuses on the distribution of different categories of text elements within a tweet, and review if it intends to artificially amplify or manipulate information on Twitter, which is defined as Spam by Twitter~\cite{twitter2021rules}. As one of the guidelines suggested by Twitter, the misuse of URLs, tags and mentions are violating the Twitter policy by using Twitter service with abusive or disruptive behavior, and some typical examples are \emph{(i)} including excessive unrelated hashtags, cashtags and @mentions of influencer accounts that are different from the topic domain to drive traffic or expand influence \emph{(ii)} adding misleading links that are irrelevant to the main topic of the tweet. In general, these spamming behavior can be detected easily via analyzing the textual composition, including the distribution of each feature (i.e., words, punctuation, tags, emojis, and links) and the correlation among features (e.g., the average semantic similarity between pairs of hashtags). For DeepTrust, the system uses a threshold $\beta$ to exclude tweets contain more than $\beta \%$ of tags, links and mentions, as the remained text can be reasonably assumed not sufficient for constructing a valid argumentation structure. In addition, DeepTrust leverages a profanity filter to remove tweets composed of offensive or explicit languages that are certainly irrelevant to the financial anomaly. As discussed, the objective of DeepTrust is to identify tweets that are reliable by either having an argumentation structure with supporting evidence or being an objective statement, and none of these two forms should contain profanities, which are subjective representations of offensive words. Therefore, tweets are firstly vectorized using a BoW model, and a pre-trained linear SVM classifier calibrated with logistic regression is used to estimate the probability of them being offensive or explicit. For tweets that have a probability higher than a user-specified threshold, they are marked as unreliable by the feature-based filter. \smallbreak

For tweet-meta features, metrics that capture tweet impression and engagement are used for assessing credibility. In the design of DeepTrust, financial anomalies are not detected in real-time, which means at the time of information retrieval, tweets with enriched information are likely to have certain numbers of interactions in the forms of tweet-meta features listed above. For tweets with zero engagement metrics, the ranking algorithm is likely to assign a lower score because it is likely to be spam information with unimportant words that are merely correlated with the financial anomaly. Thus, these tweets are often placed in lower positions of the Search Engine Results Page (SERP), and most Twitter users are used to interact with tweets from the top hundreds of organic positions. Among all metadata associated with the Tweet, the four most prominent features that collectively capture the most popular approaches users may interact with tweets are selected, namely number of retweets, replies, likes and quotes\footnote{Quotes refers to retweet with a comment, and considered as a standalone tweet by itself.}. In the Twitter sample discussed in Section~\ref{sec:irm}, a typical feature-filtering algorithm that removes tweets with zero public interactions (i.e., value $0$ for all four features) can remove approximately $52\%$ tweets from the collection. \smallbreak

For user features, a similar logic is applied to filter unreliable authors. Four important attributes (i.e., number of followers, following accounts, posted tweets, and listed accounts) of the social networking profiles are selected based on their correlation with the reach of influence and level of engagement, and used collectively to evaluate user reliability. In addition to public metrics, the creation time of the account is also taken into account, which is effective in filtering bot-generated accounts created recently for the sole purpose of spreading disinformation or other malicious intentions. \smallbreak

To further improve the re-usability of the DeepTrust framework for potential real-time reliability assessment tasks, the core feature-based filter component is implemented as a module that can be configured by domain experts with user-specified thresholds and customized evaluation logic. For each tweet in the collection, the reliability flag \emph{'feature-filter'} is updated based on the output of feature-based filter, with \emph{true} being reliable and \emph{false} being unreliable. For all other filters mentioned later, each filter corresponds with a separate reliability flag, and only tweets have \emph{true} for all reliability flags are deemed as reliable by the DeepTrust framework. \smallbreak

\subsection{Synthetic Text Screening using Discriminative Model and Classifiers}

To identify and filter out fabricated tweets generated by language models, three countermeasures (i.e., RoBERTa, GPT-2-XL and BERT-large-cased detectors) are employed along with Plurality voting in reaching a consensus on textual content authenticity. In short, synthetic text detection is a sequence classification learning task, in which the objective is to determine a category based on a sequence of textual information as expressed in the tweet. Two mainstream approaches adopted in DeepTrust, which are fine-tuning based with discriminative model and supervised-learning based with generative model and classifier, are discussed in the following sections. \smallbreak

\subsubsection*{Fine-tuning Based Detection using RoBERTa Discriminative Model} \label{sec:roberta-model}

The first detection mechanism used by DeepTrust is via fine-tuning a language model, typically called a discriminative model that detects synthetic text generated by itself or another model with similar architecture. The GPT-2D, which is a RoBERTa model trained on the output of WebText and GPT-2-XL generated output as discussed in Section~\ref{sec:gpt}, is selected based on its high performance and flexibility in detecting a diverse range of fake information accurately. GPT-2D uses sequence classifiers of RoBERTa-base\footnote{RoBERTa-base (125M): \url{https://openaipublic.azureedge.net/gpt-2/detector-models/v1/detector-base.pt}.} and RoBERTa-large\footnote{RoBERTa-large (356M): \url{https://openaipublic.azureedge.net/gpt-2/detector-models/v1/detector-large.pt}.} fine-tuned by OpenAI, which are the robustly optimized version of BERT by removing the NSP objective and focusing solely on MLM objective, and training for larger mini-batches, learning rates, and iterations~\cite{liu2019roberta}.  \smallbreak

Instead of constructing and training a customized neural fake news detector from scratch, DeepTrust uses transfer learning to improve fake news detection performance by leveraging knowledge learned from previous relevant tasks which are transferable to other downstream NLP tasks. This technique is beneficial by both reducing development time and avoiding the demand for intense computation resources throughout the training phase. DeepTrust uses model checkpoints released and published on HuggingFace, which are fine-tuned using general corpora and synthetic text generated by GPT-2-XL (1.5B) under a paired setting. The training data consists of three sources, namely \emph{(i)} 250K samples selected from WebText\footnote{WebText is an internal OpenAI corpus consists of documents from 45 million web pages for a total around 40 GB in size.} labelled as "Human" \emph{(ii)} 250K samples generated by GPT-2-XL with temperature $1.0$ and no $K$ truncation labelled as "Machine" \emph{(iii)} 250K samples generated by GPT-2-XL with $K=40$ truncation labelled as "Machine". The temperature is a hyperparameter that governs the randomness of sampling in the Boltzmann distribution. The lower the temperature, the more conservative the model is in incorporating randomness in the output. For instance, a temperature of $0.0$ forces the model to output deterministic completion for a given prompt, whereas a temperature of $+\infty$ forces the model to uniformly sample words regardless of the confidence of predictions. The $K$ value is used to control prediction diversity by sorting predictions by confidence and removing tailing predictions if their ranks are lower than $K$. This is usually to avoid models from sampling irrelevant words frequently under a high-temperature environment, and improve text generation quality by reaching an optimal balance between random and deterministic. Overall, discriminative models like RoBERTa has proven to be more robust than generative models in detecting synthetic text generated by itself as it focuses on learning detection while less relevant to the generation itself, with on average $96.4\%$ accuracy\footnote{Accuracy achieved by a RoBERTa-large model trained with synthetic text generated by GPT-2-XL using nucleus sampling.} when detecting GPT-2 neural fake text under three sampling conditions (i.e., temperature, K-truncation and nucleus sampling)~\cite{solaiman2019release}. \smallbreak

Regarding the design and configuration of the RoBERTa model, there are several adjustments made to further enhance its performance on shorter text with informal language styles and the usage of non-standard lexicons such as emoticons and hashtags. Firstly, byte-level Byte Pair Encoding (BPE) tokenizer is used for preparing input for the model, and is well-known for handling vocabulary issues such as representing out-of-vocabulary words. This is particularly useful for tweets due to the presence of emojis and hashtags. For instance, since hashtags prohibit adding spaces between words, the majority of them take the form of combined words without spaces in between them, such as \emph{"\#technicalanalysis"}. Traditional BPE can easily tokenize these neologisms as a combination of separated known tokens of \emph{"technical"} and \emph{"analysis"}, whereas the byte-level BPE is an enhanced version, in which the base vocabulary set contains all Unicode characters. In other words, the byte-level BPE can tokenize every possible word, emoji or punctuation without needing the \emph{$<$unk$>$} tokens. Lastly, a sequence classification top layer has been added to the model, which directly outputs the probability of a given input text being synthetic or human-written. The result is easily interpretable, and a decision can be made by using a minimum confidence threshold of $0.5$. \smallbreak

\subsubsection*{Supervised Learning Based Detection using GLTR, GPT-2 Generator and SVM Classifier}

The second detection mechanism is to distinguish synthetic text using a classifier trained on the language model probability distribution of domain-specific tweets and neural tweets generated by a fine-tuned language model. It consists of three components, namely a generator model for producing synthetic text for training, two language models to reproduce the language modeling process based on GLTR, and a classifier that is trained on both human-written tweets and synthetic tweets. \smallbreak

\textbf{Phase 1: Generator Model Training Data Preparation} \smallbreak

Since the task is supervised learning, the first phase is to annotate a collection of domain-specific human-written tweets for training, one naive approach is to use all retrieved tweets from the Information Retrieval module with the assumption that all of them are written by a human. Although the feature-based screening should effectively remove the majority of trivial bot-generated tweets, more deceptive synthetic text written by complex language models such as BERT or XLM may be inaccurately labeled as \emph{"Human"}. To tackle this issue, a chain of trust is established in which the output from the RoBERTa discriminator is trusted as the root entity. In other words, with a confidence threshold of $0.5$, only tweets with a real probability of $p > 0.5$ deemed by the RoBERTa discriminator are used for fine-tuning the generator model. For the Twitter financial anomaly example discussed before, only $88,936$ out of $241,714$ tweets satisfy this threshold, in which all of them are implicitly trusted to be written by a human after filtering by both interaction metrics (i.e., feature-based screening discussed in Section~\ref{sec:feature-filter}) and neural discriminative model (i.e., RoBERTa discriminative model). \smallbreak

\textbf{Phase 2: Generator Models Evaluation and Fine-Tuning} \smallbreak

The second phase is to fine-tune a language model for generating synthetic text, in which the collection of generated synthetic text is later used along with implicitly trusted tweets under a paired setting to train the classifier. The objectives of the generator model are \emph{(i)} able to generate domain-specific synthetic text \emph{(ii)} able to generate topic-related content (e.g., controlled text generation of GROVER) \emph{(iii)} consists of Twitter-specific features such as emoticons, net slang or cashtags outside of the general corpora used for training the pre-trained model. The following is a list of candidate generator models for consideration: \smallbreak

\begin{itemize}
    \item \emph{GROVER}: GROVER is a fine-tuned GPT-2 that is trained over a news-only dataset. With its controllability and quality of text generation, the synthetic text generated by GROVER can be considerably deceptive for even humans to distinguish. However, GROVER is powerful in generating lengthy paragraphs with a formal writing style from a third-person viewpoint, which is uncommon in Twitter except for posts from news agencies. In addition, the time needed to generate one piece of text is approximately $6$ minutes\footnote{Estimation based on experiments of generating one sample with controlled parameters using Google Colab standard GPU runtime.}, significantly longer than using a standard GPT-2 language model. Lastly, there is no mature text truncation mechanism other than a simple character-based cut-off to force the generated text to satisfy the character limit required by Twitter, which can severely damage the coherence of the GROVER text. 
    \item \emph{DeepTweets}: DeepTweets\footnote{DeepTweets is another fine-tuned version of GPT-2 language model specialized in generating account-specific fake tweets using historical posts as training data. Available at \url{https://lexfridman.com/deeptweets/}.} is an experimental project by Lex Fridman to generate fake tweets that mimic the writing style of individuals. It is based on GPT-2-medium (345M), and is fine-tuned using historical tweets of the targeted individual. This approach is insightful for DeepTrust, and a similar approach is adopted to fine-tune a GPT-2 generator as discussed in the following section. 
    \item \emph{GPT-2-medium Generator}: As the state-of-the-art few-shot learner that is powerful in language modeling, GPT-2 is the ideal choice of synthetic text generator for generating artificial tweets. The reason why DeepTrust only uses GPT-2-medium (345M) is two folds: \emph{(i)} smaller model requires less computing resources and dedicated GPU memory to fine-tune, along with shorter training time \emph{(ii)} comparing to the WebText used for training the general GPT-2, for each financial anomaly there are only around \numrange{30}{200}K tweets available. The lack of training samples may cause overfitting, especially if using a large model like GPT-2-XL that can simply memorize tweets instead of generalizing useful knowledge. Therefore, for optimal fine-tune performance, a smaller model like GPT-2-Medium is the ideal candidate.
\end{itemize}

With the above evaluations, GPT-2-medium is selected as the generator model for generating training data for the DeepTrust neural text classifier. To fine-tune this generator model, instead of focusing on the MLM objective adopted by RoBERTa discriminator in Section~\ref{sec:roberta-model}, the causal language modeling (CLM) loss is used to enforce unidirectional language modeling behavior. Both MLM and CLM model the probability of the masked token given context words, whereas MLM provides both words on the left and right to predict the masked token, and CLM provides only words that occurred to the left for prediction. In general, MLM is commonly used in a scenario in which a good representation of the document is required, while CLM is usually adopted for training a model in generating coherent synthetic text. For the generator model, CLM is the better objective to focus on, and the cross-entropy loss is then calculated and backpropagated to update model weights. \smallbreak

Regarding the fine-tuning parameters, the learning rate is specified to $5e^{-5}$ as a smaller value can prevent the model from updating its pre-trained weights drastically while losing its generalizability from its knowledge of the general corpora. The optimal number of epochs is decided to be $1$, as it is proven sufficient for the model to stabilize at a constant range of CLM loss during evaluation for both training and testing data. \smallbreak

\textbf{Phase 3: Synthetic Text Generation using GPT-2-medium} \smallbreak

With a fine-tuned generator model, the next phase is to apply tokenizer and configure sampling parameters (e.g., temperature, $K$, etc.) for synthetic text generation. The same byte-level BPE mentioned in Section~\ref{sec:roberta-model} is used for tokenizing the input text. For each generation, a \emph{prompt} or a hint of the main theme is provided to the model as the start of the sentence. For instance, if the beginning of sequence symbol \emph{$<$BOS$>$} is given as the \emph{prompt}, the language model will generate a text randomly based on its predictions and sampling preference. Instead of generating random synthetic sentences as training samples, DeepTrust leverages information from the implicitly trusted tweets, and use the first $n$ words (i.e., $n$ is a random integer between $2$ to the $1/3$ of the length of the tweet) as the \emph{prompt} to guide the language model during neural text generation. In other words, each synthetic text is generated under controlled settings similar to the mechanism adopted by GROVER. In addition, the temperature is set to $1.0$ for introducing a reasonable amount of randomness in language modeling, and Top-p (nucleus) sampling is adopted with $p=0.9$. The nucleus sampling, unlike the K-truncation technique discussed in Section~\ref{sec:roberta-model}, only samples from the smallest set of predictions whose cumulative probability exceeds the $p$ value, and the probability mass is redistributed among this collection of predicted words. In general, nucleus sampling is more flexible than K-truncation by dynamically selecting the set of words, but may sometimes ignore low-ranked keywords that may otherwise be included in a Top-K sampling scheme, thus a combination of both sampling techniques is also common in practice. For DeepTrust, the parameter may be overridden by users in the configuration file, but the default parameter is set to only use nucleus sampling with $p=0.9$, as it yields the best result for the Twitter financial anomaly as evaluated by the author. \smallbreak

\textbf{Phase 4: GLTR Evaluation with GPT-2-XL and BERT-Large-Cased Models} \label{sec:neural-phase-4} \smallbreak

For both collections of tweets (i.e., implicitly trusted human-written tweets and synthetic tweets generated by GPT-2-medium), DeepTrust applies the forensic tool GLTR to gather statistical information about predictions during the process of language modeling. As discussed in Section~\ref{sec:gltr}, GLTR aims to detect synthetic text by using an identical model or models with similar architecture to reproduce the language modeling process. For synthetic text, each word in the text should be more likely to match the most confident predicted word proposed by the language model used in GLTR, whereas for human-written text, words should be less predictable or not included in the subset of sampling due to the uncertainty in human writings. DeepTrust leverages two most commonly used models or referenced architecture for the task of language modeling: \emph{GPT-2} and \emph{BERT}, and customized the GLTR tool so to support using the latest, most powerful GPT-2-XL and BERT-Large-Cased models for evaluation. \smallbreak

For each text, the fractional probability score $frac(p)$ is calculated, which is the ratio between the probability of the most confident prediction by the GLTR model and the actual probability of the next word in the probability distribution. Formally speaking, for $i^{th}$ word $X_i$ in the input text with probability $p(X_i|X_{1:i-1})$,

\begin{equation}
  frac(p_i) = p(X_i|X_{1:i-1}) / p(Y_1|X_{1:i-1})  \quad \quad frac(p_i) \in [0, 1]
\end{equation}

where $p(Y_1|X_{1:i-1})$ represents the probability of the top predicted word from the language model used by GLTR. For a synthetic text with $i$ words generated using GPT-2-medium with $K=1$ and temperature = $0.0$ (i.e., deterministic prediction by always select the top predicted word), the $frac(p_i)$ for every word should equal to $1.0$ when another GPT-2-medium model is used by GLTR to calculate the fractional probability score. In other words, in the distribution of $frac(p_i)$ for a given text, a synthetic text should have a mean value closer to $1.0$ compared with a human-written text, as the language model in GLTR is more confident of each prediction when the generator model is using a sampling strategy like K-truncation or nucleus.  \smallbreak

\begin{quote}
    Human-Written: 
    \emph{"Twitter's stock price dropped following the social-media giant's removal of President Donald Trump's account \#Biz \#BizNews \#Business \#Economy \#Market \#News \#StockMarket \#Stocks \#Tech \#Trending \#Trump \#Twitter"} \smallbreak
    Synthetic: 
    \emph{"Here are today's Stock Market Leaders: 1. Twitter (TWTR) 2. Discord (DF) 3. Tesla (TSLA) 4. Google (GOOGL) 5. Facebook (FB) 6. Amazon (AMZN) 7. StifelMcQuinn"}\footnote{The synthetic tweet is generated using fine-tuned GPT-2-medium based on this retrieved tweet: "Here are today's Stock Market Leaders: 1. \$TSLA 2. \$AMZN 3. \$AAPL 4. \$MSFT 5. \$MVIS 6. \$AMD 7. \$UPS 8. \$NIO 9. \$GOOGL 10. \$FB See the full list at https://t.co/wBMiVBND4t"}
\end{quote}

\begin{figure}[!htbp]
\centering
\includegraphics[width = 1\hsize]{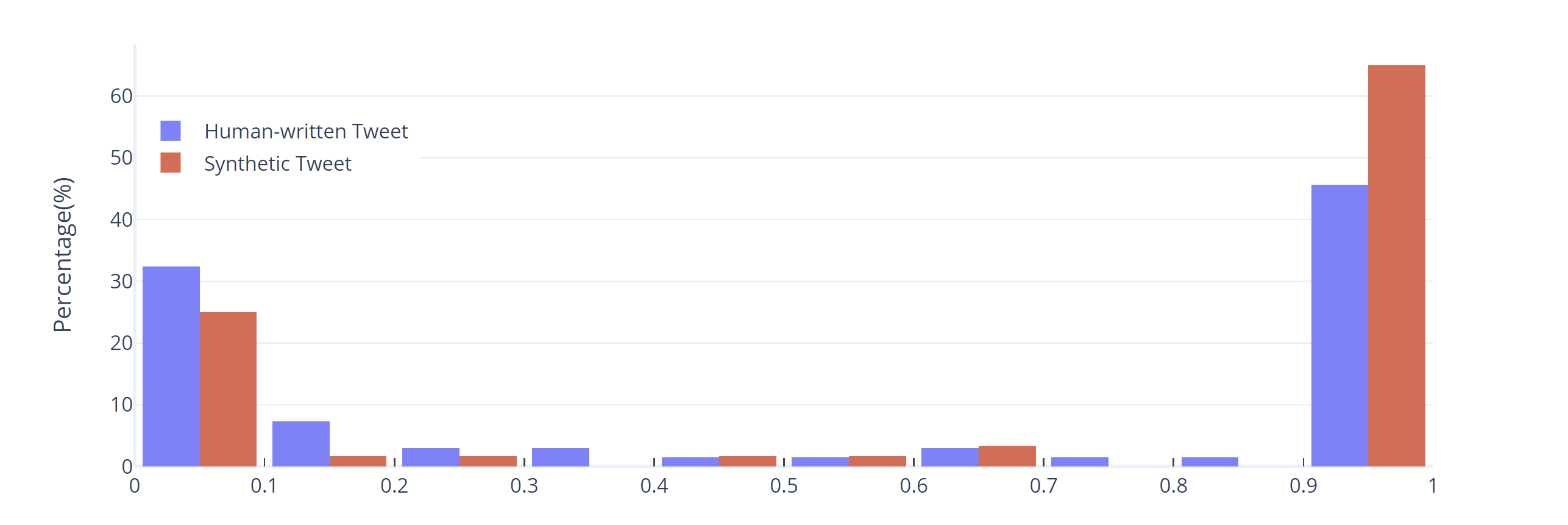}
\caption{The $frac(p)$ percentage distribution histogram of a human-written tweet and synthetic tweet.}
\label{fig:gltr-comparision}
\end{figure}

In addition, the $frac(p)$ distribution is further converted to a frequency domain and can be visualized as a histogram. An example is provided in Figure~\ref{fig:gltr-comparision}. For each bar, instead of counting the absolute value of occurrences, the percentage of words in each bin is used, equivalent to applying normalization with a normalizing constant. Instead of using min-max normalization, a direct conversion to the percentage scale is selected because min-max normalization requires a known min and max value (i.e., minimum and maximum amount of words may present in a tweet) are required, whereas converting to a percentage scale avoids fixing the scaling magnitude while still normalizing the data to a standard scale with all essential information preserved. For instance, the above section presents two tweets that are \emph{(i)} written by a human and validated by the author \emph{(ii)} generated by GPT-2-medium generator model using the configuration specified in Phase 3. When applying the GPT-2-XL model with GLTR to calculate the $frac(p)$ percentage distribution, the resulting histogram can be visualized in Figure~\ref{fig:gltr-comparision}, and the two distributions are distinguishable. Specifically, both distributions are convex-shaped whereas synthetic tweet is more right-skewed than the human-written tweet, primarily due to the difference of sampling strategy adopted by a generator model and a human. Therefore, the finalized input feature for the classifier is a matrix $X \in {\rm I\!R}^{n \times 10}$ where $n$ is the number of tweets in the collection that has passed the feature-based screening test mentioned in Section~\ref{sec:feature-filter}, and $10$ is the number of bins in the histogram. \smallbreak

\textbf{Phase 5: SVM Classifier Fine-Tuning and Detection} \smallbreak

The last phase is to train and fine-tune a classifier to perform binary classification on synthetic and human-written text. A non-linear SVM classifier with a kennel, which applies dimension reduction technique or a transformer on the input features, is used to approximate a non-linear decision function for classification. In DeepTrust, two kernels, namely the polynomial (poly) and radial basis function (RBF), are selected as they outperform others in terms of accuracy and weighted average F1 score. For each kernel, three parameters are fine-tuned by grid search algorithms, namely $C$, gamma, and function degrees for the poly kernel. The $C$ is a regularization parameter that defines the soft margin of classification (i.e., tolerance on misclassification), and a larger $C$ value means a smaller margin accepted for incorrect predictions. The gamma $\gamma$ governs the influence of features on the decision boundary, in which a larger $\gamma$ value means a greater influence. The degree parameter is only used for the polynomial kernel, which defines the degree of the polynomial function used to generate new features by applying a polynomial combination on existing features for dimension reduction. \smallbreak

For optimizing the fine-tuning, fitting and predicting speed, Intel(R) Extension for Scikit-learn\footnote{sklearnex available at: \url{https://github.com/intel/scikit-learn-intelex}.} (sklearnex) is used to accelerate the workflow. It leverages the oneAPI Data Analytics Library that is natively built by Intel(R), and invokes customized functions to reduce execution time. The fine-tuning operation is completed by running the grid search algorithm on 80+ servers over different combinations of parameters in parallel. Once the optimal parameters have been determined, instances of the SVM classifier are instantiated and calibrated so the classifier can also generate the confidence level, as represented in probabilities, of its decision on observations. The calibration process leverages ensemble learning along with cross-validation, by fitting a copy of the base classifier in each split and evaluating against the testing dataset. The final estimator is an ensemble of calibrated SVM classifiers, and the output of each observation equals the mean predicted probabilities of all classifiers. For each tweet, the predicted probability of class $0$ (i.e., Class $0$ for \emph{"Human"} and Class $1$ for \emph{"Machine"}) is updated to the database. \smallbreak

\subsubsection*{Rule-based Consensus on Tweet Authenticity using Results from Detectors} \label{sec:neural-rule}

To reach a consensus on whether a tweet is synthetically generated or human-written, the predicted probability from three detectors, namely the RoBERTa-based discriminative model and SVM classifiers (i.e., GPT-2-XL and BERT-Large-Cased), are referenced for making the decision. \smallbreak

The first option is via plurality voting with thresholds. For example, with a minimum threshold of $0.8$ on the predicted probability of class \emph{"Human"}, only tweets that have at least two out of three detectors satisfy this condition are concluded as human-written. However, this approach is problematic as the results from these three detectors should not be interpreted in a parallel relationship. In other words, for a tweet that is generated using a BERT-base model, only the SVM classifier trained over GLTR-BERT $frac(p)$ distribution can identify it as synthetic text with great confidence, whereas the other two classifiers that are trained over GPT-2-medium synthetic text and WebText may output arbitrary probability of it being true. Consequently, DeepTrust uses a weighted average score with a rule-based system to conclude text authenticity. For each tweet, the weighted probability of it being human-written or synthetically generated is calculated, while the weight is determined based on the average accuracy of the detector on 10-fold cross-validation. For the Twitter financial anomaly example, the average accuracy for RoBERTa ($0.7$ threshold), GLTR GPT-2-XL ($0.5$ threshold) and GLTR BERT ($0.5$ threshold) detectors are $0.92$, $0.81$ and $0.69$, respectively, thus weights are distributed as $38\%$, $33\%$ and $29\%$ for these three detectors. DeepTrust also employes a rule-based system to exclude unreliable tweets. Tweet are concluded as human-written only if \emph{(i)} RoBERTa discriminative model reports it being real with $p_{real} > 0.7$ and \emph{(ii)} weighted fake probability from GLTR detectors $p_{fake} < 0.4$; Tweets are concluded as synthetically generated only if \emph{(i)} RoBERTa discriminative model reports it being fake with $p_{fake} > 0.7$ or \emph{(ii)} weighted fake probability from GLTR detectors $p_{fake} > 0.9$. Thresholds are determined by the author using empirical evaluation\footnote{The author first manually selects a subset of tweets (i.e., 200 tweets from the Twitter example) that are deemed correlated and agreed by three annotators, and is treated as the ground truth. Output from detectors are then reviewed and a threshold is determined that can most accurately separate these tweets from the remaining ones.}, and may be overridden by end-users whenever suitable. The remaining tweets are left inconclusive, and are handled differently under the two detection modes DeepTrust supports.  \smallbreak

DeepTrust supports two modes of synthetic text detection, namely \emph{"Recall"} and \emph{"Precision"}. For recall mode, the objective is to relax synthetic text filtering rules and include more uncertain tweets into the collection, whereas for the precision mode, strict filtering rules are applied to ensure only human-written tweets with high confidence are included. The mode choice is depending on use cases, and is left to users for deciding. For instance, for large financial data providers like Refinitiv, a team of 100 experts may prefer the recall mode as they have the resources to manually review a large collection of tweets, whereas for individuals like retail investors, a refined collection with less than $100$ tweets is usually preferred to quickly obtain insights into the anomaly cause. The decision of pursuing a recall-focused or precision-focused approach should not be decided universally for all use cases, and thus DeepTrust leaves the option to users. For both modes, tweets that are concluded as human-written are presented to users, whereas tweets with inconclusive tags are only included when recall mode is enabled. Tweets are deemed fake by detectors are ignored under either mode. \smallbreak

\subsection{Argumentation Structure Screening using Sequence Labeling Model}

The objective of AM component is to identify if a given tweet contains a valid Twitter-specific argumentation structure, as defined in Section~\ref{sec:twitter-specific-as}. Unfortunately, due to the anonymization requirement from Twitter, DART, the widely adopted annotated dataset for learning arguments and relations on Twitter, is not available for use currently, thus training a binary classifier to filter tweets into factual and opinionated categories is infeasible. As an alternative, DeepTrust leverages another state-of-the-art generalized argument tagging tool TARGER\footnote{TARGER sequence labeling tool is available at: \url{https://github.com/uhh-lt/targer}.} with 7 pre-trained sequence taggers, as developed by Chernodub et al.~\cite{chernodub2019targer}. To accommodate changes in the training dataset, DeepTrust also proposes an alternative Twitter-specific argumentation structure that distinguishes reliable and unreliable tweets. \smallbreak

\subsubsection*{Argumentative Units Detection using Sequence Tagger TARGER}

The TARGER is an end-to-end sequence labeling system with models trained on large argument detection datasets, and is designed to identify different argumentative units (i.e., claim and premise) and entities (e.g., person, location, organization, etc.) in text from diverse domains. To label a sequence, the tweet is firstly transformed to word embeddings with FastText n-gram representations, and processed by a pre-trained sequence tagging neural model on IBM Debator - Claim Evidence Search dataset\footnote{IBM Debator - Claim Evidence dataset is available at \url{https://www.research.ibm.com/haifa/dept/vst/debating\_data.shtml}.}. The training dataset comprises $1.5$M sentences retrieved from Wikipedia Snapshot in 2017, and is scattered across $150$ different topics covering politics, social media and other generic domains. During the training phase, the optimal training parameter discussed in~\cite{chernodub2019targer} is used, in which the Adam optimizer, $0.5$ dropout rate and $0.001$ LR yields the optimal results across all pre-trained models. Regarding the model specification, TARGER uses the state-of-the-art model architecture to achieve optimal performance, which is a combination of CNN + Bi-LSTM + Conditional Random Field (CRF) as proposed in~\cite{ma2016end}. The following list discusses the details of the model:
\begin{enumerate}
\item \emph{CNN - Character-level Embedding}: Instead of using pre-trained word or sequence embeddings, TARGER uses character-level CNN to encode lexical and morphological information of words into character-level representations, which may carry enriched information comparing to a less-granularity representation. A typical construction of character-level CNN consists of \emph{(i)} input layer \emph{(ii)} convolution layer for learning combinations of characters \emph{(iii)} max-pooling layer for dimensionality reduction \emph{(iv)} output layer with character-level representations. 
\item \emph{Bi-LSTM - Contextual Information}: To obtain contextual information from a sequence of characters, each sequence is presented both forward and backward to the Bi-LSTM model, and are captured with two hidden layers. The output is the concatenated vectors of both hidden states.
\item \emph{CRF - Sequential Data Modeling}: Instead of modeling a sequence of words independently, CRF considers the correlation among neighbors using a feature function, and output another sequence considering the contextual information seen before. In other words, the sequence of text is decoded using contextual information instead of independently on a per-word basis. 
\end{enumerate}

\subsubsection*{Updated Definition of Tweet Argumentativeness and Consensus on Argumentation Structure} \label{sec:argument-mining-definition}

Although the definition of Twitter-specific argumentativeness proposed in DART~\cite{bosc2016dart} cannot be used due to the absence of annotated training dataset, DeepTrust proposes an alternative argumentation structure based on the Toulmin argument model, which is mentioned in Section~\ref{sec:argument-mining}. Initially, the argumentation filter deems a tweet as argumentative if it contains a claim-evidence pair, in which the claim should be reasonably derived and supported by the associated evidence. To formalize the terminology using components in the Toulmin argument model, an argumentative tweet should at least contain a \emph{claim}, expressed as a conclusion of the argument, and a \emph{grounds}, which is the foundation of the claim. As an example, the following tweet is a typical argumentative tweet that contains two essential argumentative units \textcolor{red!60}{\emph{claim}} and \textcolor{ForestGreen!60}{\emph{grounds}}. 
\begin{quote}
    We \textcolor{red!60}{[should sell the Twitter stock now](claim)}. The \textcolor{ForestGreen!60}{[released figures are not looking good](grounds)}.\footnote{This tweet is written by the author for demonstration purposes, and is not retrieved from Twitter. Argumentative units are labeled by IBM FastText pre-trained model.}
\end{quote}
in which the \emph{grounds} is equivalent to the \emph{premise} in TARGER. However, when applying this filter on evaluation datasets, all tweets are labeled as \emph{unreliable} because the sequence tagger cannot identify any claim from them. More analysis on factors that impact the performance of TARGER on Twitter are discussed in Section~\ref{sec:argumentation-filter-discussion}, but it is important to redefine the criteria of argumentativeness for this project, or this filter is equivalent to a baseline classifier. \smallbreak

The finalized definition of argumentativeness of tweets is a relaxed version of the prior one. Instead of requiring both \emph{claim} and \emph{grounds} to be presented in a tweet, tweets with only \emph{grounds} are also treated as argumentative, thus labeled as \emph{reliable}. The rationale behind this definition is two folds: \emph{(i)} relaxing the definition can significantly improve the recall by treating more low-confidence tweets as reliable \emph{(ii)} including exceptional cases in which an implicit argumentation structure is used (e.g., annotation sample $2$ in Section~\ref{sec:data-annotation-guideline}). For instance, there may exists tweets that only mention the \emph{grounds} (i.e., \emph{"released figures are not looking good"}) while using Twitter-specific features like cashtag (e.g., \emph{"\$TWTR"}) to replace the \emph{claim} in above example. If enforcing the stricter definition, the diversity of argumentative units on social media platforms is not taking into consideration when evaluating tweet reliability. Therefore, DeepTrust decides to use the relaxed definition considering its impact on recall and the variety of argumentation structures in Twitter. \smallbreak

\subsection{Subjectivity Screening using Classifiers and Subjectivity Lexicon}

In order to analyze text subjectivity, tweets are transformed into three different representations, namely context-independent lexical list, context-dependent word embeddings and context-dependent sentence embeddings. Classifiers (i.e., LSTM for Word Embedding and Single-Layer Perception for Sentence Embedding) or subjectivity lexicon (i.e., TextBlob) are then used to evaluate tweet subjectivity based on those representations. All classifiers are trained and evaluated using the SUBJ dataset. A consensus is reached via adopting a rule-based system similar to the one in synthetic text filter. \smallbreak

\subsubsection*{Additional Tweet Pre-Processing for Subjectivity Analysis} \label{sec:subj-preprocess}

Due to the absence of an annotated subjectivity dataset composed of  financial tweets, all classifiers for SA are trained on the SUBJ dataset, which is introduced in Section~\ref{sec:lm-sa}. However, the movie reviews from the SUBJ dataset are written in different linguistic styles, and missed most of the Twitter-specific features, such as tags and emoticons. Since it is infeasible to fine-tune classifiers using an annotated collection of tweets, it is necessary to applying pre-processing techniques on the retrieved tweets to transform them into a similar format as the movie reviews in the SUBJ dataset. For DeepTrust SA, a series of pre-processing techniques are applied in addition to the general text pre-processing steps in Section~\ref{sec:general-preprocess}, namely as follows:

\begin{enumerate}
\item \emph{Contraction Replacement}: Contractions are combination of words shortened by replacing with an apostrophe (e.g., \emph{You are} to \emph{You're}). DeepTrust uses a dictionary mapping, which consists of an aggregated list of most commonly used English contractions, to replace contractions in tweets. For contractions that are ambiguous, they are ignored and replaced with an empty string (e.g., \emph{David's} to \emph{David}). 
\item \emph{Punctuation Removal}: Functional punctuation with no concrete meanings are removed from tweets, such as hyphen, semicolon and braces, which are proven effective in reducing noisy tokens during the tokenization phase. In DeepTrust, only \emph{exclamation mark (!)}, \emph{question mark (?)}, \emph{comma (,)} and \emph{period (.)} are kept, as the presence of these punctuation may indicate subjectivity (e.g., sentence ends with multiple exclamation marks is more likely to be subjective than sentence ends with a period). 
\item \emph{Spelling Corrections}: As tweets often contain misspelled words, a collection of unigrams and bigrams Twitter word statistics annotated by EkPhrasis\footnote{EkPhrasis Twitter word statistics contains $330$M English Twitter messages, available at \url{https://github.com/cbaziotis/ekphrasis}.} is used to perform spell correction. The spelling corrector works by firstly identifying misspelled tokens in the tweet using a dictionary mapping, and replace them with the most probable spelling corrections based on sequence similarity. 
\item \emph{Tags and Emoji Unpacking}: Twitter-specific features such as hashtags, cashtags and emojis are transformed to interpretable tokens that can be analyzed on their subjectivity. For instance, complex expressions, such as tags with combined words (e.g., \emph{\#technicalanalysis"}), are properly split using a social tokenizer fine-tuned with the EkPhrasis Twitter dataset. This word segmentation step on tags is necessary to preserve the subjectivity within the combined words by converting them into known tokens of the vocab list used by neural models. 
\item \emph{Text Cleaning}: Irrelevant tokens, including \emph{email}, \emph{numbers}, \emph{phone}, \emph{date and time}, \emph{url}, \emph{user (@mentions)} are all omitted in tweets as they are assumed not bringing subjectivity to the textual content. Excessive white space are also properly repaired using regular expression to avoid issues during the tokenization phase.
\end{enumerate}

\subsubsection*{Context-Dependent Word Embedding based Analysis using BERT and LSTM Model} \label{sec:wordemb}

Distributed word representations, commonly known as word embedding, is a widely adopted technique in representing a sequence of words in higher-dimensional space by dense vectors in lower-dimensional space, and captures semantics by representing words with similar meanings in vectors that are closer in distance. Recent advanced models, such as BERT as discussed in Section~\ref{sec:event-representation}, capture both static semantic meanings and contextualized meanings with innovative embedding schema when representing sentences (e.g., token, position, and segment embedding layers in BERT). DeepTrust references the leading model architecture with the highest accuracy on the SUBJ benchmark, the BERT Embedding + LSTM approach proposed in~\cite{nandi2021empirical}, and fine-tuned on SUBJ dataset with 5-folds cross-validation before transfer learning to the Twitter domain. \smallbreak

Regarding implementation details, a batch of tweets is firstly pre-processed and tokenized using the BERT-base vocab list. All tweets are padded to a standard sequence length of $128$ tokens with post-truncation mode, and [CLS] and [SEP] tokens are inserted accordingly. Tweets are represented by the last hidden layer $h$ of transformer encoders in the pre-trained BERT-base model, which are then used as the input of an LSTM model trained with categorical cross-entropy loss and Stochastic Gradient Descent (SGD) optimizer with $0.9$ learning rate. Cyclical learning rate (CLR), proposed in~\cite{smith2017cyclical} as an effective approach to identify optimal global learning rate and avoid stuck at local minima, is used to dynamically adjust the learning rate cyclically within a pre-defined range in each epoch. A dropout layer of rate $0.2$ is adopted to prevent overfitting, and the optimal regularization parameter is determined by running the grid search algorithm. The last layer of the LSTM model uses the ReLU activation function, to introduce non-linearity and overcome the vanishing gradient problem, and a fully connected layer is concatenated to the output of the LSTM model, in which a sigmoid activation function is used to classify tweets as subjective (i.e., Class $1$) or objective (i.e., Class $0$).  Lastly, the entire model with the optimal set of parameters is trained again using the whole SUBJ dataset, stored as a checkpoint for generating predictions on the pre-processed tweets. \smallbreak

\subsubsection*{Context-Dependent Sentence Embedding based Analysis using InferSent and Single-Layer Perception}

Although the aforementioned approach with word embeddings has yielded the best performance on the SUBJ dataset, there are several limitations of using word embeddings to represent a sentence. The first issue is the sentence may be truncated to the maximum sequence length seen during the training phase, as the inner cell dimension of the LSTM model is fixed during the training phase. In addition, aside from representing static semantic meanings, the effectiveness of word embeddings in capturing interactions and relationships among words is still unknown in most researches. On the other hand, learning a universal representation of sentences, with either unsupervised or supervised learning approaches, can effectively capture enriched information embedded in phrases and interaction among entities, then subsequently encode this knowledge into a single sentence embedding as opposed to a list of word embeddings. \smallbreak

In DeepTrust, the model architecture proposed by Facebook Research in~\cite{conneau2017supervised}, Bi-LSTM with Max-Pooling (InferSent), is referenced, and its second generation InferSent2 is used to encode sentences into a vector of dimension $2048$. Regarding the model specification, for each sequence of $n$ words $X = (x_1, \dots, x_n)$, the word embedding value (i.e., the input of LSTM model) is firstly initialized with FastText n-gram representations, and the sentence embedding representation $h$ is the concatenated output of forward and backward LSTM with max-pooling sampling:

\begin{equation}
    h = max(\overset{\longrightarrow}{LSTM}(x_1, \dots, x_n)_i, \overset{\longleftarrow}{LSTM}(x_1, \dots, x_n)_i) \quad \quad \quad i \in {1 \dots n}
\end{equation}

in which for each hidden unit $h_i \in h$, its value is the maximum value between the forward and backward LSTM output. The InferSent model is trained using a supervised learning task of Natural Language Inference (NLI) with the Stanford Natural Language Inference dataset. The objective of the SNI task is to analyze pairs of premise and hypothesis in English sentences, and classify if they are in the relationship of \emph{entailment}, \emph{contradiction} or \emph{neural}. Texts are encoded in forms of sentence embedding with a sentence encoder (i.e., Bi-LSTM with Max-Pooling), and the knowledge captured by the sentence encoder is transferred into another domain for other downstream NLP tasks like subjectivity classification. \smallbreak

To evaluate tweet subjectivity, a single-layer perceptron (SLP) model is used for the binary classification task. The model is implemented using PyTorch with one linear layer of dimension $(2048, 2)$, in which $2048$ represents the dimension of sentence embedding and $2$ represents the number of classes (i.e., Subjective as Class $1$ and Objective as Class $0$). The loss function is the categorical cross-entropy, and the optimizer is RMSprop, which uses normalization on the gradient to prevent its value from vanishing or exploding. The sigmoid activation function is used to generate an output between $[0, 1]$ for binary classification. Similar to the approach discussed in Section~\ref{sec:wordemb}, the model is trained and fine-tuned with the SUBJ dataset of movie reviews, and the best performing model is stored for transfer learning into the Twitter domain. For each pre-processed tweet, the sentence is first encoded using the InferSent2 model and represented as a vector of dimension $2048$, and the SLP classifier is then used to generate predictions on whether it is subjective or objective. \smallbreak

\subsubsection*{Context-Independent Lexicon based Analysis using TextBlob}

Textblob, as discussed in Section~\ref{sec:textblob}, is a lexicon-based subjectivity analysis method that tags each tweet as subjective or objective using dictionary mapping. Tweets are firstly tokenized using the standard Punkt sentence tokenizer from NLTK, and then the POS tagging function from TextBlob is used to identify adjectives within the tweet. These adjectives are then matched against a curated subjectivity lexicon that contains around $2,900$ adjectives and their subjectivity scores are represented as a float value between $[0.0, 1.0]$. A threshold-based system is then employed to determine the degree of subjectivity of tweets, based on the average subjectivity scores of known adjectives. Subjective tweets are those with an average subjectivity score greater than a certain threshold, and the optimal threshold value is determined as $0.5$ by referencing the experiment results from Sahni et al. on another collection of $1.6$M tweets in~\cite{sahni2017efficient}. \smallbreak

\subsubsection*{Rule-based Consensus on Tweet Subjectivity using Results from Classifiers}

Similar to the rule-based system used for synthetic text filter, a list of pre-defined rules are used to reach consensus on tweet subjectivity based on outputs from three different analytical tools aforementioned. Firstly, both context-dependent based approaches (i.e., with sentence embeddings or word embeddings) have yielded an outstanding performance on the SUBJ dataset, and are assumed to be equally reliable when transferring knowledge from movie reviews to tweets in the forms of encoded representations. Secondly, TextBlob is less reliable, primarily due to the insufficient coverage of adjectives or other words with different POS tags. For example, in the Twitter financial anomaly example, around $20.74\%$ tweets have a subjectivity score of $0.0$ from TextBlob, and $71.23\%$ of them are because there are no known adjectives in the tweet that matches with records in the TextBlob subjectivity lexicon. In other words, scores derived by TextBlob may carry multiple meanings, as a score of $0.0$ may either indicate the tweet being perfectly objective, or represent a subjective sentence that does not contain any known adjectives from the TextBlob subjectivity lexicon. Therefore, the decision on subjectivity is primarily based on outputs from context-dependent analysis, and the output from context-independent approach TextBlob is only used as a moderator when there are disagreements. A tweet is classified as subjective only if \emph{(i)} both results from classifiers using sentence embeddings and word embeddings are subjective and \emph{(i)} in case of disagreement between the two results, the subjectivity score from TextBlob is greater than the threshold value of $0.5$, vice versa. The final results are updated to the database in the field of \emph{"subj-filter"}, with \emph{"subj-filter"} equals to true means objective, vice versa. \smallbreak

\subsection{Sentiment Signal Insights using Domain-Specific Language Model}

Polarity analysis, which classifies text as positive, negative or neutral, may introduce insights to financial analysts on the overall market sentiment towards the interested equity. Instead of relying on a sentiment classification model trained on general corpora, DeepTrust utilizes a domain-specific language model to drive all potential of deep learning models in sentiment analysis. The following section introduces the FinBERT model architecture proposed in~\cite{araci2019finbert}, and discusses some fine-tuning techniques DeepTrust has adopted to transfer this language model into the Twitter domain. \smallbreak

\subsubsection*{Sentiment Analysis using Domain-Specific Language Model FinBERT}

Although there exist a large number of high-performance deep learning models for sentiment analysis, they are mostly trained on general corpora such as IBDB Movie Reviews\footnote{IBDM Movie Reviews dataset is a binary sentiment classification collection with 50,000 reviews annotated as positive or negative.} or the MR Movie Reviews\footnote{MR Movie Reviews dataset is a binary sentiment classification collection with 2,000 reviews annotated as positive or negative.}, and the knowledge may not be effectively transferred to the financial domain on Twitter. For instance, financial tweets may use specialized language with domain-specific terminologies (e.g., EBIT (Earnings before Interest and Taxes), AH (After-Hours)) to convey information, and consist of complex expressions (e.g., rhetorical questions, sarcasm) that its sentiment cannot be identified trivially. Therefore, DeepTrust leverages FinBERT, a domain-specific language model that is pre-trained with massive financial corpora from Reuters, and fine-tuned for downstream NLP tasks using task-specific datasets, to offer analysts an effective way to understand the sentiment indicator from a large collection of unstructured data. \smallbreak

One novelty of FinBERT is the specially crafted list of pre-training tasks that allows the BERT model to continuously capture knowledge in different depths, in which comprised six unsupervised learning tasks\footnote{The six unsupervised pre-training tasks are namely \emph{(i)} span replace prediction from spanBERT \emph{(ii)} capital letters prediction from ERNIE2 \emph{(iii)} token-from-passage prediction from ERNIE2 \emph{(iv)} sentence order rearrangement from ERNIE2 \emph{(v)} multi-class NSP \emph{(vi)} self-supervised question-answering relation from ERNIE.} that are inspired from the latest pre-training frameworks spanBERT~\cite{joshi2020spanbert} and ERNIE 2.0~\cite{sun2020ernie}. The embedding layer has also been extended for the pre-training tasks, with an additional task embedding that specifies that task ID (i.e., 0 to 5). The pre-training is firstly done with a large collection of financial information on social media platforms (i.e., Reddit and Yahoo Finance) and financial forum (i.e., FinancialWeb), in addition to the general corpus of Wikipedia and BooksCorpus adopted in the original BERT paper. The language model is further fine-tuned with another subset of Reuters TRC2\footnote{Reuters TRC2 contains $1,800,370$ financial news between \numrange{2008}{2009}, available at \url{https://trec.nist.gov/data/reuters/reuters.html}.}, to improve its knowledge on financial news using high-quality curated news collection from professional news agencies. For the sentiment analysis task, the FinBERT is further fine-tuned with Financial Phrasebank\footnote{Financial Phrasebank is a polar sentiment dataset on financial news, available at \url{https://huggingface.co/datasets/financial\_phrasebank}.} comprises annotated sentiment scores between $[-1, 1]$ on $4.8$K English financial news, and training parameters are learning rate $2e^{-5}$ and maximum sequence length $64$. The loss function is defined as categorical cross-entropy loss, which is widely adopted for multi-class classification tasks. \smallbreak

\begin{figure}[!htbp]
\centering
\includegraphics[width = 1\hsize]{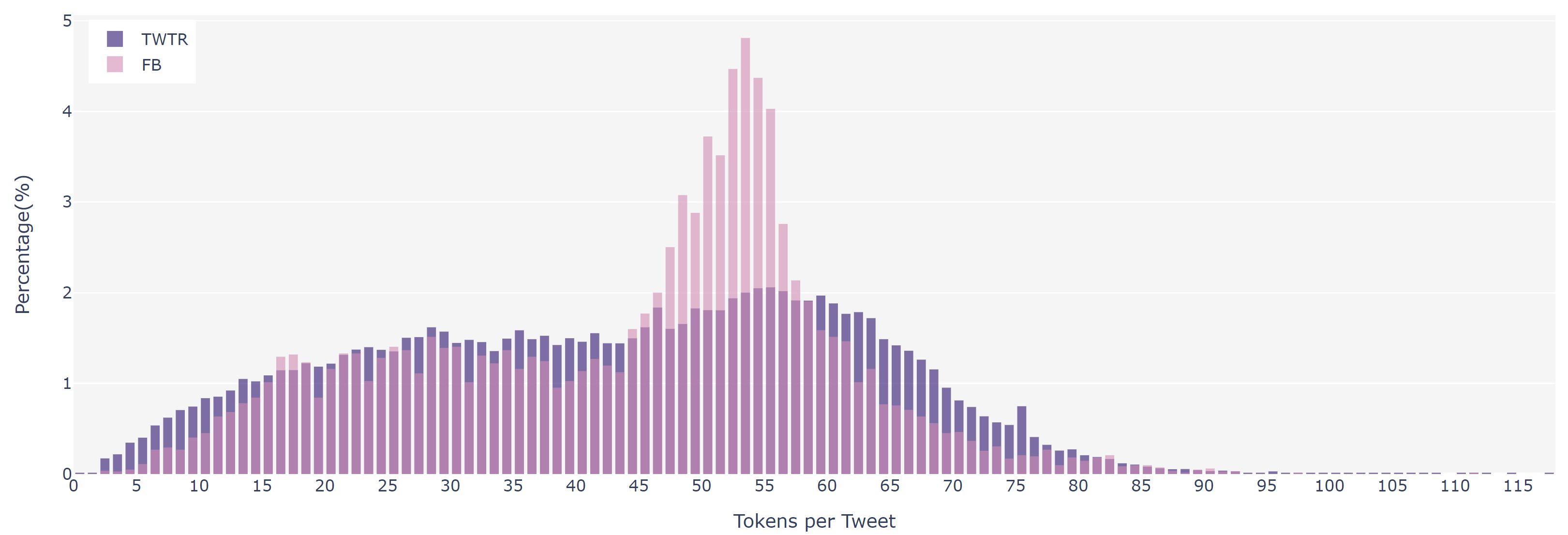}
\caption{Tokens per Tweet histogram using BERT-base-uncased vocab on Twitter and Facebook financial anomalies.}
\label{fig:sentiment-histogram}
\end{figure}

For DeepTrust, the model checkpoint trained by the ProsusAI team is used for its exceptional performance both on Financial Phrasebank and our evaluation dataset. Firstly, the maximum sequence length is specified to be $64$ as the fine-tuning process conducted by the ProsusAI team is primarily focused on sentence-by-sentence prediction, instead of per-document prediction. Consequently, DeepTrust can either use text truncation to ensure a given tweet is complying with the sequence length restriction, or compute averaged sentiment score of all sentences within the tweet, in which the former option is adopted. The rationale behind this choice is two-fold: \emph{(i)} with $280$ characters limit constraint from Twitter, the likelihood of a tweet contains more than $64$ words is negligible. As shown in Figure~\ref{fig:sentiment-histogram}, $87.3\%$ of tweets in the TWTR evaluation dataset and $93.2\%$ of tweets in the FB evaluation dataset are within the maximum sequence length $64$ after converting into a token list with BERT-base-uncased vocab, which proves that the probability of tweet exceeding the sequence length restriction is small \emph{(ii)} unlike formal writing, the absence/misuse of punctuation (e.g., "\dots" and emoticons) is common in Twitter, thus it is challenging for a sentence tokenizer to properly split the tweet into sentences. Therefore, DeepTrust uses the mid-truncation technique as advised by the original paper, to randomly select a substring between the first and third quartile (i.e., $25^{th}$ and $75^{th}$ percentile) section of the tweet, and remove the substring with minimal length to comply with the sequence length constraint. \smallbreak

In addition to text truncation, tweets are also preprocessed using the identical text pre-processing techniques mentioned in Section~\ref{sec:subj-preprocess}. The rationale is similar, as the objective is to transform tweets into a similar format as the training dataset of the language model. In the case of FinBERT, Twitter-specific features such as tags and emoticons are less frequently appearing on forums and other social media platforms. These features are required to transform into interpretable formats that resemble the knowledge of FinBERT while preserving its embedded meanings. Besides, a batch size of $5$ is specified for faster predictions. The attention mask and token type IDs are configured accordingly based on the per-batch padding. The batch input is then passed to FinBERT for evaluation. As the sentiment analysis task defined here is a multi-class classification task (i.e., positive, negative, and neutral), the softmax activation function is used to convert logits output from the last layer into probabilities, which is then converted to labels. Logits are also used for calculating a sentiment score, which is based on the absolute difference between the probability of positive and negative. \smallbreak


\chapter{Evaluation and Discussion} \label{cha:evaluation-and-discussion}

\section{Dataset and Annotation}

\subsection{Evaluation Dataset and Acknowledgement of Bias} \label{sec:bias}

Two financial anomalies that are deemed representative by the author are used as the source of evaluation dataset, namely \emph{(i)} Twitter Inc. stock price crashed on 30 April 2021, with 15.16\% decrease when closed \emph{(ii)} Facebook Inc. stock price surged on 29 April 2021, with 7.3\% increase when closed. Both extreme price movements are caused by the difference between expectations from financial analysts and the disclosed quarterly financial reports, affecting the post-earning stock price drastically. The below is a summary of the two financial anomalies:  

\begin{table}[!htbp]
\centering
\begin{tabular}{@{}lll@{}}
\toprule
                                                     & \textbf{Evaluation Dataset 1 (TWTR)} & \textbf{Evaluation Dataset 2 (FB)} \\ \midrule
\multicolumn{1}{l|}{\textbf{Stock Ticker / RIC}}     & TWTR / TWTR.K                        & FB / FB.O                          \\ \cmidrule(l){2-3} 
\multicolumn{1}{l|}{\textbf{Anomaly Date}}           & 30 April 2021                        & 29 April 2021                      \\ \cmidrule(l){2-3} 
\multicolumn{1}{l|}{\textbf{Price Movement}}        & Stock Price Plunged 15.16\%          & Stock Price Surged 7.3\%           \\ \cmidrule(l){2-3} 
\multicolumn{1}{l|}{\textbf{Possible Causes}} &
  \begin{tabular}[c]{@{}l@{}}1. Weak Future Guidance\\ 2. Slow User Growth\\ 3. Anticipated Quarterly Performance\end{tabular} &
  \begin{tabular}[c]{@{}l@{}}1. 48\% Revenues Growth over Q1\\ 2. 10\% MAU Growth during Pandemic\\ 3. 30\% Price per Ad Increase\end{tabular} \\ \cmidrule(l){2-3} 
\multicolumn{1}{l|}{\textbf{Original Dataset Size}}  & 241,714 (Tweet) / 115.690 (Author)   & 32,983 (Tweet) / 11,003 (Author)   \\ \cmidrule(l){2-3} 
\multicolumn{1}{l|}{\textbf{Subset Search Query}} &
  \begin{tabular}[c]{@{}l@{}}1. \$TWTR\\ 2. Twitter \&\& stock \&\& price\end{tabular} &
  \begin{tabular}[c]{@{}l@{}}1. \$FB\\ 2. Facebook \&\& stock \&\& price\end{tabular} \\ \cmidrule(l){2-3} 
\multicolumn{1}{l|}{\textbf{Annotated Dataset Size}} & 659 (208 Reliable / 451 Unreliable)            & 1,052 (216 Reliable / 836 Unreliable)                              \\ \bottomrule
\end{tabular}
\caption{Evaluation Datasets (TWTR and FB Financial Anoamlies) for DeepTrust Framework.}
\label{tab:evaluation-dataset}
\end{table}

The annotated dataset is a subset of the original collection of retrieved tweets output by the information retrieval module. For each evaluation dataset, the subset is selected using two search queries that the author deem are most effective in identifying correlated tweets that can explain the cause of the financial anomaly, which are shown in Table~\ref{tab:evaluation-dataset}. Firstly, the Cashtag, stock ticker symbol with a dollar sign prefix, is the most common approach for Twitter users to communicate financial information on Twitter, and it contains less noise compared with other keywords such as hashtag (i.e., \#TWTR or \#TSLA) or company names (i.e., Twitter Inc.). In the study of~\cite{hentschel2014follow} on Cashtag in 2014, the percentage of non-spam tweets of all tweets mentioning large corporations with cahstags like \emph{\$AAPL}\footnote{There are $81.2\%$ ($49,602$ out of $61,068$ tweets) non-spam tweets within all tweets that contain the cashtag of \emph{\$AAPL} in April 2013.} and \emph{\$FB}\footnote{There are $80.7\%$ ($9,817$ out of $12,169$ tweets) non-spam tweets within all tweets that contain the cashtag of \emph{\$FB} in April 2013.} are approximately $81\%$, which is a relatively reliable approach to retrieve a subset of financial-related tweets that are associated with the target stock. In addition, a combination of case-insensitive stock-related keywords (i.e., company name + \emph{"stock"} + \emph{"price"}) is defined as the second search query. The rationale behind this query is that \emph{(i)} it is inevitable to mention the company name either in forms of tags or text (e.g., Twitter or Facebook) when discussing its stock performance and \emph{(ii)} an explanation of anomalous stock price movement usually contains keywords of \emph{"stock"} and \emph{"price"}, regardless of describing the price movement or explaining the cause. Although there are exceptional cases that address the stock price movement with a single keyword \emph{"stock"} or \emph{"price"} instead of both, relaxing the condition introduce significantly more noise to the subset, from $108$ to $3,603$ tweets in the TWTR evaluation dataset. Therefore, the second search query strictly requires the presence of both \emph{"stock"} and \emph{"price"} keywords. \smallbreak

Considering the available resources for data labeling, DeepTrust has made compromises to label bias when sampling and annotating subsets, and it is important to acknowledge these biases before interpreting the evaluation result on the DeepTrust framework. \smallbreak

Firstly, DeepTrust suffers from sample bias, which arises when the annotated subset is not representative of the universe as reliable tweets are systematically more in this subset than another random subset. In other words, since the subset is selected using queries that are customized for extracting the most amount of correlated financial tweets as regards the cause of anomalous price movement, more reliable tweets are included than a randomly sampled subset. Consequently, the precision of the RA module may be overestimated (i.e., predict all samples as reliable yields a greater precision in annotated subset than a randomly sampled subset) while the recall may be underestimated when evaluating using this biased sample. To avoid sample bias, the most effective solution is to enforce strictly random sampling, but the random number in computer science is not truly random as those numbers are generated using mathematical formulas to simulate randomness. An alternative is via selecting multiple pseudo-random subsets and use the averaged evaluation metrics as the final output, which requires an extensive amount of annotation effort to label all included tweets. Considering the available resources for this project, DeepTrust acknowledges the sample bias and its implication on evaluation metrics, though still selects an annotated subset using a customized query. \smallbreak

In addition, annotators may introduce confirmation bias by including subjective thoughts either consciously or unconsciously when labeling data, which leads to inaccurate annotation for evaluation. For instance, when annotating the TWTR evaluation dataset, tweets that directly contradict the known ground truth (e.g., \$TWTR stock down 13\% as CEO won’t admit banning Trump hurt their numbers) are often labeled as \emph{unreliable} by annotators. However, the validity of evidence should be reviewed by financial analysts or domain experts, while the DeepTrust framework is solely responsible for examining whether a tweet is reliable or not in terms of its linguistic features, instead of the correctness of reasoning. Therefore, annotators are required to follow the data annotation guideline specified in Section~\ref{sec:data-annotation-guideline}, and use annotation samples as references whenever possible. \smallbreak

Lastly, recall bias, as a form of measurement bias, introduces systematic error to the annotated subset by annotating data inconsistently across similar data points, and it occurs more often when a complex decision process is required. For instance, for similar tweets such as \emph{"\$TWTR Twitter price target lowered to \$66 from \$71 at Piper Sandler"} and \emph{"\$TWTR Twitter price target lowered to \$70 from \$85 at Oppenheimer"}, they should be labeled consistently as they have no substantial difference. However, the decision process (i.e., explained in detail as annotation sample 5 in Section~\ref{sec:data-annotation-guideline}) is extremely complex, and the annotator may label inconsistently when two samples are presented separately with a long interval in-between. Therefore, to reduce the recall bias, annotators are required to review all annotations at least once before submission. \smallbreak

\subsection{Data Annotation Guideline and Annotation Samples} \label{sec:data-annotation-guideline}

As a guideline for annotators, the following criteria are listed as a sample decision process they should follow when evaluating tweet reliability: 

\begin{enumerate}
\item \emph{Entity}: Tweet should contain a direct reference to the company regardless of forms (e.g., cashtags, hashtags, company name), or mention indirectly using its associated entities (e.g., industrial trend, name of the CEO) that are adequate for a human annotator to establish a link between the target entity and its indirect references.
\item \emph{Structure}: Tweet should contain a valid argumentation structure, at least with a simplified claim-evidence pair. The evidence can either be a sentence or Twitter-specific features like URLs. Tweets that only objectively describe the stock price movements (e.g., stock price updates, news updates) should be treated as unreliable because no constructive information is contained to shed light on the cause of abnormal price movements. However, there are special cases that the tweet itself is the evidence of an implicit claim that can be associated using commonsense (i.e., annotation sample 2), and it is at the judgment call of the annotator to decide.
\item \emph{Argument or Evidence Validity}: Tweet should not be assessed based on the validity of its argument or reasoning, because the objective of DeepTrust is to refine the collection into a smaller subset of reliable tweets that are further verified by domain experts in terms of its correctness. Tweets that contain a valid reason or statement that can support the stance of its claim should be regarded as reliable. 
\item \emph{Topic}: Tweet should be centered around the topic of possible sources of abnormal price movement, and it is depending on the judgment calls of the annotator to determine the relevancy using commonsense or domain knowledge. 
\end{enumerate}

The following list consists of some annotation samples extracted from the TWTR evaluation dataset for reference: \smallbreak

\begin{enumerate}

\item \textbf{Tweet}: \emph{"Twitter stock plunges on user miss and low guidance"} \smallbreak
\textbf{Explanation}: Reliable. This is a typical example that explains the cause of a financial anomaly using an argumentation structure. 

\item \textbf{Tweet}: \emph{"\$TWTR Twitter price target lowered to \$66 from \$71 at Piper Sandler"} \smallbreak
\textbf{Explanation}: Reliable. This is a case in which an implicit argumentation structure is assumed. On the surface, the tweet is objectively describing a fact that investment bank Piper Sandler has lowered Twitter stock price target, and does not address the stock price plunge. However, as an annotator with some understanding of trading and finance, the evidence mentioned in the tweet may be linked with the implicit claim of \emph{"Twitter stock price is plunging because..."}, as a lower stock price target from a financial institution can negatively impact the confidence investors are having on the company, results in an increasing amount of sell transactions and a lower stock price. Therefore, this tweet should still be considered reliable with all conditions from the annotation guideline satisfied.

\item \textbf{Tweet}: \emph{"\$TSLA's price moved below its 50-day Moving Average on April 29, 2021."} \smallbreak
\textbf{Explanation}: Unreliable. Firstly, the TSLA entity is not directly associated with TWTR in terms of its business nature or industry, and the author cannot relate this tweet to the Twitter stock price in any means. In addition, the tweet only objectively describes the behavior of TSLA stock price using technical indicators, instead of making a claim based on this evidence. 

\item \textbf{Tweet}: \emph{"\$TWTR Is Twitter a Buy or Sell After Its Post-Earnings Plunge?"}  \smallbreak
\textbf{Explanation}: Unreliable, due to the absence of a valid argumentation structure. It is merely raising a question on possible price movement instead of providing any constructive information. 

\item \textbf{Tweet}: \emph{"Price is signal, Twitter is noise"} \smallbreak
\textbf{Explanation}: Unreliable, because the tweet is off-topic. Although it mentions Twitter Inc, with a claim-evidence structure by using the latter part as supporting evidence for the former claim, it is irrelevant to the topic of stock price plunge. 

\item \textbf{Tweet}: \emph{"\$TWTR I think earnings the selloff is overdone. And to everyone hating on this stock: Why are you still on the platform if you dislike it so much?"} \smallbreak
\textbf{Explanation}: Ambiguous. Although it is not following a standard argumentation structure as others, it still offers an opinion on the price movement (i.e., the selloff is overdone), which may indicate that the abnormal price movement is the consequence of mass irrational/emotional sell transactions from investors, and should be corrected by the market soon. However, the association aforementioned is made by the author using trading knowledge that may differ among annotators, thus it is ultimately a judgment call from the annotator. 

\end{enumerate}

\subsection{DeepTrust Evaluation Objective and Scope} \label{sec:evaluation-objective}

The defined objectives for DeepTrust performance evaluation are \emph{(i)} screening unreliable information and \emph{(ii)} identifying reliable information that explains the anomalous price movement, while the emphasis is on the former one. Nevertheless, it is important to firstly distinguish the recall metric for evaluation and the recall mode mentioned in Section~\ref{sec:neural-rule}. As stated, for large financial data providers like Refinitiv, recall mode is preferred because they have sufficient resources to manually validate a large collection of tweets, and they would want to maximize the recall in classifying reliable tweets. However, when evaluating the DeepTrust framework, the primary objective is to filter spamming or irrelevant information from the ocean of tweets, to assist analysts only focusing on a concise collection of high-quality tweets. In other words, since the RA module is less effective in examining the correctness of reasoning or validity of evidence, it is wiser to use the framework to exclude unreliable information instead of identifying reliable information. Formally speaking, with the null hypothesis of \emph{"Tweet is unreliable"}, the evaluation objective is to minimize Type II error (i.e., accept the null hypothesis while it is true/reliable - False Negative FN), but tolerate Type I error (i.e., reject the null hypothesis while it is false/unreliable - False Positive FP), and the precision of classifying unreliable tweets, calculated by FN / (FN + TN), is therefore determined as the primary objective for evaluation. \smallbreak

The performance of the DeepTrust framework can either be evaluated using a per-module approach, or with the framework as an entirety; While both approaches may be infeasible considering the resources available. Firstly, to evaluate the AD module, there must exist a standardized paradigm that defines financial anomaly, which has been discussed and proven infeasible in Section~\ref{sec:adeabt}. Secondly, to evaluate the IR module, evaluation metrics such as precision and its variant average precision may be adopted, which requires annotating the entire collection of tweets for evaluation. However, due to limited time and annotators available, it is infeasible for the author to annotate both FB and TWTR dataset with $267$K tweets in total. In addition, to evaluate the framework as an entirety, not only do the aforementioned two challenges need to be resolved first, but the author needs to manually execute the entire pipeline of DeepTrust from Anomaly Detection to Reliability Assessment, which are significantly more challenging than the per-module evaluation approach. Considering the scale of this project, trade-offs are made to ensure the evaluation feasible for completion, and can be finished with high-quality using the available annotating resources. \smallbreak

The finalized evaluation scope is restricted to components within the RA module, which are feature-based filter, neural text filter, argumentation filter, and subjectivity filter. The evaluation is defined as a binary classification task that focuses on excluding unreliable tweets, with \emph{reliable} as the positive class and \emph{unreliable} as the negative class. Evaluation metrics associated with the \emph{unreliable} class are used for evaluating the performance of DeepTrust. Considering the stated challenges for AD and IR modules, empirical evaluation is conducted on the IR module to understand factors that may affect the quantity and quality of retrieved tweet collections, while the AD module is not evaluated in this project. \smallbreak

\section{Experimental Results}

\subsection{Summary of Experimental Results}

\subsubsection*{Notation Clarification}

\emph{F}: feature-based filter; \emph{N}: neural text filter (default with P mode); \emph{A}: argumentation filter; \emph{S}: subjectivity filter; \emph{R Mode}: recall mode for N filter; \emph{P Mode}: precision mode for N filter. \smallbreak

\subsubsection*{Evaluation Metrics}

\begin{table}[!htbp]
\centering
\begin{tabular}{@{}lllllllll@{}}
\toprule
                                       & \multicolumn{4}{c}{\textbf{TWTR Dataset}}                   & \multicolumn{4}{c}{\textbf{FB Dataset}} \\ \midrule
\textbf{} &
  \multicolumn{1}{c}{\textbf{Precision}} &
  \multicolumn{1}{c}{\textbf{Recall}} &
  \multicolumn{1}{c}{\textbf{F1}} &
  \multicolumn{1}{c}{\textbf{F0.5}} &
  \multicolumn{1}{c}{\textbf{Precision}} &
  \multicolumn{1}{c}{\textbf{Recall}} &
  \multicolumn{1}{c}{\textbf{F1}} &
  \multicolumn{1}{c}{\textbf{F0.5}} \\ \midrule
\multicolumn{1}{l|}{\textbf{Baseline}} & 68.44\% & 100.00\% & 81.27\% & \multicolumn{1}{l|}{73.05\%} & 79.46\%  & 100.00\% & 88.56\% & 82.87\% \\ \midrule
\multicolumn{1}{l|}{\textbf{F}}        & 75.06\% & 68.74\%  & 71.76\% & \multicolumn{1}{l|}{73.70\%} & 83.94\%  & 71.89\%  & 77.45\% & 81.22\% \\
\multicolumn{1}{l|}{\textbf{N (P)}}    & 76.39\% & 48.78\%  & 59.54\% & \multicolumn{1}{l|}{68.62\%} & 90.51\%  & 49.04\%  & 63.62\% & 77.42\% \\
\multicolumn{1}{l|}{\textbf{A}}        & 0.00\%  & 0.00\%   & 0.00\%  & \multicolumn{1}{l|}{0.00\%}  & 0.00\%   & 0.00\%   & 0.00\%  & 0.00\%  \\
\multicolumn{1}{l|}{\textbf{S}}        & 65.12\% & 49.67\%  & 56.35\% & \multicolumn{1}{l|}{61.30\%} & 76.53\%  & 35.89\%  & 48.86\% & 62.40\% \\ \midrule
\multicolumn{1}{l|}{\textbf{F+N (R)}}  & 73.85\% & 71.40\%  & 72.60\% & \multicolumn{1}{l|}{73.35\%} & 83.54\%  & 73.44\%  & 78.17\% & 81.30\% \\
\multicolumn{1}{l|}{\textbf{F+N (P)}}  & 73.43\% & 83.37\%  & 78.09\% & \multicolumn{1}{l|}{75.23\%} & 83.56\%  & 82.66\%  & 83.10\% & 83.37\% \\
\multicolumn{1}{l|}{\textbf{F+A}}      & 75.06\% & 68.74\%  & 71.76\% & \multicolumn{1}{l|}{73.70\%} & 83.94\%  & 71.89\%  & 77.45\% & 81.22\% \\
\multicolumn{1}{l|}{\textbf{F+S}}      & 69.86\% & 85.81\%  & 77.01\% & \multicolumn{1}{l|}{72.55\%} & 81.75\%  & 86.24\%  & 83.93\% & 82.61\% \\ \midrule
\multicolumn{1}{l|}{\textbf{F+N+A}}    & 73.44\% & 83.37\%  & 78.09\% & \multicolumn{1}{l|}{75.23\%} & 83.56\%  & 82.66\%  & 83.10\% & 83.37\% \\
\multicolumn{1}{l|}{\textbf{F+N+S}}    & 70.03\% & 93.79\%  & 80.19\% & \multicolumn{1}{l|}{73.77\%} & 81.62\%  & 91.87\%  & 86.44\% & 83.48\% \\
\multicolumn{1}{l|}{\textbf{F+A+S}}    & 69.86\% & 85.81\%  & 77.01\% & \multicolumn{1}{l|}{72.55\%} & 81.75\%  & 86.24\%  & 83.93\% & 82.61\% \\ \midrule
\multicolumn{1}{l|}{\textbf{F+N+A+S}}  & 70.03\% & 93.79\%  & 80.19\% & \multicolumn{1}{l|}{73.77\%} & 81.62\%  & 91.87\%  & 86.44\% & 83.48\% \\ \bottomrule
\end{tabular}
\caption{Evaluation metrics of classifying unreliable tweets (i.e., negative class) with different combinations of filters.}
\label{tab:classification-report}
\end{table}

\begin{table}[!htbp]
\centering
\begin{tabular}{@{}lllllllll@{}}
\toprule
                                       & \multicolumn{4}{c}{\textbf{TWTR Dataset}}                  & \multicolumn{4}{c}{\textbf{FB Dataset}} \\ \midrule
\textbf{} &
  \multicolumn{1}{c}{\textbf{Precision}} &
  \multicolumn{1}{c}{\textbf{Recall}} &
  \multicolumn{1}{c}{\textbf{F1}} &
  \multicolumn{1}{c}{\textbf{F0.5}} &
  \multicolumn{1}{c}{\textbf{Precision}} &
  \multicolumn{1}{c}{\textbf{Recall}} &
  \multicolumn{1}{c}{\textbf{F1}} &
  \multicolumn{1}{c}{\textbf{F0.5}} \\ \midrule
\multicolumn{1}{l|}{\textbf{Baseline}} & 46.84\% & 68.44\% & 55.61\% & \multicolumn{1}{l|}{49.99\%} & 63.15\%  & 79.47\%  & 70.38\% & 65.86\% \\ \midrule
\multicolumn{1}{l|}{\textbf{F}}        & 64.84\% & 62.97\% & 63.71\% & \multicolumn{1}{l|}{64.34\%} & 72.88\%  & 66.73\%  & 69.06\% & 71.19\% \\
\multicolumn{1}{l|}{\textbf{N (P)}}    & 64.19\% & 54.63\% & 56.01\% & \multicolumn{1}{l|}{60.02\%} & 77.85\%  & 55.42\%  & 59.27\% & 68.32\% \\
\multicolumn{1}{l|}{\textbf{A}}        & 9.96\%  & 31.56\% & 15.14\% & \multicolumn{1}{l|}{11.54\%} & 4.22\%   & 20.53\%  & 6.99\%  & 5.01\%  \\
\multicolumn{1}{l|}{\textbf{S}}        & 53.38\% & 47.34\% & 49.19\% & \multicolumn{1}{l|}{51.41\%} & 64.67\%  & 40.30\%  & 44.64\% & 54.04\% \\ \midrule
\multicolumn{1}{l|}{\textbf{F+N (R)}}  & 63.85\% & 63.13\% & 63.46\% & \multicolumn{1}{l|}{63.70\%} & 72.54\%  & 67.40\%  & 69.44\% & 71.45\% \\
\multicolumn{1}{l|}{\textbf{F+N (P)}}  & 65.72\% & 67.98\% & 66.25\% & \multicolumn{1}{l|}{65.76\%} & 73.70\%  & 73.29\%  & 73.49\% & 73.61\% \\
\multicolumn{1}{l|}{\textbf{F+A}}      & 64.84\% & 62.97\% & 63.71\% & \multicolumn{1}{l|}{64.34\%} & 72.88\%  & 66.73\%  & 69.06\% & 71.19\% \\
\multicolumn{1}{l|}{\textbf{F+S}}      & 60.13\% & 64.95\% & 60.98\% & \multicolumn{1}{l|}{59.96\%} & 71.60\%  & 73.76\%  & 72.55\% & 71.95\% \\ \midrule
\multicolumn{1}{l|}{\textbf{F+N+A}}    & 65.72\% & 67.98\% & 66.25\% & \multicolumn{1}{l|}{65.76\%} & 73.70\%  & 73.29\%  & 73.49\% & 73.61\% \\
\multicolumn{1}{l|}{\textbf{F+N+S}}    & 63.42\% & 68.29\% & 61.36\% & \multicolumn{1}{l|}{60.44\%} & 72.81\%  & 77.09\%  & 74.09\% & 73.03\% \\
\multicolumn{1}{l|}{\textbf{F+A+S}}    & 60.13\% & 64.95\% & 60.98\% & \multicolumn{1}{l|}{59.96\%} & 71.60\%  & 73.76\%  & 72.55\% & 71.95\% \\ \midrule
\multicolumn{1}{l|}{\textbf{F+N+A+S}}  & 63.42\% & 68.29\% & 61.36\% & \multicolumn{1}{l|}{60.44\%} & 72.81\%  & 77.09\%  & 74.09\% & 73.03\% \\ \bottomrule
\end{tabular}
\caption{Weighted averages of classification metrics with different combinations of filters.}
\label{tab:weighted-report}
\end{table}

Table~\ref{tab:classification-report} shows a summary of RA module performance on predicting negative class (i.e., unreliable) tweets using the annotated subset, and Table~\ref{tab:weighted-report} shows weighted performance metrics for comparing filter performance.

\subsection{Experiment Setup} \label{sec:experiment-setup}

Following the discussion in Section~\ref{sec:evaluation-objective}, evaluation metrics such as precision, recall and F-beta measures are included for analysis. For F-beta measures, both F1 (i.e., balanced weight on precision and recall) and F0.5 (i.e., more weight on precision than recall) scores are recorded for analysis under different scenarios. \smallbreak

In addition, to establish a referencing point, a baseline classifier is used by classifying all tweets in the annotated subset as unreliable, in which the objective of DeepTrust RA module is to outperform the baseline model using different combinations of filters. For all four filters in the DeepTrust RA module, output of individual and combinations of filters are compared with the ground truth (i.e., annotated labels by the author). For neural text filter, both \emph{recall} and \emph{precision} modes are evaluated, and it is evident that the precision mode has yielded a better performance in almost all metrics. For all other combinations of filters that include neural text filter (i.e., \emph{F+N+A}, \emph{F+N+S}, \emph{F+N+A+S}), precision mode is used without explicit mentioning. \smallbreak

\subsection{Interpretation of Evaluation Metrics}

Firstly, the recall for negative class (i.e., unreliable), commonly known as specificity, is less reliable due to bias introduced during the information retrieval stage and sampling stage. When retrieving tweets from Twitter, the retrieved tweet collection does not represent the whole universe of relevant tweets about the financial anomaly, as doing so requires downloading millions of tweets that are posted during the anomaly period. In addition, less unreliable tweets are included in the annotated subset using a customized search query, leading to an overestimated recall. For instance, when using the feature-based filter on the TWTR evaluation dataset, among all $451$ unreliable tweets, only $68.74\%$ of them are correctly identified as unreliable; Whereas this value may be overestimated because with another subset with systematically more unreliable tweets, the recall should be lower with a greater denominator. \smallbreak

Nevertheless, the precision metric is the primary performance indicator of DeepTrust, which essentially tells what percentage of tweets are correctly predicted. With the goal of screening as many unreliable tweets from the collection as possible while ensuring a minimum amount of reliable tweets are incorrectly excluded, filter effectiveness is mainly compared and analyzed using both weighted averaged precision and per-class precision from the listed tables. Similar to recall, bias may inevitably influence the validity of precision metrics, and underestimate the actual precision for filters. With another random sample consists of less reliable tweets, the likelihood of FN should be lower, leading to a smaller denominator and larger precision. \smallbreak

Accuracy is generally useful with \emph{(i)} symmetric dataset with a similar number of samples in both classes \emph{(ii)} identical costs for Type I and Type II errors, which does not fit the profile of TWTR and FB evaluation datasets. Instead, the F-measure score, as the harmonic mean of the precision and recall, is effective for unbalanced class distribution by considering both FP and FN into accounts, and suitable for highly skewed classes like the FB evaluation dataset. For DeepTrust, the cost of FP is lower than FN because with FP, analysts can still manually filter these unreliable information using domain knowledge, whereas with FN, mispredicted information is disregarded by the RA module and cannot be retrieved back for further validation. Consequently, more emphasis should be given to precision instead of recall. As presented in both Table~\ref{tab:classification-report} and Table~\ref{tab:weighted-report}, F0.5-scores that allocate more weights on precision than recall are calculated and used for analysis in later sections. \smallbreak

Lastly, for unbalanced annotated subsets like TWTR and FB, it is essential to reference the weighted average of evaluation metrics when comparing different filter setups. In other words, with per-class unweighted metrics, the performance of the baseline classifier on negative class (i.e., unreliable) is significantly better than on positive class (i.e., reliable), which are $0.0\%$ for all precision, recall, F1 scores on both TWTR and FB datasets. With weighted average metrics, issues caused by class imbalances are alleviated by the support of each label. Therefore, when comparing filter performance, metrics should be averaged using the support-weighted mean of each class, which are summarized in Table~\ref{tab:weighted-report}. \smallbreak

\subsection{Interpretation of Experimental Results}

Firstly, as presented in Table~\ref{tab:weighted-report}, all filters except the argumentation filter have outperformed the baseline classifier in terms of precision and F0.5-scores. When used individually, the feature-based filter has achieved the best overall performance in terms of weighted metrics, while the synthetic text filter is most powerful in excluding unreliable tweets with high precision at the cost of a lower recall. The subjectivity filter is less reliable, results in a comparable performance with the baseline classifier. When filters are used in combinations, only the combination of feature-based filter and synthetic text filter outperforms the best individual filter on all metrics. When using subjectivity together with other filters, both per-class and weighted metrics exhibit a similar pattern, resulting in a higher recall and a lower precision. In other words, the subjectivity filter tends to classify more tweets as unreliable, while a large proportion of them are false negatives, which are undesirable for this project. \smallbreak

In addition, for synthetic text filter, the precision mode is generally better than the recall mode as it enforces a stricter rule-based consensus to classify low-confidence tweets as unreliable, at the cost of a slightly lower precision on the negative class. As shown in Table~\ref{tab:classification-report}, when the synthetic text filter is used together with the feature-based filter, using the precision mode allows more tweets to be classified as unreliable, resulting in a higher recall for the negative class, while the precision is only affected by less than $0.5\%$. It is evident that the stricter version of the synthetic text filtering rule is more effective than the relaxed version, as observed in both evaluation datasets. \smallbreak

Based on Table~\ref{tab:weighted-report}, the combination of the feature-based filter and synthetic text filter with precision setting yields the optimal precision and F0.5-score on both evaluation datasets, which are primary performance metrics used to evaluate the performance of the RA module. In addition, when using this combination on the whole dataset, there are $51,121$ tweets from TWTR and $3,353$ tweets from FB evaluation datasets concluded as reliable, with more than $79\%$ unreliable tweets excluded from the collection. In practice, this amount of tweets are acceptable for large financial firms like Refinitiv to validate its authenticity manually, and are effective in saving a large number of full-time equivalent (FTE) for the corporation. \smallbreak

\section{Discussion and Implications}

\subsection{Impact of Query Enhancement on Tweet Collection Quality}

As discussed in Section~\ref{sec:query-enhancement}, both named-entities identified in Reuters News and keywords collected by PRF are used to expand the final Twitter search query, with the objective to \emph{(i)} include as many potentially relevant tweets as possible (i.e., improve recall) \emph{(ii)} enhance query quality with dynamic keywords that are frequently mentioned by the public. However, the effectiveness of these query enhancement techniques on improving tweet collection quality is debatable, which are discussed in detail in the following sections. \smallbreak

\begin{table}[!htbp]
\centering
\begin{tabular}{@{}lll@{}}
\toprule
 &
  \textbf{Evaluation Dataset 1 (TWTR)} &
  \textbf{Evaluation Dataset 2 (FB)} \\ \midrule
\multicolumn{1}{l|}{\textbf{Reuters NER}} &
  \begin{tabular}[c]{@{}l@{}}{[}'Twitter', 'India', 'S\&P 500 INDEX - CBOE', \\ 'United States', 'NASDAQ COMPOSITE INDEX', \\ 'Thomson Reuters Foundation', 'Greta Thunberg', \\ 'DOW JONES INDUSTRIAL AVERAGE INDEX', \\ 'United Kingdom'{]}\end{tabular} &
  \begin{tabular}[c]{@{}l@{}}{[}'Facebook', 'Canary Islands', \\ 'S\&P 500 INDEX - CBOE', \\ 'iPhone', 'Apple', 'United States', \\ 'Thomson Reuters', 'Reuters', \\ 'Facebook Inc', 'Instagram'{]}\end{tabular} \\ \cmidrule(l){2-3} 
\multicolumn{1}{l|}{\textbf{Twitter PRF}} &
  \begin{tabular}[c]{@{}l@{}}{[}'twitter', 'tatar', '\$eth', '\$twtr', 'weber', \\ '\#technicalanalysis', '\#stock', '\#business', '\$amzn'{]}\end{tabular} &
  \begin{tabular}[c]{@{}l@{}}{[}'\$amzn', 'facebook', '\#investing', \\ '\$aapl', '\#stock', 'apple', '\$fb'{]}\end{tabular} \\ \bottomrule
\end{tabular}
\caption{An overview of Query Enhancement Techniques from the Information Retrieval Module.}
\label{tab:effectiveness-ner}
\end{table}

\subsubsection*{Query Enhancement using Named-Entities from Reuters News}

Firstly, it is evident that the named entities extracted from historical Reuters News generally cover a much broader domain, and most of them are irrelevant to the cause of the financial anomaly. For instance, as shown in Table~\ref{tab:effectiveness-ner}, within the top 10 retrieved named entities of the TWTR evaluation dataset, only entities \emph{"Twitter"} and \emph{"Facebook"} are correlated, whereas other keywords about countries (e.g., \emph{"India"} and "United States") and market indexes (e.g., \emph{"S\&P 500 Index - CBOE"} and \emph{"NASDAQ Composite Index"}) are ambiguous and hard to establish an association with the anomalous event. Similar phenomena are observed in the FB evaluation dataset, and including entities from Reuters News can multiple the size of retrieved tweets, introducing excessive noisy information to the collection. With further investigation, several factors have been identified that limit the effectiveness of NER in Reuters News, which are listed as follow:

\begin{enumerate}
\item \emph{Missing News Stories}: Due to uncontrollable factors (e.g., access right of the current user group), the news stories retrieved from Reuters consist of on average \numrange{10}{15}\% missing data. In particular, when querying news stories using the identifier (i.e., storyID from Eikon API) specified in the news headline, some are not associated with any accessible news stories even after several retries. Specifically, $47$ out of $298$ stories from TWTR and $36$ out of $391$ stories from FB evaluation dataset are missing or presumably cannot be accessed with the current level of privilege. 
\item \emph{Multimedia News Stories}: Unlike traditional article-based news, the financial news stories from Eikon API are primarily comprised of multimedia content from partnered news agencies. Particularly, $135$ out of $298$ stories from TWTR and $108$ out of $391$ stories from FB evaluation dataset only contain an URL to Refinitiv Newscasts\footnote{Refinitiv Newscasts is an online video platform that offers high-quality financial news from Reuters and trusted third-party partners in formats of multimedia (e.g., recorded videos, live shows and audio podcast).} in formats of multimedia content, which require additional steps (i.e., automated transcription technique for speech-to-text) before applying the NER technique for extracting named entities. 
\item \emph{Irrelevant Historical News}: It is observed that most of the retrieved news headlines are irrelevant to the cause of the financial anomaly, especially for articles that are days before the anomaly date. 
Although this observation is expected because analyst reports regarding the cause of financial anomaly usually published hours after the anomalous price movements as discussed in Section~\ref{sec:irm}, such delay is much shorter on social media like Twitter. Hence, the practicality of historical Reuters News is inconclusive. 
\end{enumerate}

In addition, the benefit of including names of key decision makers is negligible, as only less than $0.05\%$ documents are mentioning the name of key decision makers without addressing the company name using words or tags. Intuitively, this phenomenon is expected because it is unusual for people to discuss the cause of an extreme price movement without addressing the company by its name. Even when the key decision makers are the source of the anomaly, it is unlikely for a human to directly link the information to the stock market without domain knowledge. For instance, in the Microsoft example mentioned in Section~\ref{sec:query-finalization}, the prerequisite of people connecting the divorce between two key decision makers (i.e., Bill and Melinda Gates) and the stock price of Microsoft is the knowledge of \emph{(i)} Bill Gates is one of the founders of Microsoft \emph{(ii)} the divorce can negatively impact the stock by diluting the shares possessed by Bill Gates. Therefore, for reliable knowledge that discusses the potential cause of a financial anomaly, an explicit association between the information and the stock price should be addressed in the text, or a trivial implicit linkage can be assumed by the general public instead of mentioning only the information without context. \smallbreak

\subsubsection*{Query Enhancement using Dynamic Keywords from Twitter}

Compared to NER on Reuters News, using PRF on Twitter is more effective in discovering correlated keywords dynamically. For instance, PRF has identified cashtag \emph{"\$amzn"} as correlated keywords for both TWTR and FB financial anomalies, in which the earning announcement of Amazon Inc. in Q1 2021 was released on 29 April 2021, same as Twitter and Facebook. The earnings report from Amazon was frequently used to compare its performance against the other two companies, partially constitute a factor that causes an abnormal price movement for TWTR and FB. In addition, for both evaluation datasets, the company name has been correctly identified using the PRF technique (i.e., Twitter for \$TWTR and Facebook for \$FB), identical to the one retrieved from Reuters News using NER. Overall, the PRF query enhancement technique has demonstrated its effectiveness in improving search query quality as expected. \smallbreak

Besides, the trade-off of precision for recall with relaxed keyword condition \emph{("stock" OR "price")} instead of \emph{("stock" AND "price")} has justified its benefit based on the analysis of the annotated subset. Statistically, with a stricter condition, the tweet collection size for TWTR and FB evaluation datasets are reduced to $8,421$ and $3,029$ tweets, respectively, with more than $96.5\%$ tweets neglected from the collection. To justify the choice of relaxed keyword condition, it depends on two factors \emph{(i)} which objective, recall or precision, is valued more importantly by the end-user \emph{(ii)} the percentage of samples that are considered as reliable, but fail to satisfy the strict keyword condition. For the TWTR evaluation dataset, there are in total $568$ tweets in the annotated subset contain cashtag \emph{\$TWTR}, while there are only $47$ of them satisfy the strict keyword condition. In addition, within the $568$ tweets, there are $175$ of them are labeled as reliable, whereas only $17$ of them contain both \emph{"stock"} and \emph{"price"} keywords. Therefore, enforcing the relaxed instead of strict keyword condition in the search query has proved its value by retrieving more reliable information from Twitter, and with recall as the primary objective for the IR module, using the relaxed version is advantageous. \smallbreak

In conclusion, instead of using named entities and names of key decision holders from Reuters News to expand the Twitter search query, the performance of the DeepTrust IR module can be better if only using dynamic keywords discovered using the PRF technique. \smallbreak

\subsection{Impact of Naming Ambiguity and Tag Abuse on Tweet Collection Quantity} \label{sec:tag-abuse}

Firstly, it is observed that some tweets are mistakenly deemed correlated by the IR module due to naming ambiguity, and two types of ambiguity are identified as the cause of this issue:

\begin{enumerate}
\item \emph{Cashtag Ambiguity}: The first type of ambiguity arises from the naming of stock tickers. For instance, when searching for the Facebook cashtag \emph{\$FB}, tweets that are mentioning similar cashtags such as Fortress Biotech \emph{\$FBIO} or Flagstar Bancorp Inc. \emph{\$FBC} that also contain the keyword \emph{\$FB} are falsely included in the tweet collection, which are irrelevant to the Facebook financial anomaly. 
\item \emph{Lexical Ambiguity}: The second type of ambiguity is harder to identify as keywords in the search query may carry different semantics. For instance, there are a large number of advertisements that frequently mention keywords \emph{"stock"}, which within the context of these promotional materials, \emph{"stock"} means the inventory that is available for sale. However, the semantics of \emph{"stock"} that DeepTrust seeks is the security issued by a company for raising capital, which is different from those advertising tweets that coincidentally satisfy the criteria of the search query.  
\end{enumerate}

Both ambiguities collectively increase the tweet collection size significantly, whereas the latter ambiguity is the primary factor causing the TWTR evaluation dataset to have a considerably larger tweet collection than the FB dataset. As shown in Table~\ref{tab:evaluation-dataset}, the TWTR dataset consists of over $241$K tweets while the FB dataset only has approximately $32$K tweets in comparison. With further investigation, the ambiguity of the keyword \emph{"Twitter"} is the source of this issue. For the term \emph{"Twitter"}, it can either refer to the company \emph{"Twitter"} with stock ticker symbol \emph{TWTR}, or represents the social media platform Twitter that people can communicate or share information via tweets. In practice, a large proportion of tweets in the TWTR dataset are using \emph{"Twitter"} to address the social media platform instead of the company itself, which severely affects the effectiveness of IR module in maintaining a concise collection of correlated tweets. \smallbreak

\begin{figure}[!htbp]
\centering
\includegraphics[width = 1\hsize]{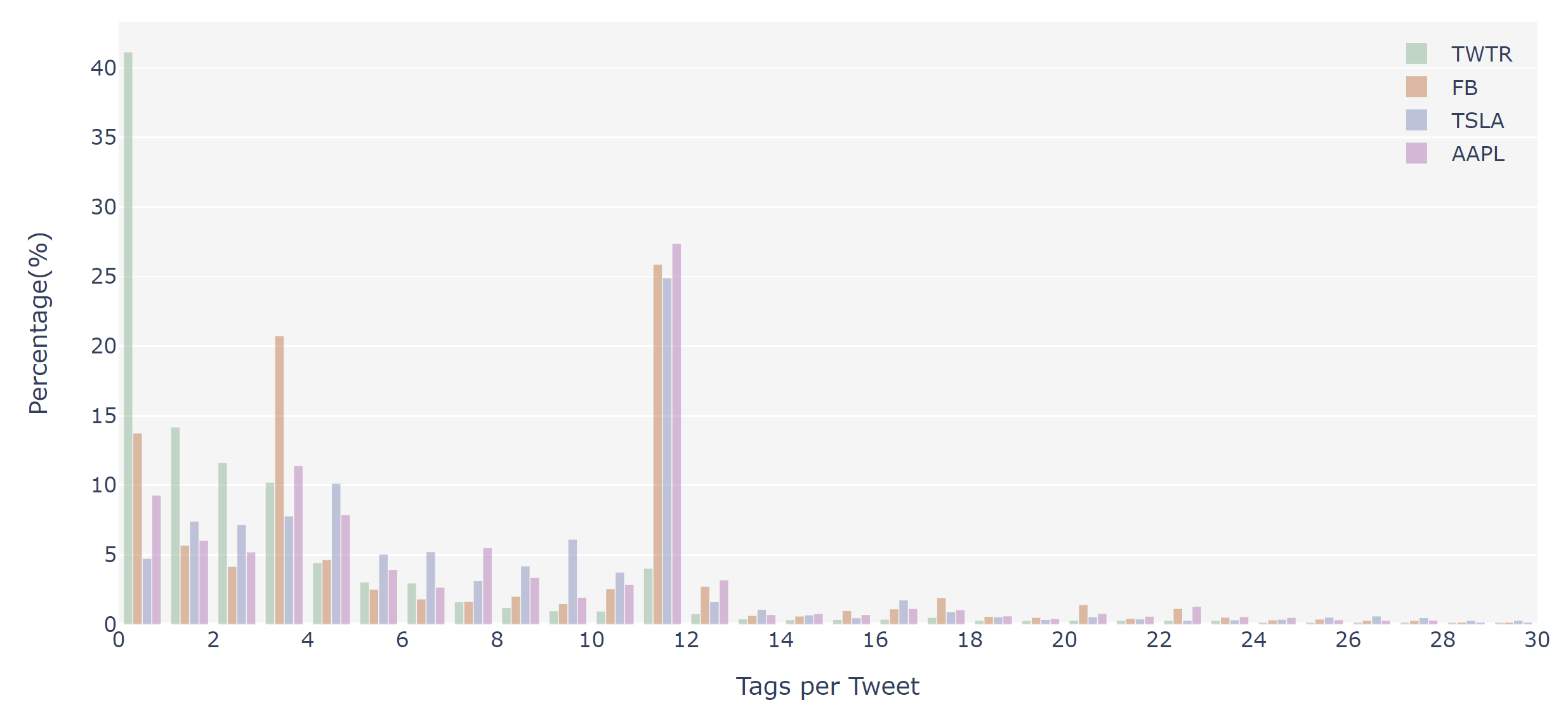}
\caption{An overview of Tags per Tweet histogram on Twitter, Facebook, Tesla and Apple tweet collections. All tweets that contain more than $10$ tags are counted for analysis.}
\label{fig:tags-histogram}
\end{figure}

In addition, the abusive behavior on Twitter hashtags and cahstags has gradually become a common practice for bot-generated content to promote automated market updates or fake news, and is considered as the greatest obstacle for DeepTrust to effectively retrieve a refined collection of high-quality tweets. Specifically, Facebook and Apple Inc., as part of the \emph{FAANG}\footnote{FAANG is an abbreviation for the five prominent American companies in the information technology industry: Facebook, Amazon, Apple, Netflix and Google.} companies, are generally more popular in terms of tags usage comparing to small and medium-sized enterprises (SMEs), while Twitter and Tesla Inc. have also attracted more attention recently due to the active engagement of their founders and CEOs on Twitter. Since these four companies are popular choices of bot accounts to increase tweet exposure, they are selected as examples to illustrate the distribution of tags per tweet using their corresponding tweet collection retrieved from their anomalies\footnote{In addition to the TWTR and FB financial anomalies, two additional anomalies, Tesla Inc. (i.e., extreme price movement on 22 Feb 2021) and Apple Inc. (i.e., extreme price movement on 4 May 2021) are used to demonstrate the tag abuse behavior.}. Statistically, as presented in Figure~\ref{fig:tags-histogram}, approximately \numrange{8}{41}\%\footnote{Percentage of tweets that have more than $10$ tags: Twitter $8.0\%$, Facebook $39.6\%$, Tesla $35.9\%$ and Apple $40.6\%$.} of them consists of more than $10$ tags per tweet, which are commonly deemed as excessive and only appear in bot-generated tweets that wish to reach a wider population with unnecessary trending tags for exposure. \smallbreak

\begin{figure}[!htbp]
\centering
\includegraphics[width = 1\hsize]{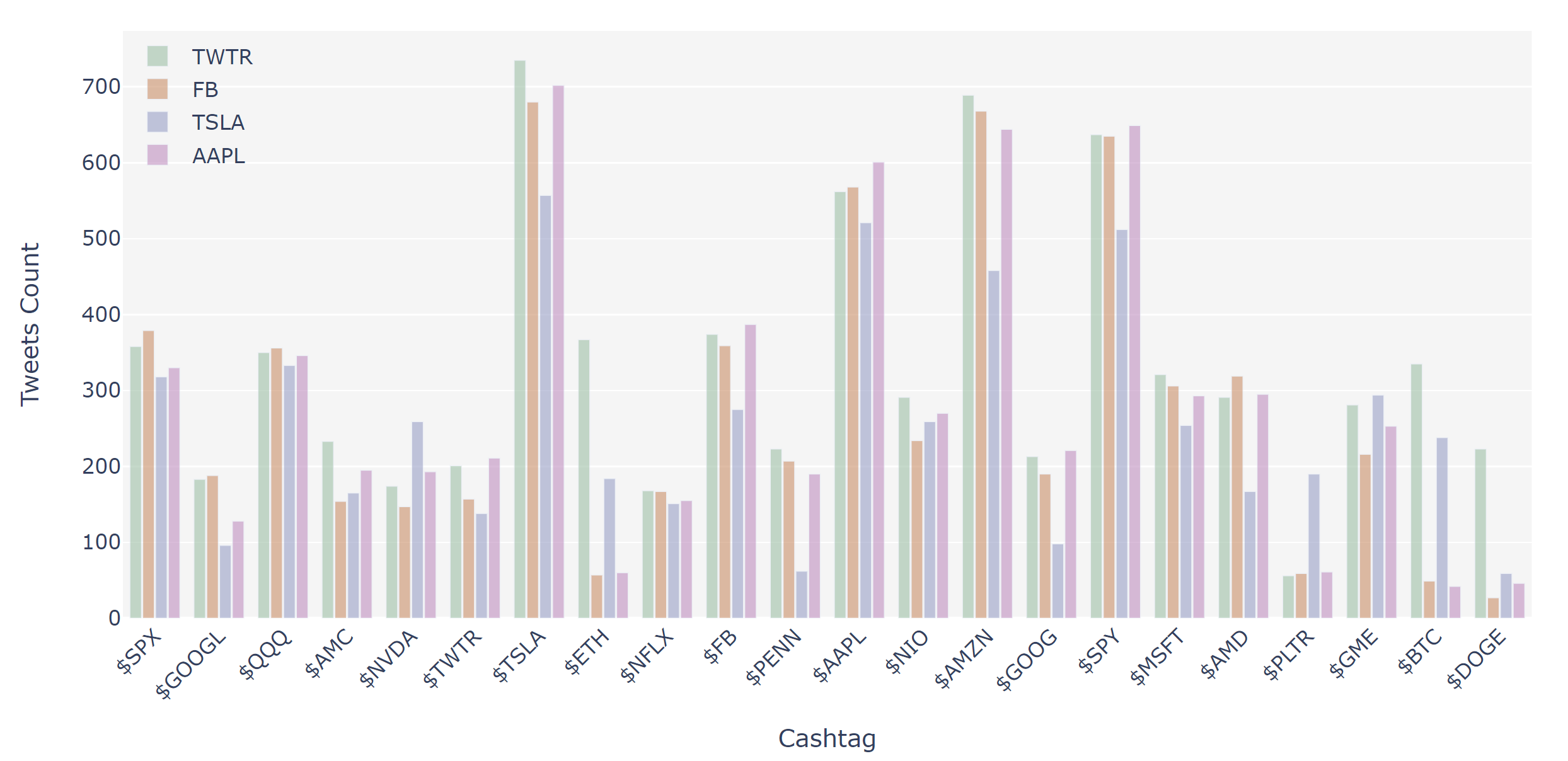}
\caption{An overview of Cashtag Abuse barchart on Twitter, Facebook, Tesla and Apple tweet Collections. All tweets that contain more than $10$ tags are collected first, and the top $15$ cashtags from each collection are counted for analysis.}
\label{fig:tags-barchart}
\end{figure}

Besides, in the context of DeepTrust, cashtag abuse is more severe than hashtags due to its popularity in the financial domain. Since cashtag is widely adopted by users to communicate financial information on Twitter, bots that monitor the market and provide live stock market updates tend to include a large list of trending cashtags to attract attention to its account. Figure~\ref{fig:tags-barchart} provides an overview on cashtag abuse. For all four companies with excessive tags (i.e., more than $10$ tags per tweet), the top $15$ cashtags from each collection are picked and counted based on their frequency of appearance. It is evident that regardless of which company is the target entity, trending cahstags such as \emph{\$TSLA}, \emph{\$AMZN}, \emph{\$AAPL}, \emph{\$SPY} have always been mentioned frequently by these speculated bot-generated tweets. Therefore, the abusive behavior on tags, especially cashtags, severely affects the effectiveness of the IR module in retrieving tweets with high precision, as these bot-generated tweets have continuously infiltrated into the tweet collection and dilute its overall quality. \smallbreak

Nevertheless, bot-generated tweets exhibit specific patterns when using cashtags, which may be leveraged as a part of the feature-based filter to screen irrelevant stock market updates, whereas the effectiveness of this approach is questionable due to its aggressiveness. Firstly, it is observed that a large proportion of stock market updates tend to state the related cashtag at the beginning of the tweet, such as \emph{"\$FB Social media giant Facebook Inc reported first-quarter revenue growth of 48\% year-over-year."}, with approximately \numrange{10}{20}K\footnote{There are $23,535$ tweets from the TWTR and $11,202$ tweets from the FB evaluation datasets that start with a cashtag.} tweets per dataset are using this writing style. To screen irrelevant stock market updates, a simple rule can be removing all tweets that start with an unknown cashtag that is not in the search query. However, such rule may be overly aggressive as tweets may \emph{(i)} contain multiple sentences, and each sentence starts with a different cashtag \emph{(ii)} start with a seemly "irrelevant" cashtag, but are correlated (e.g., the relationship between \emph{\$AMZN} and \emph{\$FB}). Therefore,  with recall as the primary objective for the IR module, these stock market updates are preserved and may be filtered based on their argumentation structure or subjectivity filter, instead of the position of cashtags. 

\subsection{Impact of Delayed Evaluation on Feature-based Filter Performance}

From Table~\ref{tab:weighted-report}, it is evident that the feature-based filter alone is effective in filtering unreliable information, with approximately \numrange{9}{18}\% boost in precision and \numrange{5}{14}\% increase in F0.5-scores when comparing with the baseline classifier. In addition, when evaluating its performance using per-class metrics from Table~\ref{tab:classification-report}, the feature-based filter individually has achieved the best precision metric among all candidates, with a slightly lower F0.5-score comparing with the combination of feature-based filter + synthetic text filter. In fact, among all three features (i.e., textual, tweet-meta and user) used for threshold-based classification, tweet-meta features that capture engagement metrics contribute the most in filtering unreliable tweets, causing approximately \numrange{50}{71}\% tweets from the TWTR and FB evaluation datasets\footnote{For TWTR evaluation dataset, $121,669$ out of $241,714$ tweets are excluded by feature-based filter based on tweet-meta features, and $23,269$ out of $32,983$ tweets for FB evaluation dataset.} excluded from the collection. A similar performance is observed in the annotated subset, with approximately \numrange{75}{84}\% unreliable tweets are correctly identified by the feature-based filter as presented in Table~\ref{tab:classification-report}. Based on statistics, the feature-based filter is the most effective coarse-grained screening solution that identifies the most amount of trivially unreliable tweets. \smallbreak

However, the achieved effectiveness may not be replicated to another scenario with a shorter interval between anomaly date and evaluation period. For this project, evaluation datasets are collected between 10 June to 15 July, 2021, more than five weeks after the anomaly occurred. One side effect of such delayed data collection and evaluation is that future knowledge is used to assess tweet reliability, as tweet engagement metrics are closely related to the popularity and perceived credibility of information on Twitter. With approximately one month buffer time, engagement metrics generally converge to a fixed number due to the tweet ranking systems employed by Twitter. As discussed in Section~\ref{sec:tweets-retrieval-and-ranking}, recency, as one factor of the ranking and retrieval model, can influence the probability of tweets appearing in the top SERP. For tweets posted months in the past, it is reasonable to assume fewer users may interact with these old tweets unless using \emph{(i)} explicit search queries with specified time ranges or \emph{(ii)} account-driven searches on tweets posted by a particular account, which are less common in practice when browsing a topic on Twitter. Therefore, it is important to acknowledge that regardless of the effectiveness of the tweet-meta feature alone in excluding unreliable tweets during this experiment, textual features and user features are equally important in distinguishing unreliable tweets and shed the light on information credible. If both datasets are collected minutes after the anomaly date, it is expected tweet-meta features would be less useful as the majority of tweets may not be interacted by users yet, while traces hidden in the user and textual features may be the key indicator for information reliability. \smallbreak

\subsection{Impact of Threshold Values on Feature-based Filter Performance}

As stated in Section~\ref{sec:feature-filter}, DeepTrust uses a threshold-based feature filtering algorithm to exclude tweets with limited public interactions (i.e., engagement metrics), while the minimum degree of engagement, expressed by four prominent features of interaction, is configured to $0$ for this project. However, in practice, the threshold value should not be determined arbitrarily without supporting evidence, and one exceptional case is identified as a counterexample of using a threshold value of $0$ for the feature-based filter. \smallbreak

In the second annotation sample included in Section~\ref{sec:data-annotation-guideline}, a typical stock price target update is deemed as reliable based on its implicit argumentation structure and its correlation with price movement. These tweets are usually generated automatically by bot accounts that constantly monitor news feeds for price target updates, or can be generated by a neural model to confuse investors with falsified information. For both cases, financial data providers like Refinitiv have sufficient tools and resources (e.g., robotics process automation solution that compares price target against curated stock fundamentals) to validate the price target using its knowledge base conveniently, thus it is safe to assume all these types of tweets are reliable first by DeepTrust, and validate its correctness later. However, among all $443$ tweets from the FB evaluation dataset that discuss stock price target updates, $257$ of them have no engagement metrics (i.e., $0$ for all four tweet-meta features), and $1,329$ out of $2,683$ tweets for TWTR evaluation dataset. In other words, with a threshold value of $0$, approximately \numrange{50}{58}\% of tweets result in false negatives if assuming all updates are authentic, which can severely impact the feature-based filter performance. \smallbreak

Nevertheless, for the feature-based filter, there is no simple solution that can preserve these stock market updates in the collection while excluding irrelevant information in large-batch simultaneously. One naive approach can be an expert-written list of exceptional cases, to ensure stock market updates or other forms of bot-generated content (e.g., earnings calendar updates) that are meaningful are always deemed as reliable by the filter. However, using a hard-coded list instead of relying on linguistic features affect the flexibility of DeepTrust, and preparing such lists also requires extensive effort in both composing and maintaining it up to date. Therefore, instead of relying on DeepTrust to include all stock market updates into the trusted collection of tweets, it is advised that analysts should reference price target updates from their proprietary data feeds such as Refinitiv Eikon, and ignore these updates from the Twitter data stream. \smallbreak

\subsection{Impact of Rule Aggressiveness on Synthetic Text Filter Performance}

For the synthetic text filter, there are several configurable parameters for users to adjust the filter aggressiveness, namely RoBERTa and SVM Classifier thresholds, GLTR-BERT and GLTR-GPT2 weights, and filter mode (i.e., precision or recall mode). Among all these parameters, weights for GLTR-related variables are determined well-founded based on the average accuracy of the detector on 10-fold cross-validation, and the choice of filter mode is determined based on use cases. However, for the RoBERTa threshold, its value $0.7$ is decided arbitrarily based on empirical analysis. To investigate the impact of filtering rule aggressiveness on synthetic filter performance, a series of experiments are conducted to evaluate the sensitivity of performance metrics on changing RoBERTa threshold values. \smallbreak

\begin{figure}[!htbp]
\centering
\includegraphics[width = 1\hsize]{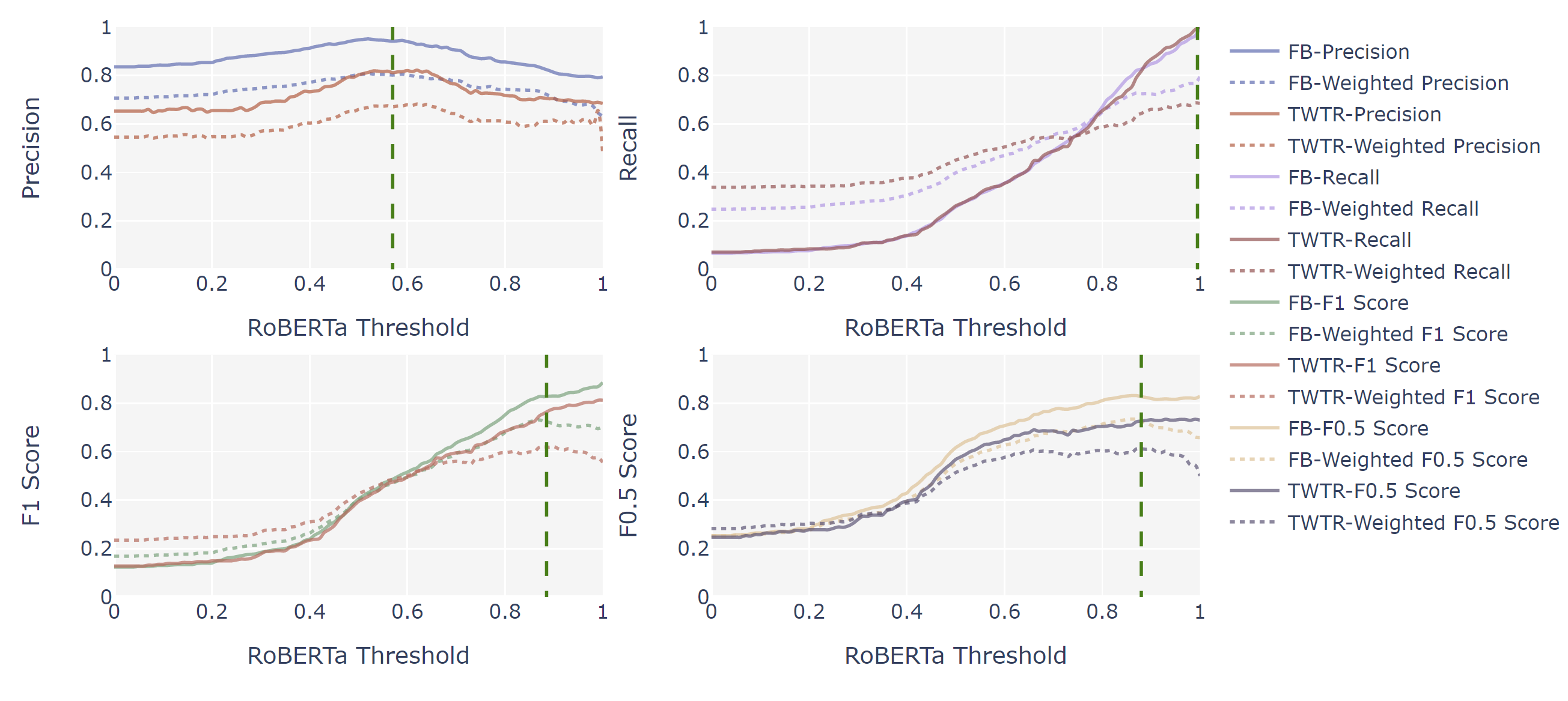}
\caption{Sensitivity Analysis on how synthetic text filter performance is affected by changes in RoBERTa threshold value that interpreted as filter aggressiveness. Classifier threshold is set to $0.9$ and filter mode is \emph{precision}. }
\label{fig:synthetic-filter-sensitivity}
\end{figure}

Firstly, for binary classification on imbalanced classes, the default threshold of $0.5$ may not represent an optimal decision boundary of the predicted probabilities, resulting in poorer performance. Although the synthetic text filter uses a rule-based system instead of a binary classifier, the rationale behind is comparable. Common techniques adopted for identifying the optimal decision threshold include visualizing a receiver operating characteristic (ROC) curve or precision-recall curve first, then select the best threshold that yields the lowest loss (e.g., categorical cross-entropy) or the greatest performance metric (e.g., F0.5-score for this project). For the synthetic text filter, instead of changing the decision threshold of the classifier, the RoBERTa threshold is selected based on its prominent impact on filter aggressiveness. As presented in Figure~\ref{fig:synthetic-filter-sensitivity}, the performance of synthetic text filter (precision mode) is evaluated with different RoBERTa thresholds, from \numrange{0.0}{1.0} with $0.01$ interval in-between. For each performance metric, per-class (i.e., negative class) and weighted metrics are calculated, including all key metrics defined in Section~\ref{sec:experiment-setup}. For each subplot, the averaged RoBERTa threshold that yields the greatest weighted metric on TWTR and FB annotated datasets are recorded and presented in the dashed green vertical line, and it is evident that a threshold greater than $0.5$ is better for precision and F0.5-scores. In particular, for TWTR annotated subset, the optimal threshold for weighted F0.5-score is $0.9$, and it is $0.86$ for FB annotated subset. However, when evaluating based on weighted precision, the optimal thresholds for TWTR and FB annotated are only $0.62$ and $0.52$, respectively, significantly lower than the optimal threshold for F0.5-scores. Nevertheless, both primary objectives require a threshold greater than $0.5$, indicating that a more aggressive filter is preferred. Therefore, for Refinitiv that entirely focuses on the precision objective, a less aggressive filter (i.e., RoBERTa threshold \numrange{0.5}{0.6}) is more suitable, whereas for SMEs who seek a balanced solution with more emphasis on precision, a more aggressive filter (i.e., RoBERTa threshold \numrange{0.8}{0.9}) is desirable.  \smallbreak

In addition, as stated in~\cite{solaiman2019release}, the $95\%$ accuracy achieved by the pre-trained RoBERTa discriminative model on its training data (i.e., WebText and synthetic texts generated by GPT-2-XL) is far from its full potential. During the training phase, both human-written and its corresponding synthetic text are presented to the model under paired settings, which is infeasible in practice. For standalone detection situations, this model may underperform due to the absence of reference texts, and result in lower performance metrics. To compensate for its weakness under standalone scenarios, one suggestion proposed by OpenAI in~\cite{solaiman2019release} is to use this model in conjunction with metadata-based filters, which is identical to the observation in our experiment. In terms of weighted metrics presented in Table~\ref{tab:weighted-report}, the combination of the feature-based filter and synthetic text filter yield the optimal performance among all candidates, and result in a \numrange{6}{7}\% boost in F0.5-score compared with individual synthetic text filter.  Therefore, our experiment reinforces the conclusion made by the OpenAI team, and successfully enhances the overall performance of the RA module by leveraging the strength of multiple filters, instead of relying on an excessively aggressive filter to improve performance metrics. \smallbreak

\subsection{Impact of Ever-Stronger Adversaries on Synthetic Text Filter Effectiveness}

It is evident that the synthetic text filter has achieved outstanding performance in filtering unreliable information, with the highest precision on both evaluation datasets. When used in conjunction with the feature-based filter, the F0.5-score has increased about $6\%$ at the cost of a slight drop in precision as presented in Table~\ref{tab:classification-report}. In fact, this combination has yielded the best F0.5-score and precision among all combinations when evaluating using weighted averages, and achieved the best metrics on the TWTR evaluation dataset when evaluating with per-class metrics, which has proved its effectiveness in filtering unreliable information. Both fine-tuning based and supervised learning based approaches have collectively contributed to the effectiveness of this filter. However, with the release of advanced few-shot learners such as GPT-3, whether this filter can maintain its sharpness against the next frontier of generative writing is unknown, and it is essential to understand which agent, the generator model or the discriminative model, in this adversarial game can win the race. \smallbreak

To understand the impact, the GPT-3 model proposed in~\cite{brown2020language}, the latest version of few-shot learner in the GPT family with ten times more parameters (i.e., $175$B parameters) than its predecessor GPT-2, is used to evaluate the performance of the synthetic text filter. For instance, the following tweet is a neural text generated by the GPT-3 model\footnote{Generated using the interface developed by Sushant Kumar, available at: \url{https://thoughts.sushant-kumar.com}.} based on the prompt \emph{"stock"}:
\begin{quote}
    "I don’t care if you think \$AAPL‘s stock is overpriced. \smallbreak
    You’re wrong. Get over it and focus on true business metrics."
\end{quote}
in which even for a human, it is indistinguishable from a human-written tweet. In fact, when applying the synthetic text filter, none of the detectors are able to identify it as a synthetic text based on its traces of language modeling. In particular, the RoBERTa detector claims this tweet being human-written with a probability of $62.09\%$, while both SVM classifiers trained on TWTR financial anomaly predict this tweet as human-written with $100.00\%$ confidence. More interestingly, the GLTR $frac(p)$ distribution for this tweet is nearly identical to the histogram of the human-written tweet presented in Figure~\ref{fig:gltr-comparision}, with a balanced convex shape instead of a right-skewed distribution. Since the OpenAI team currently only invite selected researchers at institutions to join their private beta to explore the GPT-3, it is uncommon to see GPT-3 generated synthetic text on Twitter when this report is written, and the performance achieved by the synthetic text filter still reflects its true capability in screening neural texts in present. Nevertheless, DeepTrust acknowledges that with more powerful adversaries released to the public, the performance can be overestimated by using evaluation datasets comprised of less deceptive neural texts generated by BERT or early versions of GPT models. \smallbreak

\subsection{Impact of Writing Style Difference on Argumentation Filter Performance} \label{sec:argumentation-filter-discussion}

For the argumentation filter, its performance is identical to an inversed baseline classifier, by predicting all tweets as reliable under the relaxed definition of argumentativeness. As presented in Table~\ref{tab:classification-report}, metrics for the argumentation filter are all $0.00\%$ as no tweets are predicted as unreliable, and when used in conjunction with other filters, it has no impact on the overall performance as it always outputs true for all tweets in both evaluation datasets. In addition, this filter has only two settings (i.e., depending on whether an argumentative tweet should contain only \emph{premise} or both \emph{claim} and \emph{premise}), in which the relaxed setting will predict all tweets as reliable and the strict setting will predict all tweets as unreliable. Therefore, this filter fails to derive tweet reliability based on its argumentation structure, and it is essential to understand the cause of ineffectiveness. \smallbreak

Unfortunately, though the author redefined the Twitter-specific argumentativeness and leveraged state-of-the-art sequence tagger TARGER to detect argumentative units in tweets, the absence of DART still severely impacts the effectiveness of the argumentation filter. In particular, the training dataset for TARGER comprises snippets from Wikipedia, written in a different style than posts on Twitter. For all English Wikipedia articles, a style guideline named Manual of Style (MoS) proposed by Wikipedia is enforced to improve the readability of pages, including instructions on capitalization, punctuation, grammar and other grammatical features. For instance, in the IBM Debator - Claim Evidence Search dataset, a typical financial sentence takes a formal writing style with proper usage of punctuation:

\begin{quote}
    \emph{"The stock market liked the acquisition and the price of Ariba's shares rose from $\$57$ at the time of the announcement to $\$173$ at closing on March 9, 2000, which also marked the peak of the Internet Bubble."}
\end{quote}

whereas tweets are mostly written in an informal tone, such as examples presented in Section~\ref{sec:neural-phase-4}. Although additional preprocessing techniques are applied to transform tweets into a similar format, the difference is still drastic and the sequence tagger cannot properly identify claims from them. In addition, due to the character limit, it is common to use implicit claims, or replace them with abbreviations and Twitter-specific features like cashtags as discussed in Section~\ref{sec:argument-mining-definition}. For humans, they can easily associate the tweet with its implicit claim using domain knowledge, whereas for the sequence tagger, only explicit claims can be identified. Therefore, instead of using a sequence tagger trained on Wikipedia, a classifier trained on tweets with annotated argumentative units is believed to be more effective in filtering unreliable tweets based on its argumentation structure. \smallbreak

\subsection{Implication of Subjectivity on Information Reliability}

Firstly, the objective of the subjectivity filter is to exclude opinion pieces that are usually represented by subjective statements, and it is assumed that subjective information is less reliable compared with objective statements when used to explain a financial anomaly. However, in practice, there is no visible relationship between subjectivity and reliability, as \emph{(i)} most tweets written by individual users comprised of subjective words and \emph{(ii)} reliable knowledge may be conveyed in subjective statements written in first-person perspective, such as this one from FB annotated subset:

\begin{quote}
\emph{"Wow. \$FB with an extremely good earnings report. $48\%$ YoY top-line growth and almost doubling their net income from last yr. Mega tech is so strong. \$FB is the only FAANG stock I own atm."}
\end{quote}

which is reliable because it discusses the cause of sky-rocket price using grounds (i.e., \emph{"extremely good earnings report"}) to support its claim (i.e., \emph{"Mega tech is so strong"}) with backing (i.e., \emph{"$48\%$ YoY top-line growth"} and \emph{"doubling income"}), while being subjective as it uses first-person narrative with emotional words like \emph{"Wow"} and \emph{"So"}. There are numerous cases in both annotated subsets that subjective tweets are indeed valid and reliable knowledge that can explain the anomaly, and it is inconclusive on the correlation between subjectivity and reliability. \smallbreak

\begin{figure}[!htbp]
\centering
\includegraphics[width = 1\hsize]{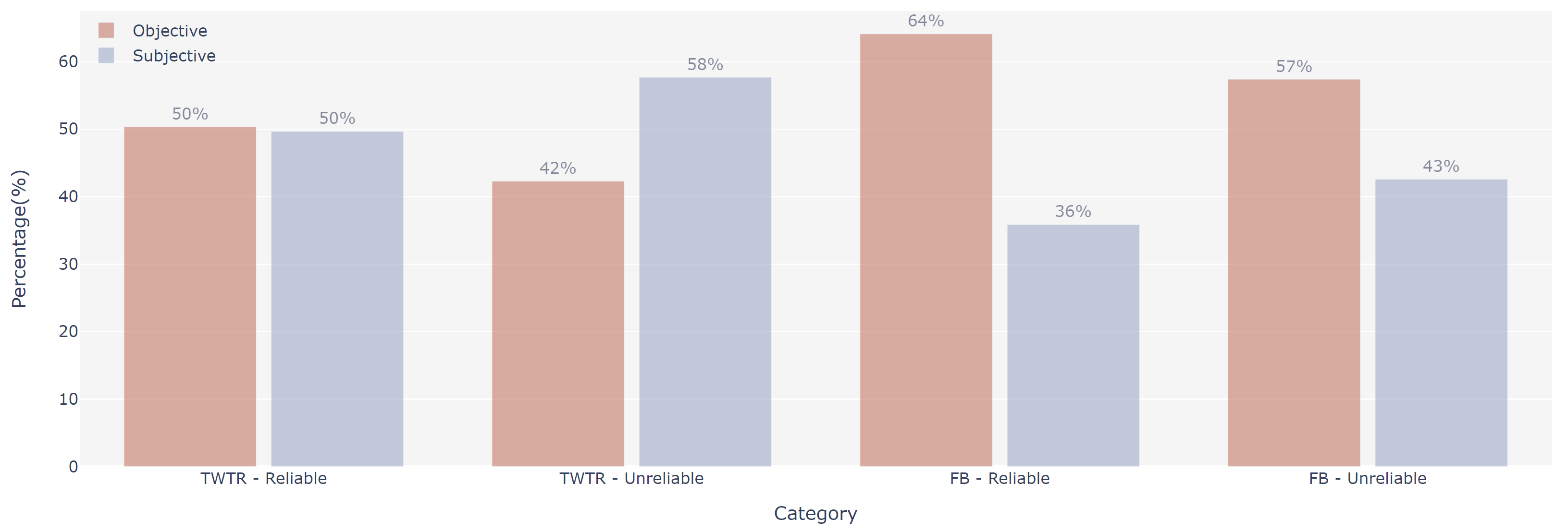}
\caption{The distribution of subjective and objective tweets of TWTR and FB annotated subset based on classification results of the subjectivity filter.}
\label{fig:subjectivity-distribution}
\end{figure}

Statistically, as presented in Figure~\ref{fig:subjectivity-distribution}, the distribution of subjective and objective tweets in TWTR and FB annotated subsets are relatively random, with no clear indication on traits that distinguish reliable tweets from unreliable ones. In particular, the subjectivity composition of unreliable tweets for TWTR (i.e., $42\%$ objective and $58\%$ subjective) is completely reversed comparing with FB (i.e., $57\%$ objective and $43\%$ subjective), and it is also unverified that reliable tweets consistently contain more objective tweets than subjective ones, partially reinforce the counterexamples aforementioned. However, it is worth noting that the distribution is based on the output of three subjectivity models using different representations (i.e., word embeddings, sentence embeddings and subjectivity lexicon), while the ground truth of subjectivity distribution is unknown as annotators are not required to label tweets based on their subjectivity. In other words, the above evaluation only describes the fact that the subjectivity filter does not identify a clear trait for assessing reliability using subjectivity, while the actual correlation between subjectivity and reliability remains a mystery, and requires additional annotation efforts to reveal the truth. Therefore, considering the complexity and inconclusiveness of the relationship between subjectivity and reliability, this feature is not included as a part of the data annotation guideline, and annotates are not required to assess reliability based on tweet subjectivity. \smallbreak

\subsection{Implication of Sentiment Signal Alignment on Reliability Assessment}

Designed as a bonus feature for equity traders to uncover insights on market, the sentiment signal is not originally considered as a valid filter to exclude unreliable information. One critical issue is the aggressiveness of such filter, and the assumption it relies on may also introduce excessive confirmation bias to the outcome. The following paragraphs use both statistics and reasoning to justify if leveraging a sentiment filter is effective or not for DeepTrust, and some lessons learned from experimenting with this filter. \smallbreak

In general, a stock price surge indicates bullish market sentiment while a stock price crash indicates bearish market sentiment. Hence, a sentiment filter operates by assuming correlated tweets should contain an aligned sentiment with the price movement direction, either because \emph{(i)} the tweet uses sentiment-aligned keywords when describing the price movement or \emph{(ii)} anomalous price movement is caused by a sentiment-aligned issue, and the tweet is discussing it using sentiment-aligned keywords. For instance, when describing the TWTR financial anomaly, common descriptive keywords such as \emph{"tanking"}, \emph{"crash"} are all associated with negative sentiment. In addition, when describing the cause of anomaly like \emph{"weak future guidance"} and \emph{"slow user growth"}, the overall sentiment of these tweets is likely to be negative, aligned with the price movement. Statistically, for TWTR annotated subset, $60.10\%$\footnote{TWTR: Out of $208$ reliable tweets, $125$ tweets have negative sentiment, $50$ tweets are neutral and $33$ tweets have positive sentiment.} tweets are aligned with the negative price movement, while $72.22\%$\footnote{FB: Out of $216$ reliable tweets, $45$ tweets have negative sentiment, $15$ tweets are neutral and $156$ tweets have positive sentiment.} tweets in FB annotated subset align with the positive sentiment brought by the stock price surge. It is evident that the majority of reliable tweets have a sentiment that is neutral or aligned with price movement direction, and validates the aforementioned assumption on sentiment signal alignment. \smallbreak

Nevertheless, there exist several counterexamples that contradict the above findings. Firstly, both TWTR and FB financial anomalies are not representative of the whole universe of equity markets, such as the bond market and derivatives market (e.g., options, futures). Considering a European put option that allows the holder to sell the associated equity at expiry with the strike price, investors may only earn profits if the stock price is far below the strike price to cover the cost of option premium. Consequently, these investors largely rely on negative news of the company to bring down the stock price before the expiry date, and earn profit in the opposite direction to traditional traders that open positions by buying stock shares. In addition, short selling is another scenario in which the overall sentiment can be the opposite of the price movement direction. In the financial field, short selling is a trading strategy that investors use to generate profit by speculating on the decline in equity price, and is legit in most countries including the United States and the United Kingdom. Unlike traditional stock trading strategies that earn profit from higher stock price, shorting shares require investors to borrow shares upfront and sell it to other investors, and earn profit by purchasing them back later with a lower stock price. Thus, experienced investors using this strategy would anticipate the stock price to decline, and express a positive sentiment when it occurred, which contradicts the assumption on sentiment signal alignment. \smallbreak

Overall, since using the sentiment signal from Twitter is less reliable and excessively aggressive, this filter is excluded from the DeepTrust framework. However, the market sentiment is essential for short-term investors to earn profit based on its impact on technical indicators. For instance, market sentiment may assist contrarian investors, who trade inversely to the prevailing consensus, to understand if equity is overvalued or undervalued. Instead of relying on Twitter sentiment signals, market sentiment indexes such as CBOEVolatility Index (VIX) are popular choices for analysts to understand investor confidence in the market. VIX, calculated based on option premium, measures the expectation for volatility over the next month, and indicates whether investors are optimistic or pessimistic about the market. Therefore, though the sentiment filter can only apply to a limited number of markets, market sentiment indexes are generally more reliable and flexible when applying to complex markets like the derivatives market. \smallbreak

\chapter{Concluding Remarks}

\section{Conclusion}

In this project we proposed DeepTrust, a practical knowledge retrieval and assessment framework for equity traders to understand the cause of pricing anomalies with non-standard data sources. The framework consists of three core modules, namely the \emph{(i)} anomaly detection module - detect extreme price moves in the securities market \emph{(ii)} information retrieval module - collect unstructured data from Twitter that are correlated to the financial anomaly \emph{(iii)} reliability assessment module - refine a quality collection of reliable information that explains sudden price changes. The objective of DeepTrust is to source and organize disparate information that uncovers insights of a sudden price surge or crash, using unstructured data from Twitter. Through extensive analysis, DeepTrust is capable of collecting a concise curated database of correlated financial information from the ocean of tweets, and distinguishing various forms of unreliable information precisely, including human-written disinformation, synthetic text and opinion pieces. In addition, as a collaborative project with Refinitiv, a customer-centric prototype is developed for end-users to experiment DeepTrust on recent pricing anomalies of Twitter Inc. and Facebook Inc, and evaluate its performance accordingly using an annotated dataset labeled by the author following a well-defined annotation guideline based on the definition of financial reliability. \smallbreak

To the best of our knowledge, DeepTrust is the first domain-specific framework that detects, gathers and understands financial anomalies using NLP. Compared with other existing reliability assessment solutions like multi-modal framework SpotFake for fake news detection and heuristic-driven ensemble Framework for COVID-19 disinformation detection, DeepTrust shows outstanding performance in screening unreliable information from different angles, including textual and tweet-meta features, traces of generative language model, argumentation structure and subjectivity. The proposed framework outperforms the baselines by a margin of $16.53\%$ precision and $11.76\%$ F0.5-score on weighted averages. Besides assessing reliability, DeepTrust can also sharply determine spikes in stock price caused by unexpected events like Mergers and Acquisitions (M\&A) rumors, and gather event-related information from Twitter using enhanced structured queries constructed in an unsupervised manner. Evaluations on both datasets devised for this framework demonstrate prominent performance on all three tasks and achieves its objectives stated in Section~\ref{sec:objectivies-and-contributions}. Although there are still room for improvement on more complex securities markets (e.g., derivatives market, bond market) and more deceptive disinformation generated by ever-stronger language models, DeepTrust offers a sophisticated yet flexible solution for financial data providers to alleviate the burden on screening information from social media feeds manually, and shed the light on pricing anomalies with a reliable yet concise collection of tweets. \smallbreak

\section{Legal, Social, Ethical and Professional Considerations}

Based on the ethics checklist\footnote{Ethics checklist is available at: \url{https://www.doc.ic.ac.uk/lab/msc-projects/ethics-checklist.xlsx}.} from Imperial College London Department of Computing, this project involves \emph{(i)} personal data collection and processing \emph{(ii)} exclusive civilian application focus. All other stated ethical concerns are not applicable to this project, thus neglected in the following discussion. \smallbreak

Firstly, DeepTrust involves collecting and processing personal data publicized on Twitter. Based on the General Data Protection Regulation (GDPR)\footnote{General Data Protection Regulation is a new regulation on data protection applicable to European Union starting from $2018$. } Article 4, \emph{personal data} refers to \emph{"information relating to an identified or identifiable natural person"} who can be \emph{"identified, directly or indirectly, in particular by reference to an identifier such as a name..."}. Tweets published on Twitter is, therefore, considered as personal data because it contains direct identifiers (e.g., author name, profile picture) and indirect identifiers (e.g., geolocation) of a natural person, and all tweets collected and processed in this project are protected by GDPR legislation. Since processing political opinions is prohibited in general based on GDPR Article 9 (i), DeepTrust intentionally neglects political polling fields when collecting information from Twitter, and focuses only on contents exempted from the stated prohibited categories. \smallbreak

In addition, based on the compliance requirement from Twitter, it is the obligation of developers and researchers to honor user intent on Twitter by respecting the changes users made to the availability of their posted content, such as modification and deletion. The rationale is that tweets are private data on public social media platforms based on ongoing consent under a contract between the viewer and author, and the viewer only has the right of data disposition instead of data ownership. To comply with this term, DeepTrust provides an update feature in the information retrieval module to assist users in updating the retained tweet collection and ensuring it is synchronized with the latest image on Twitter. \smallbreak

Besides, to ensure security for the curated database, DeepTrust originally intends to apply anonymization on personal data after the completion of this thesis. As stated in GDPR Recital 26, \emph{"The principles of data protection should therefore not apply to anonymous information"}, in which the anonymous information means it is unrelated to a natural person using direct or indirect identifiers. To be fully qualified for anonymization, the retained database should be deleted as both the tweet content and its metadata can be used to identify a natural person, which makes evaluation results presented in this thesis not reproducible. Nevertheless, DeepTrust applies pseudonymization, which replaces identifiers (i.e., metadata and all other fields except for the tweet ID) with a private number, and this safeguarding measure complies with Article 89 for \emph{"archiving purposes in the public interest, scientific or historical research purposes or statistical purposes"}, and other researchers may use this pseudonymized dataset for further experiments. \smallbreak

From an environmental perspective, training and fine-tuning ever-larger language models introduce significant environmental and financial costs to the project. Following the benchmark conducted by Strubell et al., training a BERT-base model on GPUs is equivalent to a domestic flight in terms of \si{CO_2} emission~\cite{strubell2019energy}, while DeepTrust leverages several complex models such as GPT-2 and FinBERT that are trained with a larger dataset with hyperparameter tuning. These increasing environmental and financial costs are mostly borne by marginalized communities that are less benefited by the introduced framework and more harmed by the negative environmental consequences from the electricity and \si{CO_2} emissions, as advised by Bender et al.~\cite{bender2021dangers}. Fortunately, this project largely benefits from pre-trained models and applies transfer learning to migrate models into the financial domain on Twitter, instead of retraining them from scratch. Using training logs from the Weights \& Bias training tracker~\footnote{Weights \& Bias is a developer tool for experiment tracking, available at \url{https://wandb.ai/site}.} and Python logging facility, approximately $30$ days of GPU computation is used for training classifiers and fine-tuning language models throughout this project, and all computations are done on a single NVidia GTX 1080 Ti graphics card. Using the Machine Learning Impact calculator, the estimated carbon emission is $77.76$ \si{kg} \si{CO_2}-equivalent\footnote{Estimations calculated by Machine Learning Impact Calculator~\cite{lacoste2019quantifying}, available at \url{https://mlco2.github.io/impact\#compute}.}, and may bring $77.76$ Global Warming Potential (GWP) to the Earth, which quantifies the warming brought by energy emission over the next $100$ years. Overall, DeepTrust is developed conscious of financial and environmental costs associated with model training, and utilizes computing resources efficiently to reduce carbon footprint whenever feasible. \smallbreak

\section{Future Works}

\subsection{Financial Information Retrieval from Multiple Social Media Platforms}

In addition to Twitter, the public may use other social media platforms to discuss or plan a mass collaborative strike to the stock market, causing an abnormal price movement out of blue. The most prominent example in 2021 is the story of GameStop\footnote{GameStop Corporation is a US chain gaming and trade-in retailers, traded on NYSE with stock symbol GME.}, in which a group of retail investors from Reddit purchase a massive amount of shares of GameStop, as a revenge to Wallstreet elites and hedge funds who have heavily shorted the company for its poor performance over past years. Although it is unexpected to most investors, there are several clues that can be found on the social media days before the financial anomaly happened. Weeks before the price rockets, well-informed individuals have started to widely post information on Reddit chatroom r/wallstreetbets about the \emph{Short Squeeze} action, which may trigger an unexpected stock price raise to force short-sellers to close their positions and iteratively driving up the stock price due to low-supply of tradable shares. The short squeeze action started on 22 January 2021, and drastically boosted the stock price from around \$$20$ per share to its peak at \$$469.2$ per share on 28 January 2021. This example shows that retail investors may use social media platforms other than Twitter to discuss financial information, or even plan a collective action that may impact the stock price severally with abnormal trading behavior. Therefore, the DeepTrust information retrieval module may be extended to include information from other social media networks such as Reddit and Discord, so to capture more relevant information that may inform financial analysts on the cause of abnormal pricing movements. \smallbreak

To upgrade DeepTrust and support processing information from multiple data sources, there are several major adjustments needed. Firstly, the information retrieval module needs to include additional handlers to communicate and extract information from the Reddit and Discord API, and the Pseudo-Relevance Feedback mechanism should also be customized for each portal. In addition, synthetic text filters should also be adjusted for different writing styles in each social media platform. For instance, it is uncommon to see hashtags and emojis on Reddit comparing to Twitter, and the writing styles may be different in each platform, so the SVM classifiers need to be fine-tuned separately for each data source. Lastly, Reddit is a forum in which the discussions under each standalone post are equally important as the original post itself. Therefore, a knowledge graph may be adopted to capture interlinked entities mentioned about a certain topic. \smallbreak

\subsection{Real-Time Reliable Financial Information Retrieval Framework}

Instead of understanding the cause of a financial anomaly in the past, the DeepTrust framework can be adjusted to support real-time market anomaly detection and explanation features, referencing the pioneering work done in Project Mosaic from Refinitiv Labs\footnote{Project Mosaic is a real-time analytical tool that assists financial analysts to identify and understand anomalous price movements.}. The key challenge would be completing the work of all three modules within a limited time frame, which demands a more sophisticated solution. \smallbreak

To fully support real-time analytics, all three modules in the DeepTrust framework require a certain amount of rework. Firstly, the AD module should include another mode of detection with machine learning models trained on historical pricing data, and continuously evaluate if a price movement is anomalous or expected. Low-latency prediction should be enforced so that whichever time interval (e.g., tick, 1 minute, 5 minutes, 1 hour) is used for the pricing data, abnormal events can be detected immediately for user actions. In addition, the IR module needs to be slightly adjusted to use recent search instead of full-archive search on Twitter, and pagination should also be properly configured to avoid including duplicate information into the tweet collection. Lastly, there are no adjustments needed for the RA module to support a low-latency assessment on tweet reliability as all components are implemented with a RESTful API, so pre-trained models are loaded using another spawned subprocess for generating predictions. Overall, the objective of this enhanced DeepTrust is to take a proactive approach to explain extreme price movements in real-time using reliable information extracted from Twitter. \smallbreak

\subsection{Weighted Reliability Score using Author Reputation}

In the present design, the feature-based filter primarily uses public metrics of tweets and their author to infer source credibility, while this design can be extended to a reputation-based credibility assessment engine that uses author metadata to infer tweet reliability with greater precision. For instance, the credibility analysis system proposed by Alrubaian et al. used hybrid-level feature extraction on both tweet and author information to calculate averaged sentiment score and popularity score based on event engagement, public metrics and historical tweets~\cite{alrubaian2016credibility}. Each feature is ranked based on its influence on perceived credibility, then used to calculate an aggregated credibility score for training a supervised classifier. The system was evaluated using an annotated dataset on the topic of Houthis movement in Yemen with $1.4$M tweets from $489$K authors, and has around \numrange{85}{95}\% accuracy over two testing datasets. A similar system TweetCred was proposed in Gupta et al., in which a set of $45$ features extracted from both tweet and author metadata are used to train an SVM classifier on classifying tweet credibility during the 6 famous chaotic events of 2013~\cite{gupta2014tweetcred}. \smallbreak

Therefore, DeepTrust can leverage influential features with prominent impacts on perceived source credibility, and enhance the existing feature-based filter into a comprehensive reliability assessment engine that assigns a weighted score to each author. A tweet can therefore be filtered based on a combination of its textual content and the weighted credibility score of its author, instead of relying on an absolute threshold on its public metrics. \smallbreak


\bibliographystyle{unsrtnat}
\small{
\bibliography{references}
}



\end{document}

%% file: includes.tex
\usepackage[a4paper,hmargin=2.0cm,vmargin=2.0cm,includeheadfoot]{geometry}
\usepackage{textpos}
\usepackage[numbers]{natbib}
\usepackage{tabularx,longtable,multirow,caption}
\usepackage{fancyhdr} 
\usepackage{url} 
\usepackage[english]{babel}
\usepackage{amsmath}
\DeclareMathOperator*{\argmax}{argmax} 
\usepackage{graphicx,accents}
\usepackage{dsfont}
\usepackage{epstopdf} 
\usepackage{mathtools}
\usepackage{backref} 
\usepackage{array}
\usepackage{latexsym}
\usepackage[pdftex,pagebackref,hypertexnames=false,colorlinks]{hyperref} 
\usepackage{subfig}
\usepackage{esdiff}
\usepackage{cleveref}
\usepackage{algorithm}
\usepackage{algpseudocode}
\usepackage{booktabs}
\usepackage{titlesec}
\usepackage[toc,page]{appendix}
\usepackage{adjustbox}
\usepackage{csquotes}
\usepackage[dvipsnames,table,xcdraw]{xcolor}
\usepackage{bm}
\usepackage{esvect}
\usepackage{stmaryrd}
\usepackage{setspace}
\usepackage{longtable}
\usepackage{siunitx}
\titleformat{\chapter}[display]{}{\filleft\scshape\chaptername\enspace\thechapter}{-2pt}{\filright \Huge \bfseries}[\vskip4.5pt\titlerule]
\titleformat{name=\chapter, numberless}[block]{}{}{0pt}{\filright \Huge \bfseries}[\vskip4.5pt\titlerule]

\titlespacing{\chapter}{0pt}{-15pt}{25.5pt}
\titlespacing{name=\chapter, numberless}{0pt}{16pt}{15pt}

\hypersetup{pdftitle={},
  pdfsubject={}, 
  pdfauthor={},
  pdfkeywords={}, 
  pdfstartview=FitH,
  pdfpagemode={UseOutlines},
  bookmarksnumbered=true, bookmarksopen=true, colorlinks,
    citecolor=black,%
    filecolor=black,%
    linkcolor=black,%
    urlcolor=black}

\usepackage[all]{hypcap}

\usepackage{color}
\usepackage[tight,ugly]{units}
\usepackage{float}
\usepackage{tcolorbox}
\usepackage[colorinlistoftodos]{todonotes}
\usepackage{ntheorem}
\theoremstyle{break}

\newcommand{\code}[1]{\texttt{#1}}

\def\@makechapterhead#1{%
  \vspace*{1\p@}%
  {\parindent \z@ \raggedright \sffamily
    \interlinepenalty\@M
    \Huge\bfseries \thechapter \space\space #1\par\nobreak
    \vskip 1\p@
  }}

\def\@makeschapterhead#1{%
  \vspace*{1\p@}%
  {\parindent \z@ \raggedright
    \sffamily
    \interlinepenalty\@M
    \Huge \bfseries  #1\par\nobreak
    \vskip 1\p@
  }}

\allowdisplaybreaks

\DeclarePairedDelimiter\abs{\lvert}{\rvert}%
\DeclarePairedDelimiter\norm{\lVert}{\rVert}%
\makeatletter
\let\oldabs\abs
\def\abs{\@ifstar{\oldabs}{\oldabs*}}
\let\oldnorm\norm
\def\norm{\@ifstar{\oldnorm}{\oldnorm*}}
\makeatother

%% file: titlepage.tex
\begin{titlepage}

\newcommand{\HRule}{\rule{\linewidth}{0.5mm}} 


\includegraphics[width = 4cm]{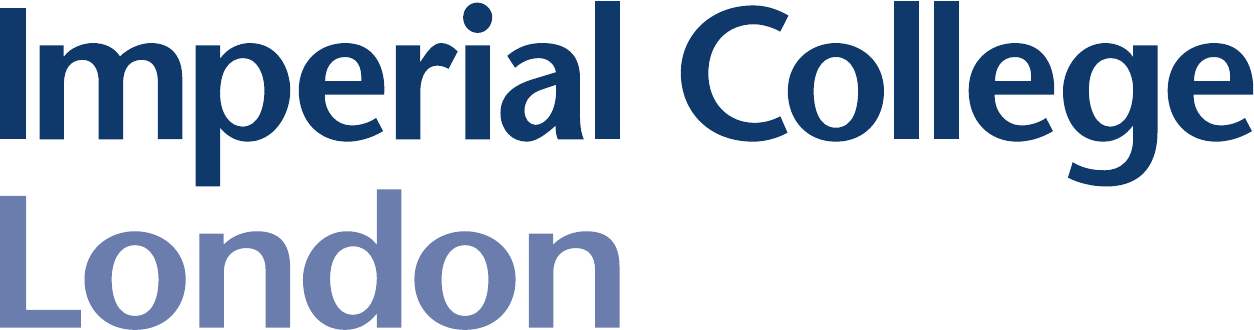}\\[0.5cm] 

\center 


\textsc{\Large Imperial College London}\\[0.5cm] 
\textsc{\large Department of Computing}\\[0.5cm] 


\HRule \\[0.4cm]
{\fontsize{19}{60} \bfseries \reporttitle}\\ 
\HRule \\[1.5cm]
 

\begin{minipage}{0.4\textwidth}
\begin{flushleft} \large
\emph{Author:}\\
\reportauthor 
\end{flushleft}
\end{minipage}
~
\begin{minipage}{0.4\textwidth}
\begin{flushright} \large
\emph{Supervisor:} \\
\supervisor 
\end{flushright}
\begin{flushright} \large
\emph{Second Marker:} \\
\secondsupervisor 
\end{flushright}
\end{minipage}\\[4cm]

\vfill 
Submitted in partial fulfillment of the requirements for the MSc degree in
\degreetype~of Imperial College London\\[0.5cm]

\makeatletter
\@date 
\makeatother

\end{titlepage}